Spring 5-2022

# Speaker Diarization and Identification from Single-Channel Classroom Audio Recording Using Virtual Microphones

Antonio Gomez

Antonio Gomez
*Candidate*

Electrical and Computer Engineering
*Department*

This dissertation is approved, and it is acceptable in quality and form for publication:

*Approved by the Dissertation Committee:*

Dr. Marios Pattichis, Chairperson

Dr. Ramiro Jordan

Dr. Sylvia Pattichis

Dr. Kim Linder

Dr. Manel Martinez-Ramon

# Speaker Diarization and Identification from Single-Channel Classroom Audio Recording Using Virtual Microphones

by

**ANTONIO GOMEZ**

BS, Electrical Engineering, Florida International University, 1986
MS, Engineering Management, Florida International University, 1997

DISSERTATION

Submitted in Partial Fulfillment of the
Requirements for the Degree of

**Doctor of Philosophy**
**Engineering**

The University of New Mexico
Albuquerque, New Mexico

**May 2022**

# DEDICATION

I would like to dedicate this dissertation to a very special person in my life, my wife Grace, for her support during this long journey. For all the hours, days, months, years she spent giving me the breath to move on, and for all the times she told me it is time to finish as well. For her, my eternal gratitude.

I would like also to dedicate this work to my children, Daniel, and Carolina, for understanding why sometimes I was not there with them. I hope this work would serve as an inspiration to them.

# ACKNOWLEDGEMENTS

I would like to sincerely acknowledge the labor of my advisor and dissertation chair, Dr. Marios Pattichis, for his continued support during all these years. Thank you for believing in me, Marios. I would have never done this without you on my side.

I would like also to thank my committee members, for taking the time to review and advise my work. Thank you, Dr. Kim Linder, for being always there as a friend.

Finally, I would like to say thank you to my managers at Honeywell and Sandia National Labs for their support and understanding.

**Speaker Diarization and Identification**

**from Single-Channel Classroom Audio Recording**

**Using Virtual Microphones**

By

**ANTONIO GOMEZ**

BS, Electrical Engineering, Florida International University, 1986

MS, Engineering Management, Florida International University, 1997

Ph.D. Engineering, University of New Mexico, 2022

## ABSTRACT

Speaker identification in noisy audio recordings, specifically those from collaborative learning environments, can be extremely challenging. There is a need to identify individual students talking in small groups from other students talking at the same time. To solve the problem, we assume the use of a single microphone per student group without any access to previous large datasets for training.

This dissertation proposes a method of speaker identification using cross-correlation patterns associated to an array of virtual microphones, centered around the physical microphone. The virtual microphones are simulated by using approximate speaker geometry observed from a video recording. The patterns are constructed based on estimates of the room impulse responses for each virtual microphone. The correlation patterns are then used to identify the speakers. The proposed method is validated with classroom audios and shown to substantially outperform diarization services provided by Google Cloud and Amazon AWS.

# TABLE OF CONTENTS

# LIST OF FIGURES

# LIST OF TABLES

# Chapter 1. Introduction

The field of speech processing, which includes speech recognition, separation, transcription, and enhancement, has undergone several transformational changes. Despite significant progress, speaker identification in crowded rooms continues to be a difficult problem. Crosstalk and large amounts of background noise make these environments particularly challenging. Most speaker identification and diarization systems rely on the use of Deep Learning methods that require pre-training on large datasets. Speech features such as formant frequencies, pitch contours, and coarticulation are extracted from the test samples and are eventually matched against a database of training samples [1]. The databases need to contain as many training examples as possible and should be updated periodically to maintain a proper performance level [2]. The accuracy of the identification depends on the size of the database: the bigger the database, the better the accuracy, but the longer the training times [3]. In addition to long training times, databases are prone to bias concerning spoken language and accent [4]. This biasing is usually unintentional and unconscious, and it is the product of the environment where the speech recognition system is developed [5].

The limitations of speech processing systems are more evident in challenging situations such as collaborative environments, meetings, or large-scale educational settings in general. These environments commonly consist of multiple speakers sitting around a table located inside a room. The speakers can take turns to speak, but it is not unusual to have two or more speakers talking at the same time. The environment can also be very noisy if we have numerous participants or groups inside the same room. These types of environments are too difficult for most speech processing systems, requiring in many cases heavy manual analysis. Manual diarization of meetings is a tedious and time-consuming task that requires many hours of processing, and it is subject to many interpretation errors.

There is a need in many educational research activities to understand how the classroom material engages the students. To understand how students interact, classroom sessions are recorded and transcribed. An important problem here is to determine which participant is speaking at a particular moment, what she or he has said, and for how long the participant spoke. Automated methods usually require multi-channel audio recordings and are prone to errors due to noise and crosstalk. Also, these systems have limitations in the number of speakers they can process, as well as the length of the audio segments. While diarization systems do not require enrollment of the speakers, they can only generate abstract labels of the speaker that is active in an audio segment. On the other hand, speaker identification systems can provide non-abstract labels by enrolling the participating speakers. The enrollment process consists of each speaker providing several seconds of noise-free speech without crosstalk. This requirement cannot be met when the data consists of audio recordings of busy meetings with noisy backgrounds. It is thus important to develop speech identification and diarization methods that do not impose any requirement to pre-enroll the speakers.

This dissertation aims to provide the foundations of a new approach to speaker identification and diarization using virtual microphones and spatial information. Simulations are never perfect, but this work shows that it is possible to use an approximation of a real room geometry to obtain the acoustic parameters necessary to simulate reception in a virtual array of microphones and use these simulated signals for speaker diarization and identification. The simulation is based on a physical model that requires no databases, it is independent of the spoken language or accent of the participants, it does not require prior speaker enrollment, and it presents high immunity to noise.

The proposed approach relies on the fact that discriminant information about the 3D geometry of each speaker is embedded in the recorded audio from a single microphone. The basic idea is to recognize speakers using acoustical simulation. As part of the simulation process, the proposed method computes the Room Impulse Response (RIR) for each of the microphones and the speakers and simulates the reception on each of the virtual microphones. The accuracy of the process of computing RIRs is verified through real-life measurements of the correlation patterns. Based on the simulated reception over the virtual microphones, the method computes correlation patterns among the virtual microphones. The recorded audio is then also used to generate different correlation patterns based on hypothesized speaker locations. A classifier is applied to the generated correlation patterns to select the most likely speaker location.

This approach has several advantages. First, we do not require databases of speech. Our system is based on physical models that are unique to the scene we are analyzing.

Because we do not have databases to train the model, our system requires capturing only about 1 to 2 seconds of audio from each speaker for both training and recognition. In contrast, state-of-the-art systems require tens of seconds of clean audio for training and several seconds of identification. Second, our system has been conceived to operate in noisy environments where microphone arrays and cross-correlation analysis have been proven to be efficient methods for speaker discrimination [6],[7]. Third, the simulation does not require multi-channel audio, but it uses a single channel recording as a reference for the simulation. Finally, our system can run on simple computers without the need to access remote computer clusters or databases.

## 1.1 Motivation

This work is motivated by the need for a reliable non-manual method of assessing the level of engagement of the students participating in the Advancing Out-of-school Learning in Mathematics and Engineering (AOLME) program at the University of New Mexico (UNM) [8]. AOLME is a collaborative learning environment where students are introduced to STEM subjects such as integrating computer programming and middle school mathematics. It forms part of the educational research activities performed at the University of New Mexico's Image and Video Processing and Communications Lab (ivPCL) [9]. AOLME sessions are video recorded for later analysis that includes students' participation and overall level of attention, as well as the facilitator's interaction with the students. The analysis consists of evaluating the activities of the participants such as hands or head movement, use of keyboard and mouse, lip movement, etc., and transcription of the sessions to determine the time when a participant is speaking and for how long. Detailed participation statistics for each participant are currently not available because manually measuring talking times is time-consuming and plagued with errors. AOLME organizers have tried several transcriptions systems currently available in the market or open-source code, all without much success. The AOLME environment is extremely challenging for any speech recognition and transcription system due to multiple groups talking at the same time and the presence of background noise and echo. Hence, there is a need for a robust system that can overcome the limitations of the current state-of-the-art methods and complements the ivPCL methods, with the application to process hundreds of hours of video recordings.

AOLME video analysis presents other challenges in addition to the presence of multiple speakers and noise. First, these videos were taken with a simple video camera using a single microphone located at the meeting table. Budget limitations restrict the purchasing and use of more advanced equipment with multi-channel audio recording capabilities. Second, there are already hundreds of hours of these video recordings that need to be analyzed. There was no previous speaker enrollment that could be used to train a speaker identification system. Furthermore, most participants only speak for several seconds at a time, which makes the identification process more difficult. Even if for future sessions it is possible to record multichannel audio and enroll the speakers, there is the need to process the existing videos, therefore the need for a flexible method that can handle new and existing recordings.

## 1.2 Related Research

Assessing the level of engagement of participants in collaborative educational sessions requires the application of tools to extract relevant information from audio and video data. This information is then interpreted and translated into statistical data for the researchers. For this end, these tools can either identify activities in a video scene that are related to attention behaviors (e.g., typing and writing), or they can identify the active speaker or speakers in an audio segment. Related work to this dissertation includes both types of tools.

In the area of activity tracking, it is important to mention the work by UNM's ivPCL lab in direct connection with the AOLME program. Darsey [10], analyzes video using color and optical flow for tracking hand movement. Teeparthi et al. [11], [12], presents fast methods of video analysis for hand and object tracking as well. Jacoby et al. [13] works in human activity detection using context-sensitive approaches, while Jatla et al [14] uses 3D Convolutional Nets. Eiliar et al. [15] provides a maintainable open-source activity system.  Detection of attention traits is investigated by Shi et al. [16], using AM-FM models to detect head direction and group interactions.

Research on speech processing covers a vast area containing different topics. Under the umbrella of speech processing, we find speech identification, speech enhancement, speaker verification, speaker diarization, and speaker identification, among others. This dissertation focus on speaker identification and diarization as part of the research labor of the ivPCL lab and AOLME programs.

Speaker identification is the process of recognizing the identity of a speaker or several speakers present in a speech segment. Speaker diarization is the process in which an audio recording that contains several speakers is dissected into segments that contain only one speaker at a time [17]. Speaker Diarization is often defined as "who said what, and when", and for "how long". Both Speaker Identification and Speaker Diarization are important mechanisms in many audio-processing tasks.

Most of the research on speech processing nowadays is focused on the use of Artificial Neural Networks and Deep Learning. Deep Belief Networks (DBN) are widely used in speech recognition [18], [19]. X-vectors are considered today state of the art in speaker recognition [20]. X-vector methods outperform classic i-vector methods in the order of 9.23%, and they have been tested with datasets such as VoxCeleb, NIST SRE 2016, and SWBD [21].

The research in this dissertation focuses on the use of spatial information and virtual microphone arrays for speaker identification and diarization. There is no attempt to cover methods that do not use spatial information or virtual microphone arrays for speaker identification and diarization. Although not as extensive as the neural network and deep learning research, it was possible to find numerous works that demonstrated the use of spatial information for speaker identification and diarization, as well as applications of virtual microphone arrays for acoustic signal enhancement and meeting diarization. The references to these works are presented in the next sections.

Most literature regarding the application of microphone arrays (multichannel audio) and beamforming is related to the implementation of spatial filters to improve the signal-to-noise ratio (SNR). Nevertheless, several researchers found ways to exploit the operational principles of microphone arrays and apply them to speaker identification and diarization. Xavier Anguera et al. [22] propose the use of beamforming algorithms as the forefront of a speaker diarization system. These beamforming algorithms take advantage of the environment commonly encountered in meetings, such as multiple microphones, to enhance a single signal of interest. Anguera et al. optimize a conventional delay and sum beamforming array to operate under the constraints of an unknown number of speakers, unknown location of both speaker and microphones, and microphone mismatching. The Time Differential of Arrivals (TDOAs) of the microphones are calculated by cross-correlation. Diarization is accomplished by agglomerative clustering where each cluster is modeled via a Gaussian Mixture Model (GMM). A separate set of GMMs is used to model the TDOA features.

In a similar manner as Anguera, Mitianoudis et al. [23] propose the use of beamforming in parallel with Independent Component Analysis (ICA) for audio source separation. The ICA for source separation requires knowledge of the parameters of the mixing matrix. If these parameters are not known, then the separation problem becomes a Blind Source Separation problem (BSS), which is an ill-posed problem (multiple solutions). Mitianoudis et al. propose the use of the directivity pattern of beamforming (use of phase information) to select signals among different possible permutations.

Both previous authors exploit the phase information of signals captured by microphone arrays. In my research, I also exploit the phase information (TDOA between microphones) as a means of determining the relative position of the active speaker and thus the identity. The previous work shows the use of cross-correlation to calculate the TDOA. Klein et al. [24] study the performance of the multi-channel cross-correlation (MCCC) coefficient method as a robust solution to calculations of TDOA under noisy and reverberant environments. Padois [25] studies the performance of time-domain beamformers based on the generalized cross-correlation functions. Padois generates a sound source map by interpolating the cross-correlation function between microphones, to generate a two-dimensional hyperbola of the spatial likelihood function. The number of hyperbolas corresponds to the number of microphones used in the array. The source position can be determined by averaging the hyperbolas and determining their maximum value as the intersecting point for the location. In general, the experimental results show that resolution improves with the number of microphones, up to a number where the performance seems to plateau.

Pasha et al. [26] present work that is closely related to our research on RIR and room geometry for TDOA estimation. Pasha et al. propose a method of source localization that utilizes RIRs amplitudes to fit a TDOA surface and an amplitude surface across a room of known geometry. The RIR is obtained from a set of microphones of an unknown location. The RIR amplitudes of the direct path impulses are higher and have a shorter relative time of arrival for the signals that are closer to the receiving microphone. The area with the maximum amplitude and minimum delay is considered the estimated source area. The center of these areas is the estimated source location. Similar work was previously presented by Tervo et al [27], but, instead of source location, this work focuses on localization of acoustic reflections using the combined TOA and the TDOA information contained in the RIR.

All the work presented so far takes advantage of the properties of beamformers, TDOA, TOA, DOA, and cross-correlation, but also requires an array of physical microphones. In this dissertation, the method depends only on the information captured by a single microphone. Research material on single microphone acoustic separation based on spatial information is more limited, as well as work on virtual microphone arrays for the same purpose. Nevertheless, there is interesting work that provided useful information for my work. Perhaps the closest work to my research that I found is presented by Hu et al. [28]. Hu et al. propose a method to utilize the reverberant information, known as the Direct-to-Reverberant Ration (DRR), from a single channel recording for Speaker Diarization. Hu et al. estimate the DRR using the algorithm from Peso Parada et al. [29] and combine it with a Mel-Frequency Cepstral Coefficient (MFCC) diarization method proposed by Vijayasenan et al. [30]. The principle is to use both MFCC and DRR features in combination so a trained system can perform a clustering type of classification. The estimates for the DRRs are computed using features such as Signal-to-Noise ratios,

MFCCs, power spectrum, and zero-crossing rates. It is important to notice that this work was tested only using simulated meeting recordings and assumes that the speakers are stationary.

Because the research work in this dissertation proposed virtual microphone simulations, it is necessary to present some relevant work in this area. Yoshioka et al [31] describe a way of linking several recording devices, such as laptops or mobile phones, to create a virtual microphone array. Once the link has been established, the multi-channel audio can be used for speaker diarization. Yoshioka et al. claim to achieve a 13.6% diarization rate when 10% of the speech duration contains more than one speaker.

The Yoshioka et al. approach is very innovative but requires the presence of several recording devices in the meeting room. More aligned with this dissertation is the work of Katahira et al. [32], Del Galdo et al. [33], and Izquierdo [34]. Here the authors propose methods to simulate arrays of microphones by interpolating the signal received by two physical microphones. The authors demonstrate that the virtual microphone arrays improve the SNR in reverberant environments, hence their potential application for speech processing devices. Even though these methods succeed in emulating a set of virtual microphones, they need at least two physical microphones as “seed”, which are not available for the method presented in this dissertation.

Finally, Tapia et al. [35] presented a bilingual speech recognition method inspired in the research presented in this dissertation. Tapia utilizes still video frames to estimate the approximate geometry of the speakers and simulate the center microphone reception using Pyroomacoustics. The simulated audio is used along with ALOME transcriptions to generate the training sets for a convolutional neural network.

## 1.3 Thesis Statement

The main objective of this dissertation is to develop a method that applies spatial information and virtual microphone arrays to identify multiple speakers in a single channel audio recording of a collaborative environment and provide activity statistics of each of the participants.

This method is aimed to succeed in challenging environments with multiple active speakers and background noise, conditions that make the current state-of-the-art methods perform poorly. For this purpose, the work in this dissertation presents the implementation of an acoustic model based on a virtual room of rough similar geometry to the actual acoustic scene being analyzed, and then the simulation of the signals received by a virtual microphone array located in the virtual scenario. The signal delay between each virtual microphone represents the relative physical position of the active source that in this case is each speaker. The research goal is to find a suitable way to extract the spatial location embedded into a single channel recording to implement the model and subsequent virtual microphone array.

## 1.4 Contributions

The contributions expected from this work include:

- A method to identify speakers in a collaborative environment by extracting spatial information from a single channel audio recording utilizing an acoustic simulation and virtual microphones.
- A solution for the limitations of current state-of-the-art speaker identification methods concerning:
    - Multiple speakers
    - Speaker gender or accent
    - Background Noise and reverberation.
- Development of speaker identification framework that is based on an explainable model developed in terms of the physical characteristics of the problem, and hence does not require large datasets to train many parameters.
- The basis for a tool for quantitative analysis of video recordings for assessing the level of interaction of participants in collaborative environments.

## 1.5 Dissertation Overview

This dissertation is divided into 6 chapters that cover background theory and other related work, a description of our method, experiments, and results, and conclusion and recommendations for future work. The dissertation is presented as follows:

- Chapter 2 gives a background of audio spatial theory and its applications in speaker diarization and identification, and how they functionally compare with other state-of-the-art methods.

- Chapter 3 presents the foundations on which the proposed method in this Dissertation is based and a block diagram of its implementation.
- Chapter 4 describes the practical implementation, including software, model implementation, simulation ions, video analysis, and audio segmentation.
- Chapter 5 presents the experimental results obtained when analyzing audio under controlled and uncontrolled environments, and the experimental comparison of our method against current Google and Amazon speaker diarization methods.
- Chapter 6 presents a summary of this dissertation and possible future work.
- Appendix A contains the scripts and pseudo-code for the Python implementation of Pyroomacoustics.
- Appendix B presents the most important LabVIEW Sub-VIs front panels and block diagrams.
- Appendix C contains the specifications of the equipment used in the audio laboratory.

# Chapter 2. Background

This chapter introduces the principles that form the foundations that define the method described in this dissertation. The section begins with basic acoustic theory, concepts, and definitions, and continues with a presentation on microphones and microphone arrays. It finalizes with an introduction to methods for speaker diarization and identification, covering both classic methods and Deep Learning methods.

## 2.1 Acoustics Principles

The perception of sound by a sound capturing device (e.g., a microphone or human ear), not only depends on the characteristics of the sound source, but it also depends on the medium where the sound propagates, the physical environment where the sound source is located, and the relative locations of the capturing devices and the source. This dissertation, considers all these factors to create models that represent the environment where the sound sources, i.e., the speakers, are active.

Chapter 1 presented a brief introduction to the AOLME program. The AOLME video recordings were taken inside rooms where the participants gather in groups sitting around tables. The exact geometry of the room is unknown, but the video recording provides clues about the location of the speakers, the separation between them, their physical height, and the location of the recording microphone. These clues can be used for modeling a virtual room which can be defined as a three-dimensional enclosed space where the acoustic event takes place. This virtual room may not be necessarily the whole space where all the AOLME participants are, but it can be the space surrounding the participants in a single table. The approximate geometrical and physical characteristics of the virtual room allow us to emulate the reception on arrays of virtual microphones.

### 2.1.1 Sound Propagation: Near and Far Fields

Consider an acoustic source such as a person speaking, a stereo system playing a song, or a running ventilation fan. Sound from these sources propagates in the form of circular air pressure waves, away from the source. They can propagate in all directions if the source is in an open field (Fig. 1a), or directionally if the source is in proximity to a non-conducting medium such as a wall (Fig.1b). In acoustic theory, the relative location of a source to a point in space determines its field location. A source is in the near field if its distance to a point is less than one wavelength of the acoustic signal it is emitting. Sources that are located at distances greater than one wavelength are located at the far-field. The field location of a source plays an important factor when modeling the perception of sounds wave at a point in space.

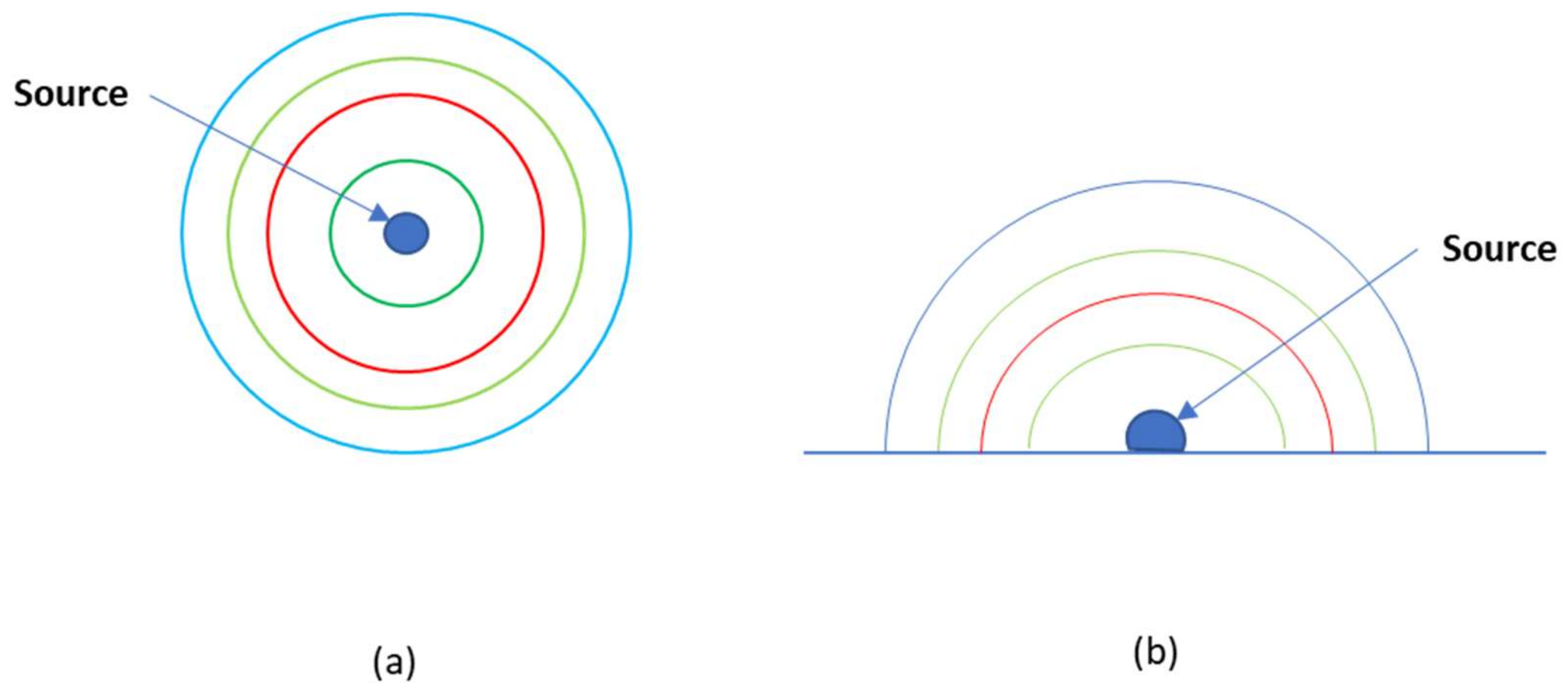

**Figure 1: Propagation of Sound Waves. (a) Free Field. (b) Directional.**

Fig. 2 shows a representation of the near field, the transition zone, and the far-field. In the near field, the sound waves behave turbulently, with more circulation than propagation. At about a distance of one wavelength from the source, the sound waves begin transitioning into propagation. At more than one wavelength, sound waves mostly propagate into the infinite. A point located at the near field perceives the sound waves as circular while one located at the far-field will consider these waves planar [36].

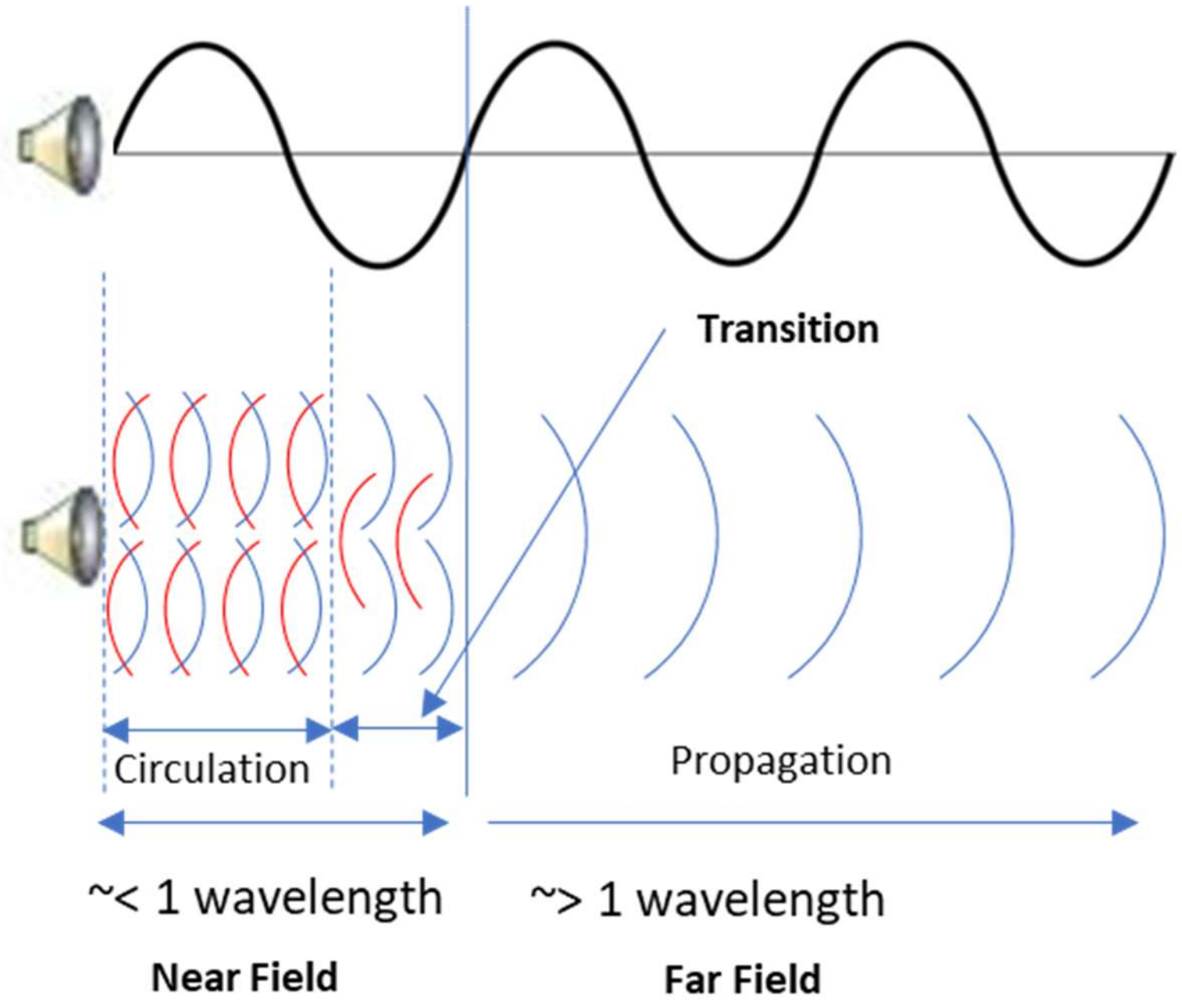

**Figure 2: Near and Far-Field Areas.**

### 2.1.2 Sound Propagation: Direct Path, Reflections, and Reverberation

The perception of sounds varies depending on whether the listener is located inside a theater room, small dormitory, or an open field. These differences in perception are the result of the behavior of the sound waves when they propagate across a medium. To visualize this phenomenon, consider for example a room where there is one acoustic source *S* (a person speaking) and one microphone *M*, as represented in Fig. 3.

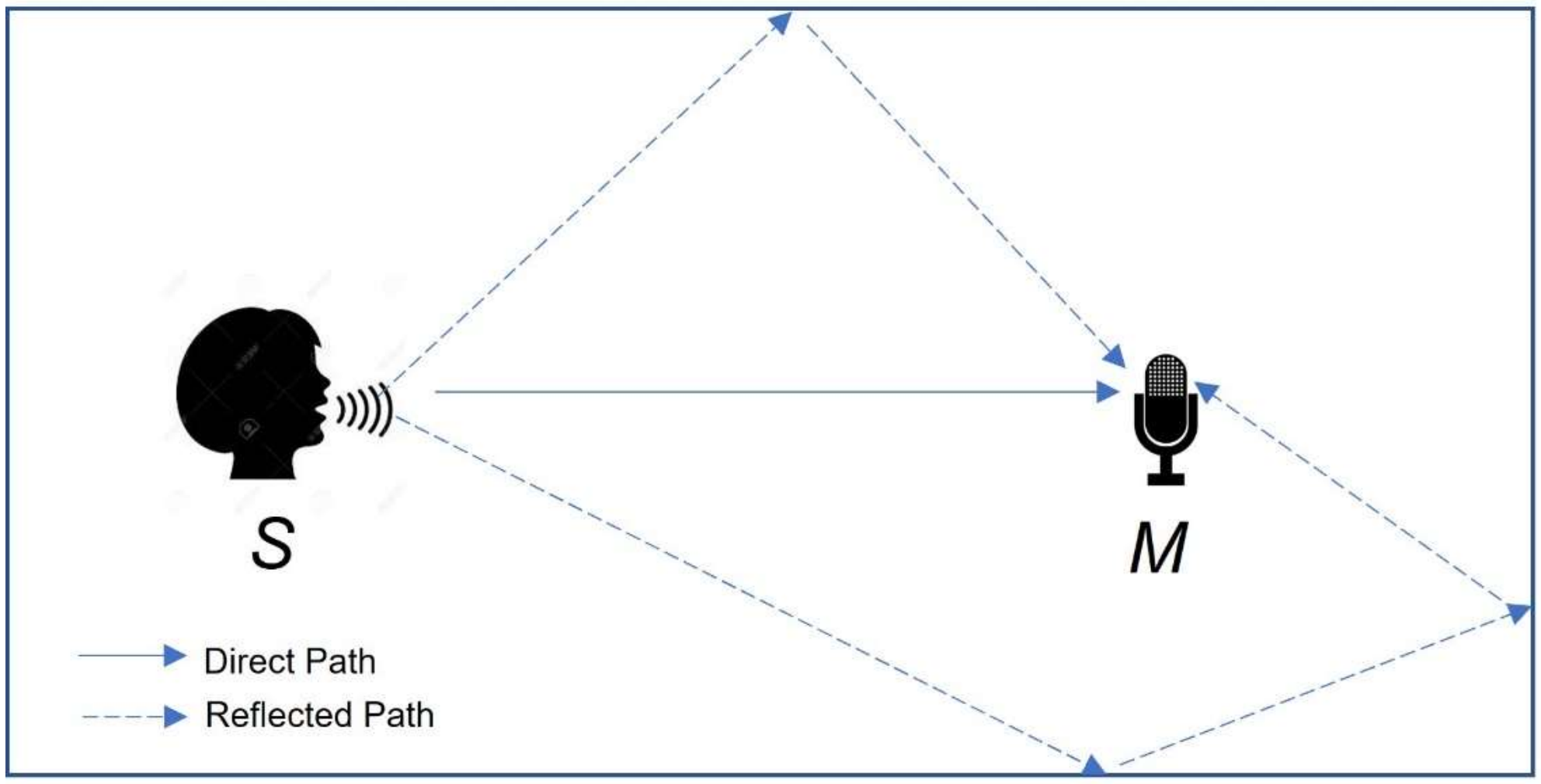

**Figure 3: Direct and Reflected Paths for Sound Propagation in a Diffuse Field.**

In Fig. 1(a) and Fig. 1(b), sound from the source will reach an observer or receiver directly, from one direction without reflections. In this case, the source is said to be in an acoustic *free field*. Fig. 3, in contrast, represents a *diffuse field*. In this case, the sound reaches the microphone from more than one direction due to reflections. As in the free field, the direct signal received at the microphone is characterized by the distance from the source to the microphone. This distance determines the sound pressure at the microphone, and the time it takes from the sound wave to reach the microphone. This time is known as the Time of Arrival (TOA), and it is a function of the speed of the sound in the room and the Euclidian distance from the source to the microphone. Each reflection contributes similarly.

The signal received at a microphone can be expressed in mathematical terms. If we consider a signal *s(t)* from an acoustic source located in the far-field, this signal is captured by a microphone as a signal *x(t)* that is the convolution of the Room Impulse Response (RIR) *h(t)* with additive noise *w(t)* as given by:

$$x(t) = s(t) * h(t) + w(t) \qquad (2.0).$$

The RIR is unique for every two points in the room and depends on the geometry of the room, the absorption of the materials in the room, and the frequency of the sources [37]. The RIR consists of three parts: the direct path, the early reflections, and the late reverberations. The direct path component is determined by the Euclidian distance of the source to the microphone, and it is a function of the Time of Arrival (TOA) or the time it takes for the signal to travel from the source to the microphone. The other two components of the RIR are related to the reflections of the sound waves at the walls and objects in the room. The early reflections usually arrive 5 ms after the direct path. The late reverberations arrive 20 or 30 ms after the early reflections begin.

The RIR can then be expressed as the summation of each of the impulse responses corresponding to the direct path and the reflections:

$$h(t) = \sum_{k=1}^{K} h_k(t) + m(t) \qquad (2.1),$$

where *K* is the number of reflections, *k* is the index number of the reflection, and *m* is the measurement noise. The RIR lasts until the reverberation energy decays to 60 dB on what is known as the $T_{60}$ time. The $T_{60}$ was calculated empirically by Sabine in 1890 and can be expressed as:

$$T_{60} = \frac{55.25 \cdot V}{c \cdot S \cdot a} \qquad (2.2),$$

where $V$ is the total volume of the room, $c$ is the speed of sound, $S$ is the total surface of the room, and $a$ is the absorption coefficient of the room (0 to 1). The reverberations are characterized by the frequency of the sources but, in the case of the early reflections, this influence is minimum [27]. Fig. 4 depicts a representation of a RIR with its three components.

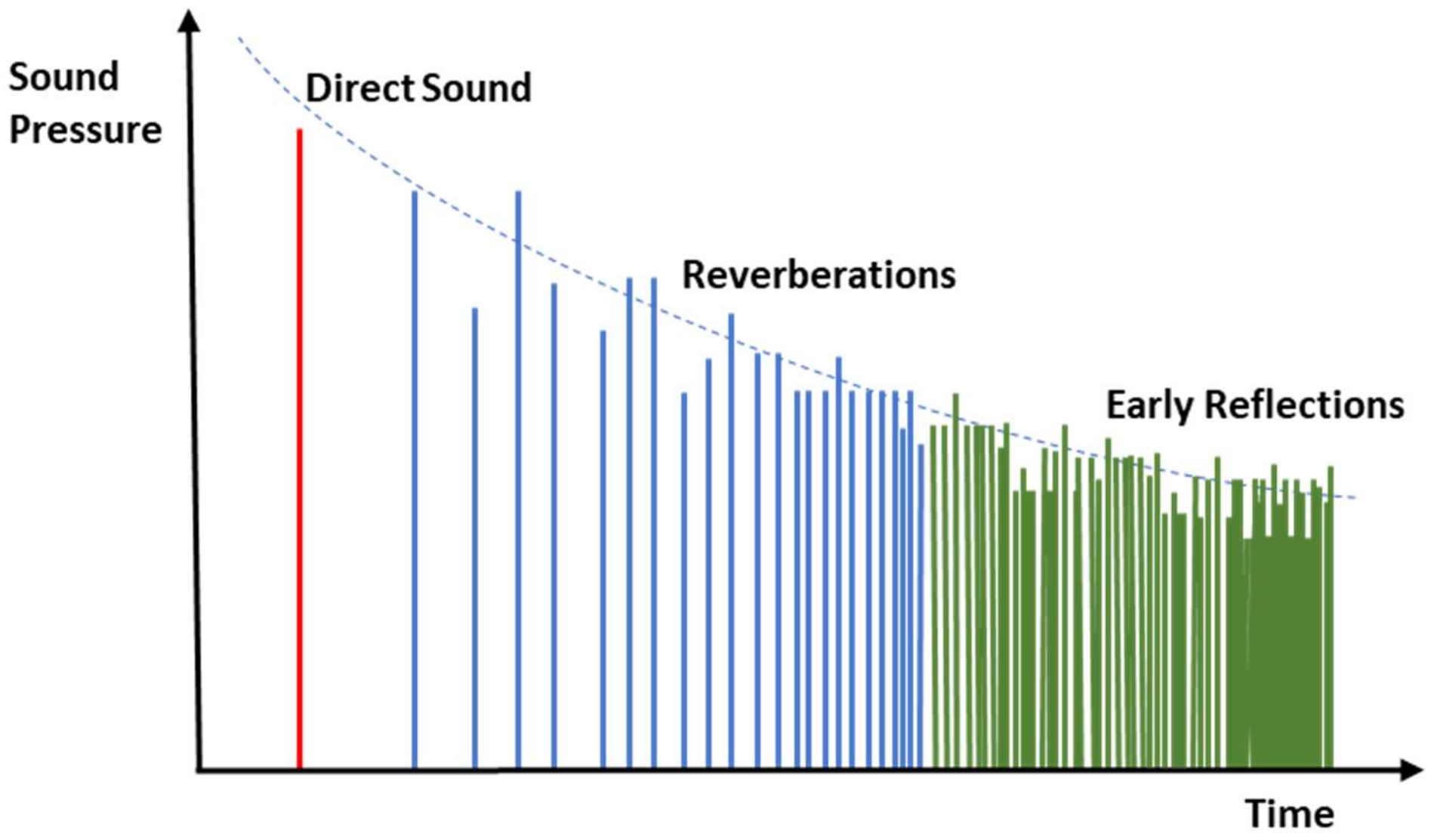

**Figure 4: Representation of the Room Impulse Response and its Components**.

The path of the reflections from the walls can be represented as direct paths coming from imaginary sources called *Images*. The signal at any microphone would be then represented by the number of contributing sources plus their image reflections. All acoustic reflections are subject to a TOA that depends on the distance of the path of the reflection. Section 2.3 presents more detail on the concept of acoustic images and their role in room simulation.

## 2.2 Microphones and Microphone Arrays

The previous section introduced microphones as devices capable of capturing sound. In general terms, microphones are sensing devices that detect changes in air pressure and convert these changes into electrical signals. Microphones are categorized by their electrical conversion type and their directionality pattern. Deep technical details for each type of conversion and directionality pattern microphone are out of the scope of this dissertation. The dissertation will only consider the type of microphones used during the research.

### 2.2.1 Classification of Microphones

This research used two types of physical microphones: Condenser omnidirectional, and condenser cardioid. The condenser term refers to the type of electrical conversion of the sound, and the terms omnidirectional and cardioid refer to the directionality of the microphone.

Condenser microphones work by utilizing a variable condenser that detects the air pressure changes. The change in pressure translates into a movement of the plates of the condenser thus changing its capacitance. The changes in capacitance are measured by the changes in the charging current in a circuit. Condenser microphones are also known by the name of *electret*. They are the most popular type of microphones today.

The directivity pattern of a microphone determines its gain or sensitivity according to the direction of the incoming sound. Omnidirectional microphones are equally sensitive to incoming sound from any direction. These microphones are simple pressure sensing devices or *acoustic monopoles*. Cardioid microphones are also known as pressure gradient microphones, and they are characterized for a directionality pattern that is like a heart (hence their name *cardioid*, from the Greek "heart"). Fig. 5 shows the typical directivity pattern for omnidirectional (a) and cardioid (b) microphones [38]. All AOLME videos were recorded using Audio-Technica ATR3350 condenser omnidirectional microphones.

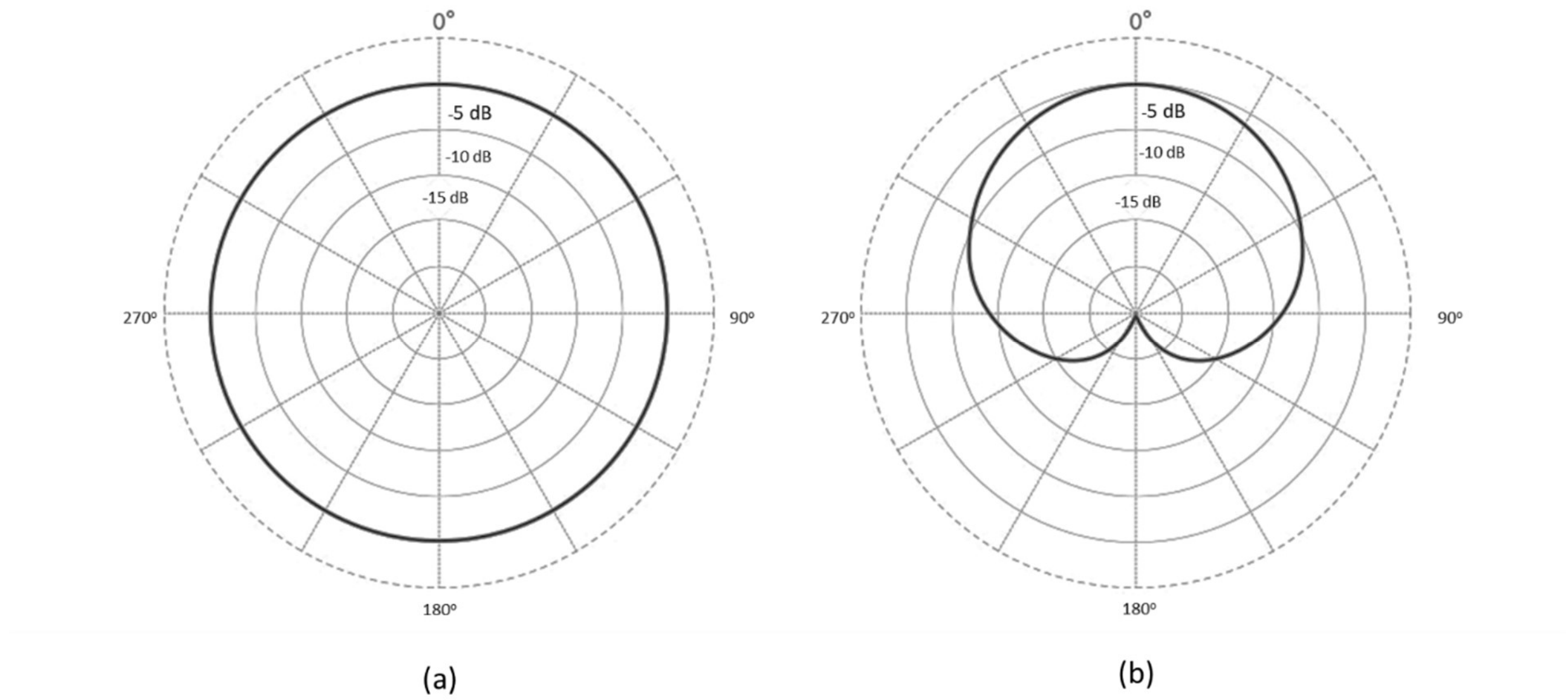

**Figure 5: Polar Pattern Plot of Directivity of Two Types of Microphones: (a) Omnidirectional. (b) Cardioid.**

Regardless of their type., all microphones generate noise. The conversion of sound pressure waves into electrical signals carries electrical noise, which has a flat spectrum [39]. Manufacturers usually indicate the electrical noise of their microphones in a Signal to noise ratio (SNR) number at a certain sound level. Appendix C contains the technical specifications of the microphones used for this research.

### 2.2.2 Microphone Arrays

When two or more microphones are arranged into a geometric pattern, they become a microphone array. Microphone arrays have important functional properties that are of interest when capturing sound in noisy environments, or when directionality is needed to discriminate between sound sources. An important part of the results of our research is based on the functional characteristics of microphone arrays.

Microphone arrays allow for the incorporation of spatial dimensionality to sound capturing. The difference between the signals captured by any two microphones separated a distance *d* provides information that can be used for source localization, tracking, and general noise reduction.

Microphone arrays can be expressed mathematically by (2.3)

$$\boldsymbol{x} = s\boldsymbol{d} + \boldsymbol{v} \qquad (2.3),$$

where $\boldsymbol{x}$ represents the vector of all microphone signals, *s* is the source audio signal, $\boldsymbol{d}$ is the propagation vector represented in (2.4), and $\boldsymbol{v}$ is the additive noise [40]. The vector $\boldsymbol{d}$ is expressed by (2.4)

$$\boldsymbol{d}(f) = [a_1 e^{-2\pi f \tau_1} \ldots . a_n e^{-2\pi f \tau_n} \ldots . a_N e^{-2\pi f \tau_N}]^T \qquad (2.4),$$

where $a_n$ represents the attenuation factor $1/_{d_s}(n)$ , $\tau_n$ is the channel delay $d_s/_c(n)$ and $d_s(n)$ is the distance between the source and a microphone n, with *c* the speed of sound.

The fine details of the theory behind microphone arrays are out of the scope of this dissertation. Nevertheless, it is important to have a basic knowledge of the properties of microphone arrays due to their applications in source localization, spatial filtering, and source separation. All of these are applications related to this research and will be discussed later in this section.

#### 2.2.2.1 Microphone Arrays Configurations

The possible geometries of microphone arrays are infinite. These different geometries are guided by the number of microphones that can practically be allocated to an array, and the type of acoustic scenario the array is intended to operate. The most common types are *linear*, *circular*, and *volumetric (3D)* [41].

*a) Linear Microphone Arrays:*

In this type of array, the microphones are linearly arranged. Fig. 6(a) represents a five-microphone array with a separation of 0.05 m between microphones. This array configuration is very popular, and it is designed to capture the sound that is in front of it. This type of array cannot distinguish from sounds that are coming from the same angle to the axis of the array, as the sound waves will arrive at the microphones with the same time delay. Fig. 6(b) shows the directivity pattern of the microphone array of Fig. 6(a), calculated at 450 Hz with the speed of sound $c$ = 343 m/s.

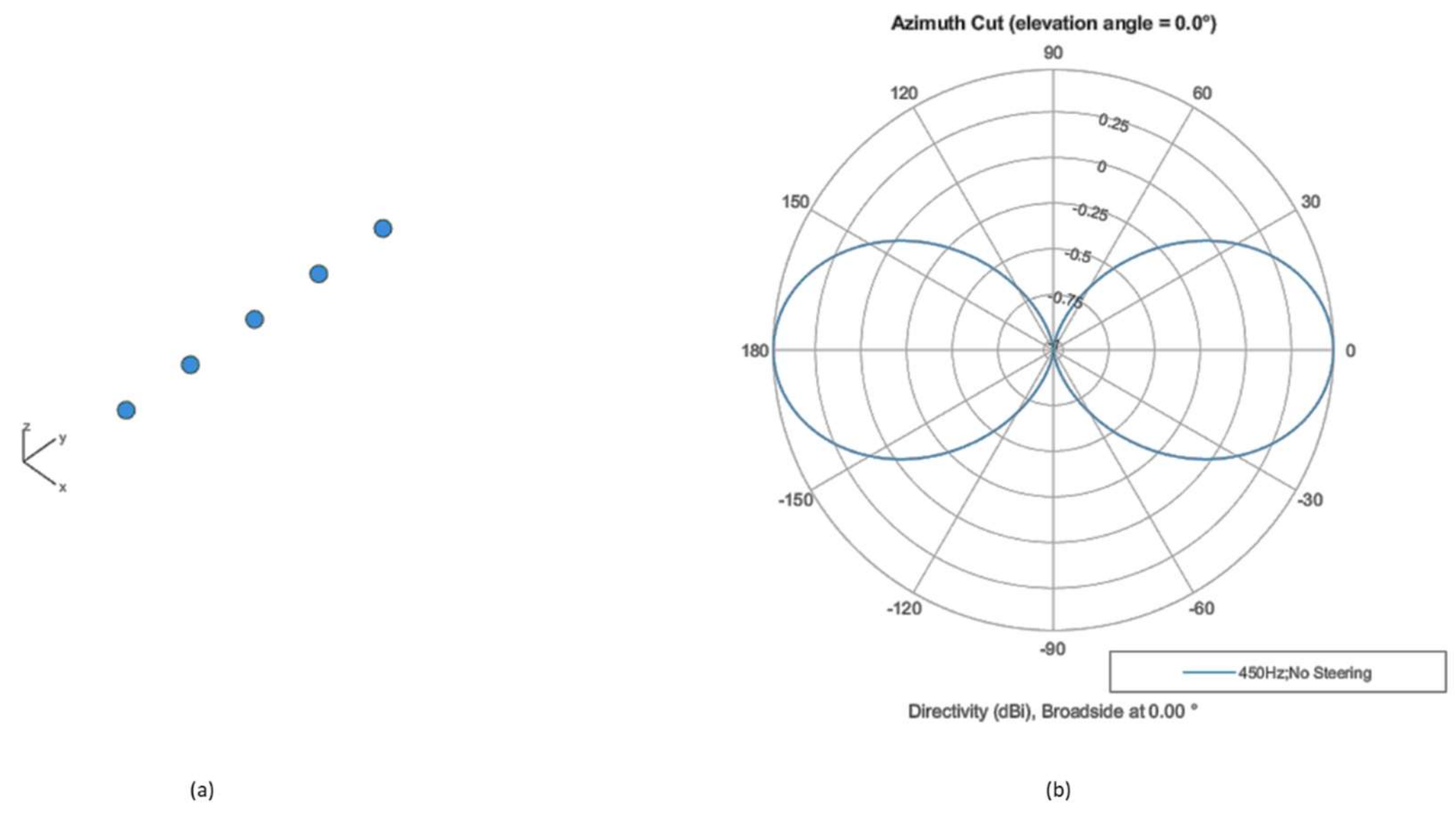

**Figure 6: Linear Microphone Array. (a) Geometry. (b) 3D Directionality Pattern.**

A variant of this type of array is a cross-linear array, also known as a planar microphone array. This type of array consists of two linear arrays perpendicular to each other, as shown in Fig.7. This is the type of array used in this dissertation for virtual microphone simulations.

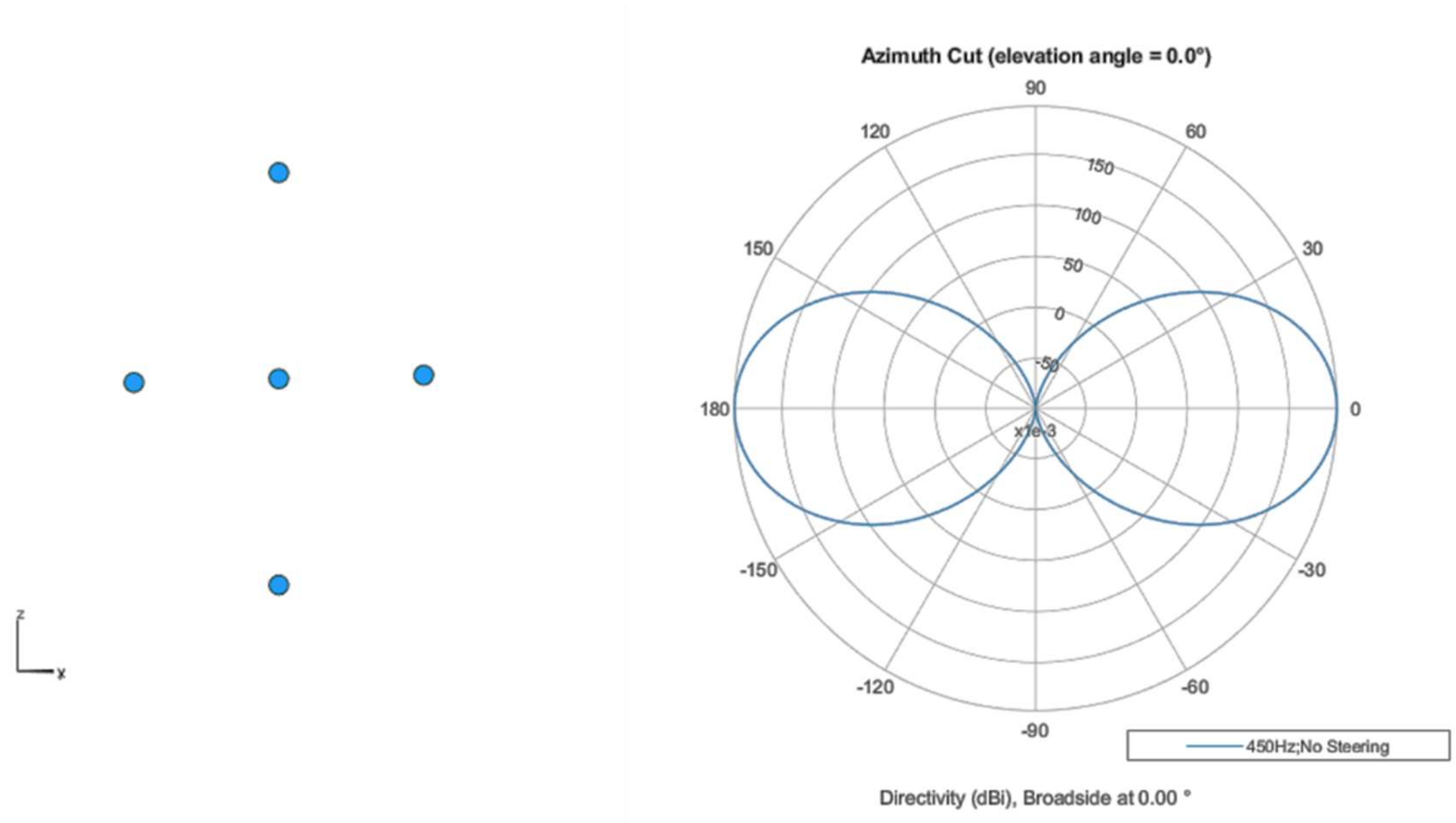

**Figure 7: Cross-Linear Array and Azimuth Directivity Pattern.**

*b) Circular Microphone Array:*

This microphone array has its elements positioned circularly. They can consist of one circle, or several concentric circles, as shown in Fig. 8. This type of array is commonly found in conference equipment that is in the center of a meeting table.

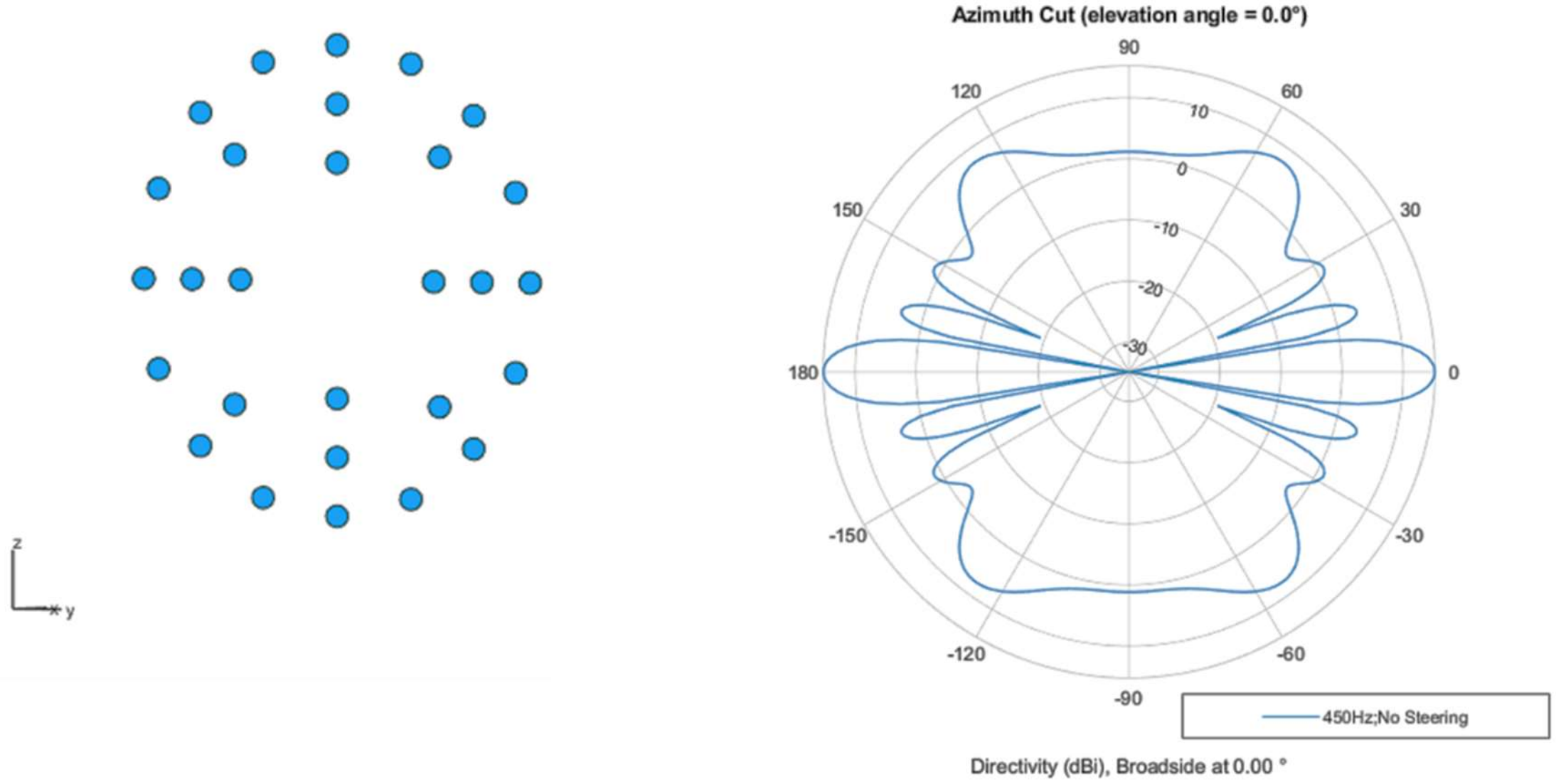

**Figure 8: Circular Microphone Array and Directivity Pattern.**

*c) Volumetric (3D) Array:*

This type of array forms a lattice with its elements, as shown in Fig. 9. They can capture sound from any direction, for as long as they are "suspended in the air" with no other interference. Their shape can vary as cubes, spheres, or cylinders.

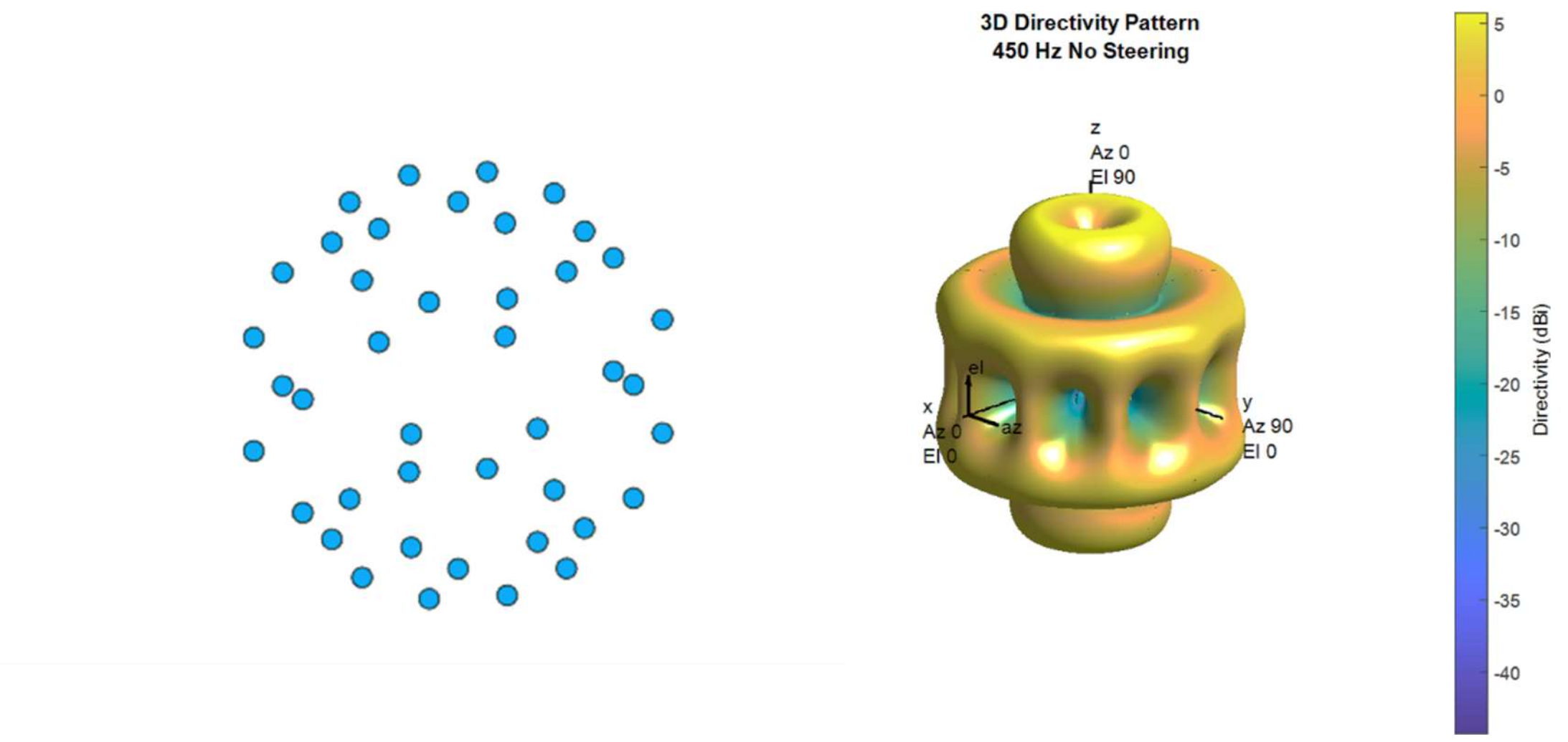

**Figure 9: Volumetric Microphone Array and 3D Directivity Pattern.**

#### 2.2.2.2 Spatial Aliasing

Signal aliasing occurs when the sampling frequency is less than twice the largest signal frequency component. When the bandwidth of the signal is greater than half of the sampling frequency, spectral overlapping happens.

Spatial aliasing occurs similarly. To reconstruct a spatial signal from a set of samples, it is necessary to have a spatial sampling period that is less than half of the signal wavelength [42]. In microphone arrays, the phase difference between two microphones should be less than $\pi$ to avoid spatial aliasing [43]. This constraint means that given a signal of frequency $f$, there is maximum distance $d$ between microphones before spatial aliasing occurs, and vice versa. For an audio signal of wavelength $\lambda$, this distance is half of the wavelength:

$$d \leq \frac{\lambda_{min}}{2} \qquad (2.5),$$

which translates to a maximum frequency of

$$f_{max} \leq \frac{c}{2d} \qquad (2.6),$$

where $c$ is the speed of sound.

#### 2.2.2.3 TDOA and Cross-Correlation

A very important property of microphone arrays is the Time Difference of Arrival (TDOA) between microphones. The TDOA is defined as the difference in time a signal takes to reach two points separated by a certain distance $d$. To understand this concept, assume there are two microphones $M_i$ and $M_j$ separated by a distance $d$, and sound source S located at distances $D_i$ and $D_j$ from microphones $M_i$ and $M_j$, respectively, as shown in Fig. 10:

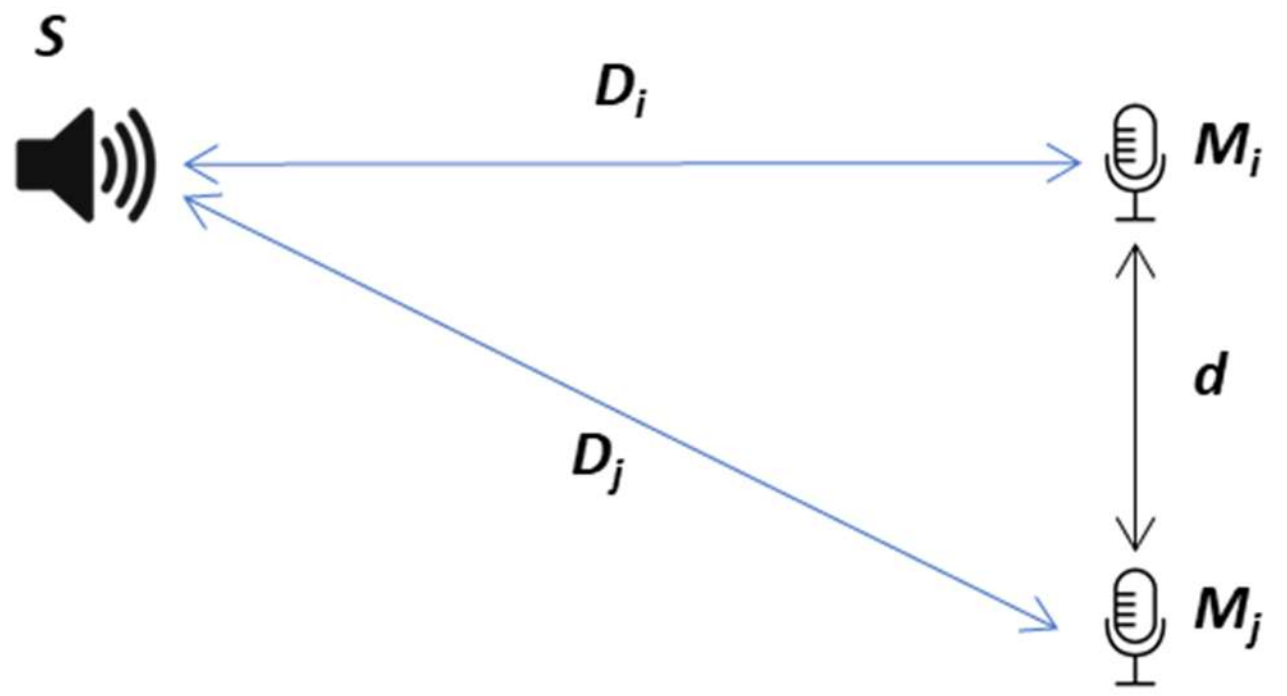

**Figure 10: Location of Source and Microphones for Time Difference of Arrival.**

The difference in the distance $\Delta\boldsymbol{D}$ between $D_i$ and $D_j$ is defined as:

$$\Delta\boldsymbol{D} = c * (\Delta t) \qquad (2.7),$$

where $c$ is the speed of sound and $\Delta t$ is the TDOA between $M_i$ and $M_{j.}$ Conversely, if $\boldsymbol{d}$ and $\Delta t$ are known, it is possible to determine $D_i$ or $D_j$ if one of them is known. From (2.7), it is also possible to infer the proximity of the source to either microphone by the sign of $\Delta\boldsymbol{D}$. Because $\Delta t = t_i - t_j$, a positive $\Delta t$ indicates that $M_i$ is closer to the sound source than $M_{J,}$ whereas a negative $\Delta t$ indicates the opposite.

The signal delay between microphones $M_i$ and $M_j$ can be also expressed in terms of their *cross-correlation* (CC). Let $r_{i,j}(t) = x_i(t) \circledast x_j(t)$ denote the cross-correlation between microphone signals $x_i(t), x_j(t)$ corresponding to the microphones $M_i$ and $M_j$. The CC $r_{i,j}(t)$ between these two signals is defined as:

$$r_{i,j}(t) \triangleq \mathrm{E}\left[x_i(t)\overline{x_j}(t)\right] \qquad (2.8).$$

The normalized cross-correlation is defined by:

$$R_{i,j}(t) = \frac{1}{a \cdot b}\, r_{i,j}(t) \qquad (2.9),$$

where the $a, b$ are defined using $a = \sqrt{\sum_t x_i^2(t)}$ and $b = \sqrt{\sum_t x_j^2(t)}$.

#### 2.2.2.4 Beamforming and Spatial Filters

The process of filtering each of the outputs of the microphones of an array into a single output is known as *beamforming.* Beamforming steers the array's directivity pattern into a particular direction using beamforming filters [40]. The combination of the signals from each microphone is governed by:

$$y = \boldsymbol{w}^{\boldsymbol{H}}\boldsymbol{x} \qquad (2.10),$$

where $\boldsymbol{w}$ represents the beamforming filters and $\boldsymbol{w^H}$ is the conjugate transpose. The beamforming filters can be estimated as a function of a propagation vector $\boldsymbol{d}$ and a noise correlation matrix $\boldsymbol{Q}$ using:

$$\boldsymbol{w} = \frac{\boldsymbol{Q}^{-1} d}{\boldsymbol{d}^H \boldsymbol{Q}^{-1} \boldsymbol{d}} \qquad (2.11).$$

The filter described in equation (2.11) is known as the Minimum Variance Distortionless Response (MVDR), and it is one of the most popular types of beamforming filter. Refer to [31] for a full explanation of beamforming filters.

If the location of the sound sources **d** is known, it is possible to construct a *spatial filter* for each of the sources. This approach is used to minimize crosstalk between channels and for noise reduction. The details of Spatial Filtering are outside the scope of this dissertation. Nevertheless, a brief introduction is presented because future work proposed in this dissertation includes a possible combination of the proposed method with spatial filtering and beamforming for speaker separation.

## 2.3 Modeling of Room Acoustics

The proposed research requires the modeling of room acoustics. The simulation of microphones and sources are all based on physical models that predict the effects of the acoustic reflections given the geometry of the room and the location of the speakers. To this end, simulations calculate RIRs to the target points. The methods to model room acoustics are dived into two categories: geometrical acoustics-based and wave acoustic-based [44]. Geometrical acoustics-based methods work by capitalizing the reflection properties of sound, i.e., sounds reflect into smooth surfaces in the same way light does, following Snell's law. These methods are relatively easy to implement but do not take into consideration the roughness of the reflective surface. On the other hand, wave acoustic-based methods take into consideration the characteristics of the sound wave, providing a more accurate simulation. In contrast with geometry methods, wave methods are more computationally intensive and are limited to low-frequency ranges [45]. The simulation package used for this research is geometry acoustic-based; wave acoustic-based methods are not considered in this dissertation.

The two more common geometry acoustic-based methods of modeling are the Ray Tracing Method and the Image Source Method. This dissertation focus on the Image Method as this method is the one used by the simulation package.

### 2.3.1 Ray Tracing Method

The Ray Tracing Method assumes that sound radiates from the source as several rays [44] whose energy is the total energy of the source divided by the number of rays. These rays propagate at the speed of sound and, when they reach a boundary surface, some of the energy is reflected in an angle $\alpha'$ equal to the incidence angle $\alpha$, as it is shown in Fig. 11 (a). The perceived sound at any point is represented by an echogram that contains the history of all the ray reflections [45] plus the direct ray. The Ray Tracing Method was introduced in the late 1960s and was widely used until the 1980s. Ray Tracing is a relatively straightforward method, but its resolution is limited [45].

### 2.3.2 Image Source Method

The Image Source Method (ISM, also known as Mirror Image Source Method MISM), is perhaps the most popular modeling method in use [46]. Image methods are used to solve physics problems, and in the late 1970s, Allen and Berkley [46], [47] introduced an algorithm to RIR related applications.

In the Image Source Method, a virtual image or specular reflection of the source is created perpendicularly to the source, as shown in Fig. 11 (b). The sound received by the sensor $M$ is the summation of the sound from the source $S$ and the image source $S'$.

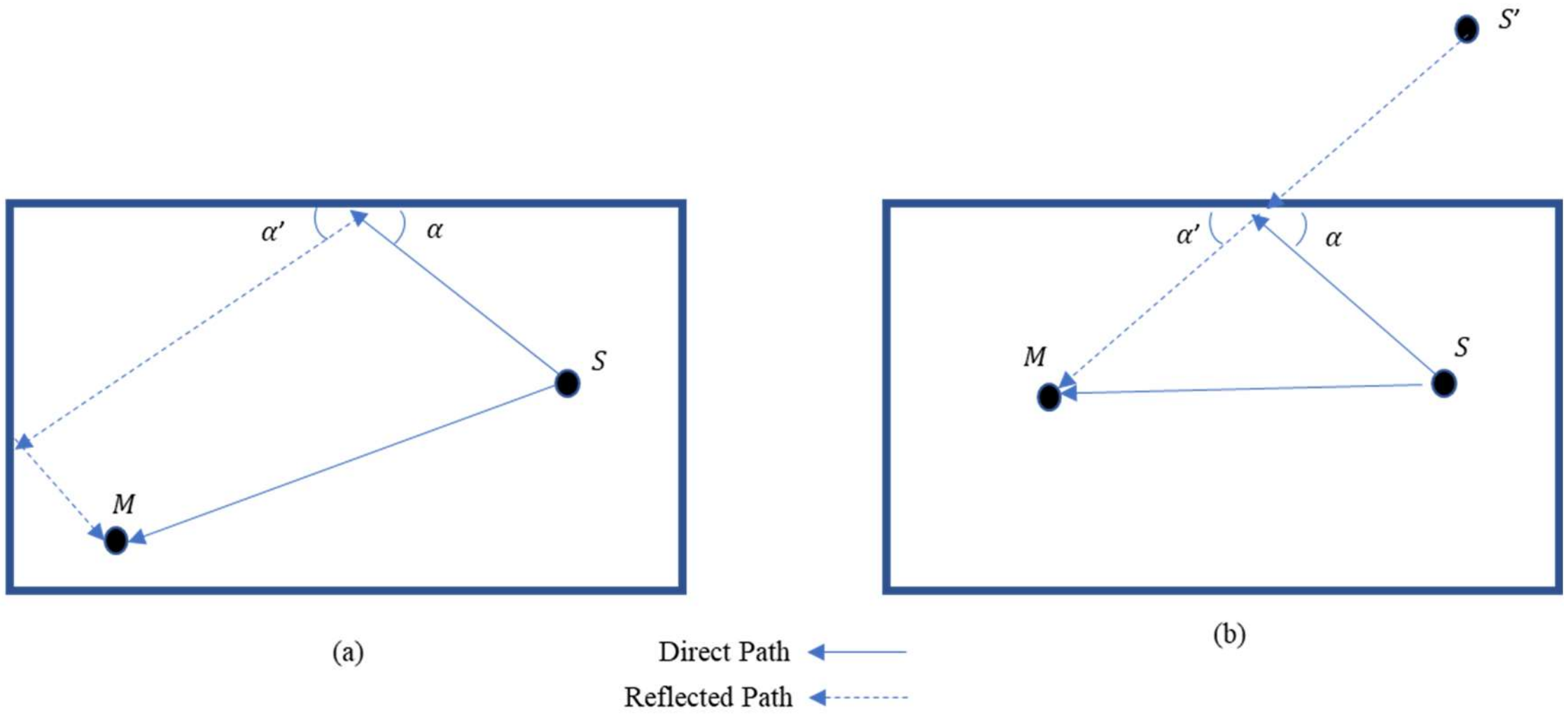

**Figure 11: Simulation Methods. (a) Ray Tracing. (b) Image Source Method.**

The ISM needs the amplitude and delay of the image sources to calculate the RIR. Because there are infinite possible reflection paths, the ISM creates a map of mirrored rooms with the position of the number of desired images, as shown in Fig. 12.

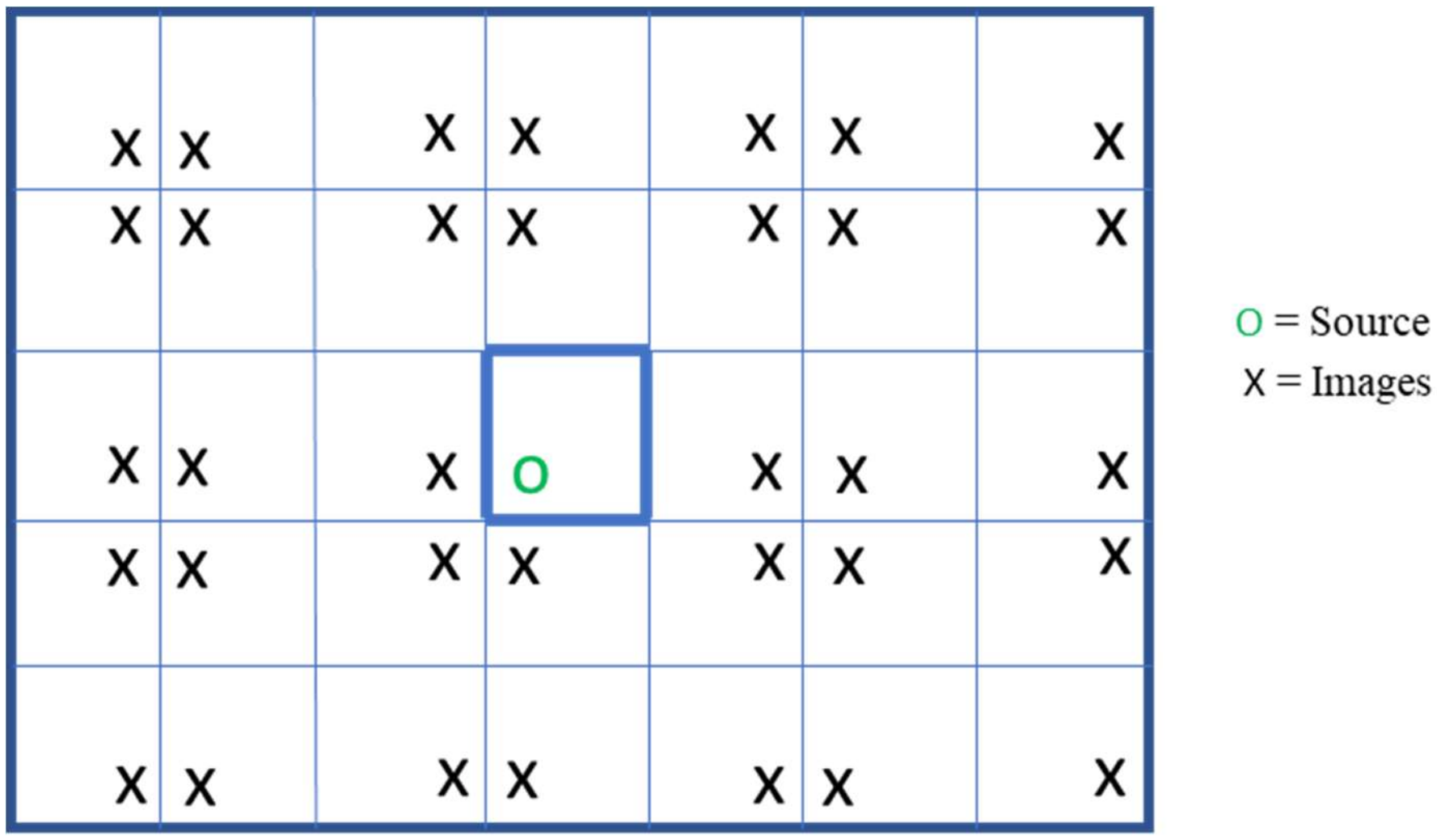

**Figure 12: Source Image Map.**

The coordinates of each of the images are calculated using the map with the corresponding room size and the position of the source. Once the position of the images is calculated, the Euclidian distance $d_n$ from the image $n$ to the source is used to calculate the delay $\tau_n = \frac{d_n}{c}$ , where $c$ is the speed of sound inside the room.

Finally, the amplitude $A_n$ of each of the signals from the images is calculated by the reflection coefficient $\beta_n$ of each of the walls crossed by the path from the image to the sensor, using (2.12) [46]:

$$A_n = \frac{\beta_n}{4\pi . d_n}. \qquad (2.12).$$

The RIR $h(t)$ is calculated using the amplitude and the delay for the images:

$$h(t) = \sum_{n \in \mathbb{N}} A_n \cdot \delta(t - \tau_n) \qquad (2.13),$$

where $\mathbb{N}$ represents the image sources and $\delta$ is the impulse function.

## 2.4 Characteristics of the Human Speech

The performance of the method described in this dissertation will improve if the acoustic models are tailored to human speech. Human speech has some characteristics that can be exploited and used to compensate for some of the deficiencies encountered with the approximation of the geometry of the room and the limitations of the modeling software.

Two characteristics of human speech: fundamental frequency and directionality are of particular importance. Speech is a non-stationary signal, or rather said, a non-stationary process, meaning that its frequency content is not unique in any given interval of time. The fundamental frequency of the human voice varies from 85 Hz to 180 Hz, with women going up to 255 Hz, and children to 300 Hz and even higher [48]. The whole spectrum of the human voice contains frequencies that go up to 8kHz. Much of the energy is found in frequencies that are below 500 Hz for males and 800Hz for females [49]. As a curiosity note, the frequency sensitivity of the human ear is very close to the frequency spectrum of the human voice. This research work focus on the fundamental frequency to develop the acoustic models. More detail is presented in the Experimental Implementation section of this dissertation.

The other important characteristic of human speech is its directionality. Speech does not propagate equally in all directions, but rather has directionality due to the location of the mouth and the shadow cast by the head and the torso [50]. Fig. 13(a) depicts the propagation of sound in the horizontal direction, while Fig. 13(b) presents the propagation in the vertical axis. Lower frequencies propagate farther from the back of the head than higher frequencies. Most propagation occurs at the front of the head. This directionality property is utilized for positioning the speakers in the simulation models.

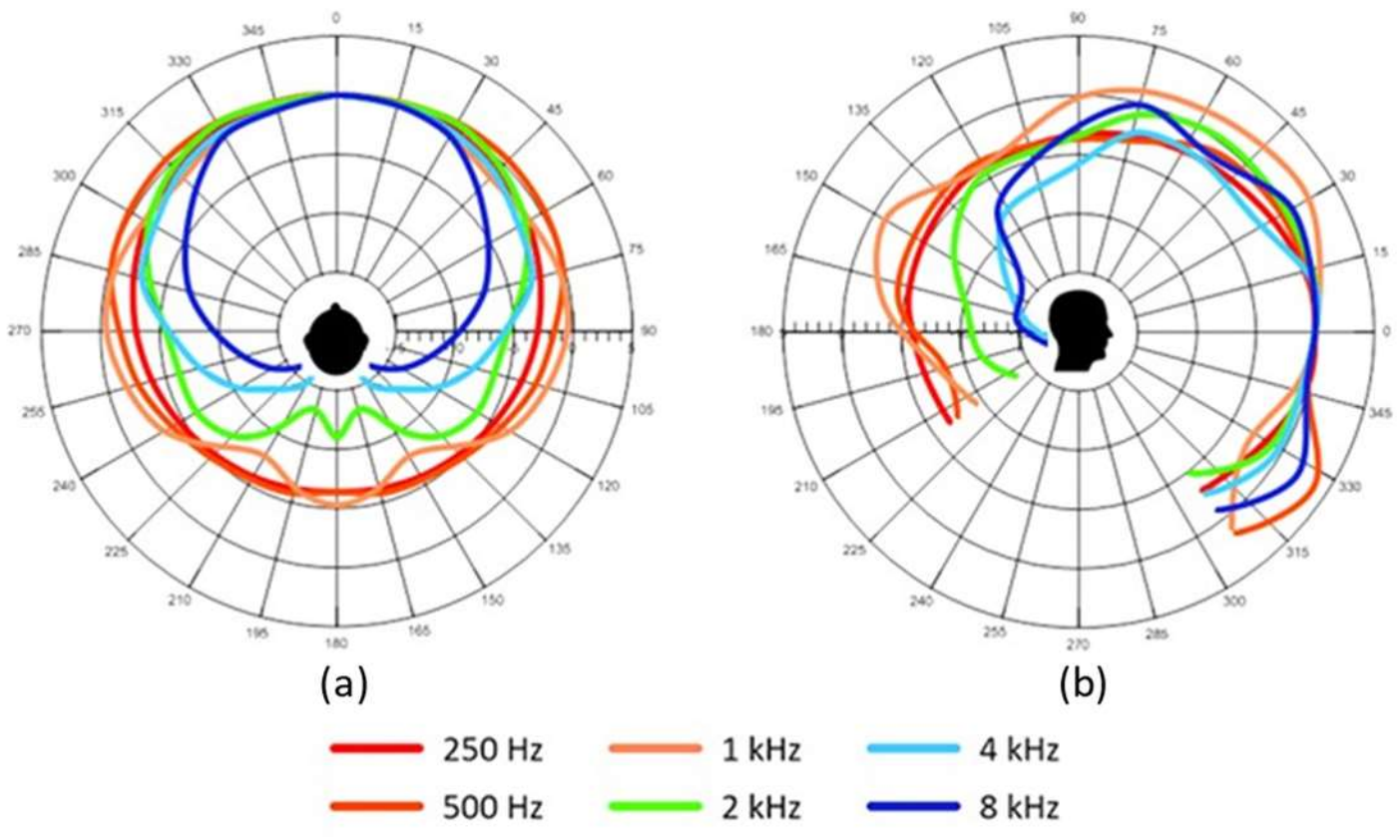

**Figure 13: Directionality of Human Head [50].**

## 2.5 Speaker Diarization and Identification

Section 1.2 presented a background on several methods for speaker diarization and identification that relate to this research. To understand the differences between these methods and the method proposed in this research, this next section reviews the fundamentals on which some of these methods are based on. Reviewing in detail all the methods presented in section 1.2 requires an effort that goes beyond the scope of this dissertation. For this reason, this dissertation only focuses on the most recent and common methods of speaker diarization and identification.

### 2.5.1 Methods for Diarization and Identification

Speaker diarization can be summarized as “who said what, and when”, and for “how long”. The task of determining for how long one speaker has been active in a multi-participant conversation requires speaker diarization and subsequent identification with non-abstract labels. Speaker identification should not be confused with *speaker verification*. A system that accepts or rejects the identity claim by a speaker is called a *speaker verification system.* This dissertation divided these methods into two categories: Classic methods and Deep Learning methods.

#### 2.5.1.1 Classical Methods for Diarization and Identification

Anyone conducting a web search for “speaker diarization and identification methods” will find thousands and even millions of documents that somehow relate to the subject (by the time of this dissertation, “speaker diarization methods” gave 66,400 hits, “speaker identification methods” about 48,000,00 hits, and “speaker diarization and identification methods” some 121,000). Nevertheless, until researchers started using Deep Learning and Neural Network methods, most speaker diarization and identification methods consisted of four basic modules or steps: A feature extraction module, a speech or voice activity detector (SAD or VAD, respectively), a segmenter or speaker change module, and finally, a clustering mechanism [51], [52]. Fig. 14 shows a block diagram of the four modules.

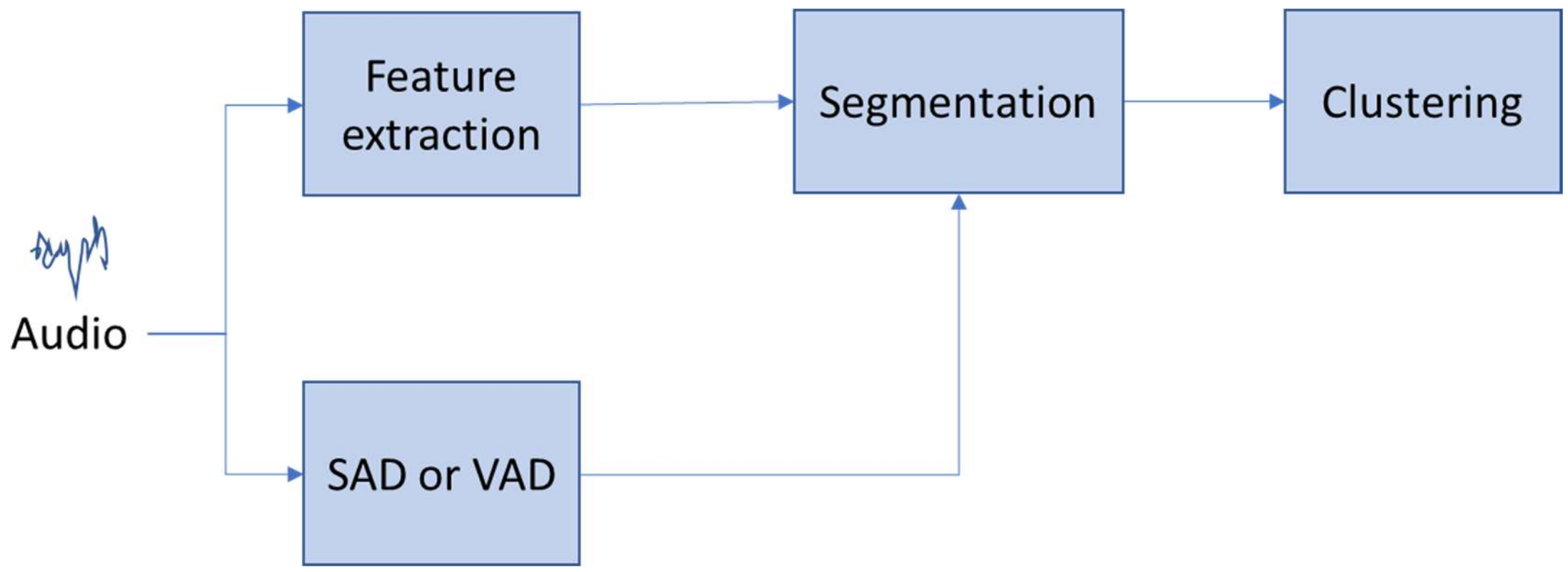

**Figure 14: Block Diagram of a Typical Speaker Diarization System.**

The feature extraction module generally uses Mel-Frequency Cepstral Coefficients (MFCC) as features. Not as popular as MFCCs, Linear Frequency Cepstral Coefficients (LFCC), and Perceptual Linear Predictive are also used as features [52].

The purpose of the speech activity module (SAD), also known as the voice activity module (VAD), is to detect the presence of speech. SADs or VADs (hereon referred to as VAD) eliminate audio segments that contain no necessary information, such as noise or music, thus improving the performance of the segmenting and clustering modules. There are several different algorithms for these detectors, that vary from just energy level detection to binary classifiers based on pre-trained speech models. This research uses a custom-made VAD to segment the audio into frames, as it will be shown in later chapters.

The next module to follow is the segmenter or speaker change detector. The segmenter detects when there is a speaker change in the audio and creates frames that ideally only contain one speaker. This is a necessary step before clustering, where the grouping of the clusters is done without previous information. A common method of segmentation is to measure the distance between two segments. Segments that belong to the same speaker are usually closer in distance than those coming from a different speaker. The models are usually Ergodic Hidden Markov Models (HMMs), where each state represents a speaker, and the probabilities are modeled by Gaussian Mixture Models (GMMs). Bayesian Information Criterion (BIC) is used to determine the nearest clusters, merging the clusters that generate the highest BIC, stopping the process when the values of the BIC are no longer positive. This algorithm was introduced by Chen et al. [53] and it is defined as (2.14) for a parametric Gaussian Mixture Model (GMM) with clusters $C$ with features [51]

$$BIC(M) = \log \mathcal{L}\,(X|M) - \frac{\lambda}{2}\#(M)\log(N) \qquad (2.14),$$

where $N$ is the number of samples, $\#(M)$ is the number of parameters of the model, and $\lambda$ is a tunable parameter.

The final clustering step groups together segments that belong to the same speaker. In most speaker diarization and identification approaches, clustering is achieved by agglomerative hierarchical clustering (AHC). Using the same distance concept, each segment is its cluster at the beginning of the process, and parts of clusters are merged until the stopping criterion is met. This criterion is ideally to get the number of clusters equal to the number of speakers. In practical terms, the stopping criterion is a threshold that is preset at the beginning of the process.

#### 2.5.1.2 Deep Neural Networks

The applications of Deep Neural Networks (DNNs) to speaker diarization have gained a lot of momentum in recent years. It is difficult to keep pace with the amount of research that is done on an almost monthly basis in this field. It is therefore of importance to have a basic understanding of how DNNs are applied to the problem of speaker diarization.

In general terms, DNN speaker diarization/identification methods are divided into four groups [54]: Stage-wise, end to end, online, and multimodal. From these groups, this dissertation will address stage-wise and multimodal groups as they relate more to the research work.

#### 2.5.1.2.1 Stage-wise diarization

Stage-wise diarization methods are based on the same blocks or stages as the GMM methods covered in the previous section, but they rely on DNNs that employ a universal background model (UBM), rather than GMMs for feature extraction and clustering. For GMMs to be computationally efficient for feature extraction, each sequence of feature vectors is converted into a fixed-length vector, or *supervector* [55]. Because this approach makes GMMs susceptible to speaker and channel variations of utterances [56], it is desirable to reduce the dimensionality of the supervectors. These lower-dimensional vectors are called *i-vectors* (the i-vectors were previously referenced in the background section). The representation of i-vectors assumes that speaker and channel-dependent variabilities reside in a lower-dimensional space [57], which is represented by a total variability matrix $\boldsymbol{T}$. For GMMs this conversion can be expressed as:

$$s = s' + Tw \qquad (2.15),$$

where $s'$ is the speaker and channel supervector and $w$ is the i-vector. Fig. 15 shows a GMM/i-vector framework [56].

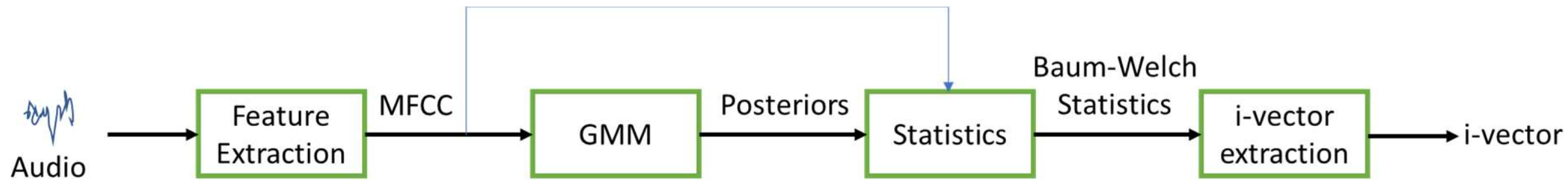

**Figure 15: GMM/i-vector Framework.**

At this point, the effort moved to replace the GMM generated i-vectors for DNN generated i-vectors. The idea behind this approach is to replace the GMM generated posteriors for the feature vectors and take a DNN trained acoustic model using senones to generate these posteriors. Fig. 16 represents this approach [54].

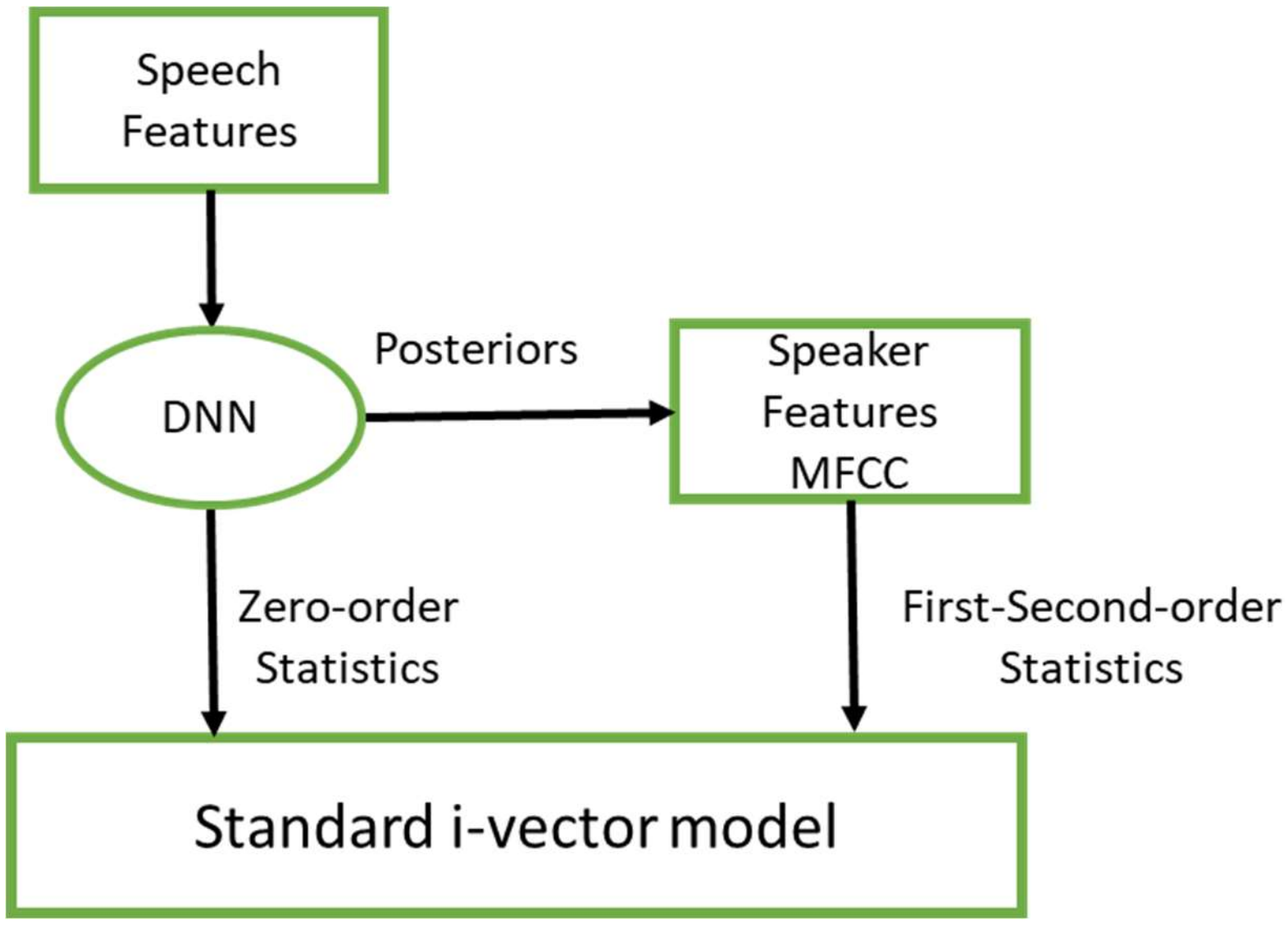

**Figure 16: DNN/i-vector Implementation.**

Although the performance of DNN based acoustic models has been proven [56], they require a large set of training data and more computational cost as well.

In addition to i-vectors, other DNN approaches include the use of d-vectors and x-vectors embedding. D-vectors were introduced by Variani et al. (2014), and they are based on assigning the ground truth training utterance to labels of the training frames to the corresponding utterance in the training stage, converting the problem into a classification one. For a more detailed description of d-vectors, refer to [54],[56].

X-vectors are derived from d-vectors. Instead of using frame-by-frame speaker labels, they use utterance-level speaker labels by aggregation. As was referred to in the background section, x-vectors outperform most i-vector and d-vector approaches. Refer to [54], [56], for details on x-vectors.

Deep Learning clustering techniques are also applied in replacement of conventional distance and similarity methods. Clustering is treated as either a supervised or unsupervised problem, by employing recurrent neural networks (RNN) or discriminative sequence-to-sequence neural networks. References to these methods can be found in [54].

#### 2.5.1.2.2 Multimodal Speaker Diarization

Related to our approach of exploiting video clues and spatial information, Deep Learning is applied to the analysis of not only visual clues, such as movement of lips [58], but also to the content of the speech of the participants [59], [60]. In this sense, multimodal methods train the networks based on the patterns that most likely belong to a genre of participants. For example, in a collaborative environment with students, the facilitator is more likely to have a calm voice, in contrast with the students. In recent publications, W.

Kang et al. [61], have presented speaker diarization based on d-vectors combined with spatial information provided by microphone arrays.

### 2.5.2 Current State-of-the-Art Methods for Diarization and Identification

State-of-the-art methods cover the speaker diarization and transcription that major technology players are offering. They keep their technologies a secret, as they compete to have the most reliable service available, thus the difficulty in obtaining detailed information on how their methods work. It is expected that they somehow use the speaker diarization approaches reviewed in the previous sections.

IBM, Google, Amazon, and Microsoft offer cloud computing that includes speech processing services based on algorithms that use Deep Learning and Machine Learning. Amazon's, Google's, and Microsoft's are all closed-source cloud services that provide an API for speech-to-text processing and speaker diarization. This dissertation reviewed Amazon's Transcribe (AWS) [62], Google's Cloud [63], and Microsoft Azure Speech Services [64], and experimentally compared Amazon's and Google's against our proposed system.

Amazon's Transcribe accepts either audio files or streaming data, single-channel, and outputs text files with speaker diarization if this option is selected, and the number of speakers is specified. Amazon's Transcribe works better with 2-5 speakers, and it is language-dependent, limited to 120 minutes of audio. Amazon's Transcribe stores the voice data to train the models [65] unless the users select the option to delete this data.

Amazon offers a highly trained set of models called Amazon Transcribe Medical which is aimed at medical transcriptions. Users can also customize the vocabulary to better fit their needs. Amazon functionality can be accessed via REST and SOAP protocol over HTTP [66].

Google's Cloud works similarly, with an interface for long speech, single-channel input for transcription purposes [65]. The optimum number of speakers is set at a maximum of 5. As with Amazon Transcribe, Google offers the option of privacy that prevents data logging that could be used to improve the models. Google's models are optimized for phone conversations or videos, accepting 16 kHz or 8 kHz audio, respectively, depending on the application [67]. It also offers vocabulary customization. Google offers good scalability, infrastructure, and payment schemes that are considered the best among the technology giants [66].

Microsoft offers speaker diarization utilizing its Cognitive Services. Microsoft's Diarization system ranked first at the VoxSRC challenge 2020 by achieving a Diarization Error (DER) of 3.71% in development and 6.23% in evaluation testing [68]. The datasets consisted of audio collected from YouTube recordings. For the challenge, the network was trained with 1500 hours of simulated mixed training audio. Microsoft Speaker Recognition [69] offers text-independent speaker recognition/verification. The speakers need to be enrolled to create a signature, which is later compared with the audio to be analyzed. The minimum requirements are 20 seconds of speech for training, and 4 seconds of speech for identification, with unlimited speaker enrollment, with only one speaker present. In the case of diarization, Microsoft can only recognize up to two speakers. Microsoft Transcription requires multi-channel audio for diarization and the signature of the participating speakers for identification, labeling each speech segment with its correspondent speaker. Microsoft does not collect users' voice tracks to train its models. Users can customize their vocabulary and the environment they are expecting to operate, meaning that customization must include noise, indoor or outdoor environments, multi-gender speech, etc. [64].

# Chapter 3. Proposed Method

The previous chapter discussed the principles of acoustics, speech, and speaker diarization that served as the foundations for this research. This chapter will cover the mathematical models that are used to estimate the virtual microphones, and how they are implemented into a working system. Finally, it will present a block diagram detailing the function of each of the elements of the proposed system and its operation.

## 3.1 Methodology

The goal of this dissertation is the *identification of speakers from single-channel recordings using virtual microphones*. This statement includes the objective (identify speakers), the data source (single channel recording), and the means to accomplish this objective (using virtual microphones). This section will begin by identifying the physical and mathematical elements of the models needed for the virtual microphone simulation.

Let us consider a collaborative environment such as the one represented in Fig. 17 (a), where we have three speakers sitting around a table with a central recording microphone. Such an environment can be represented as a simple 2D model shown in Fig. 17(b) that shows the relative location of the speakers and the recording microphone.

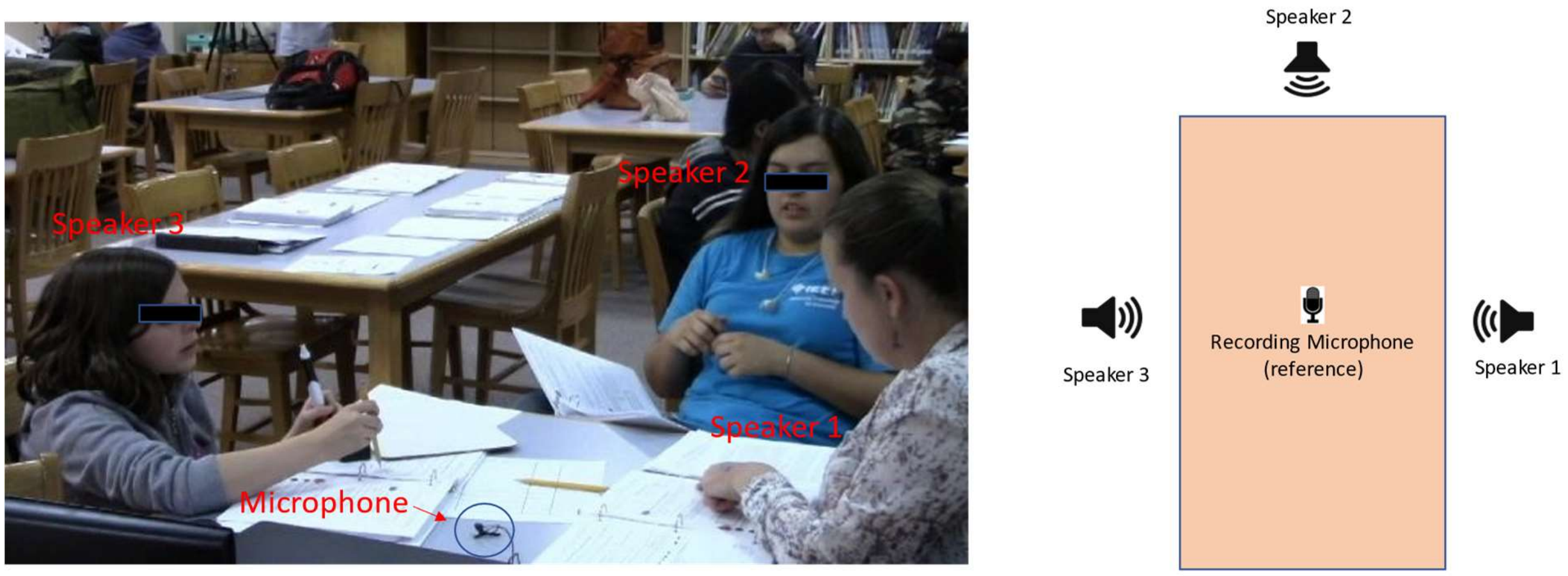

(a) (b)

**Figure 17: Collaborative Environment (a) with 2D Model (b).**

To capitalize on the properties of microphone arrays, it is necessary to find a method to simulate several virtual microphones based on the information contained in the signal captured by the central microphone. From the discussion on microphone arrays in chapter 2, it is possible to implement several different virtual array configurations. Let us consider a cross-linear array with four microphones and one central recording microphone, as shown in Fig. 18.

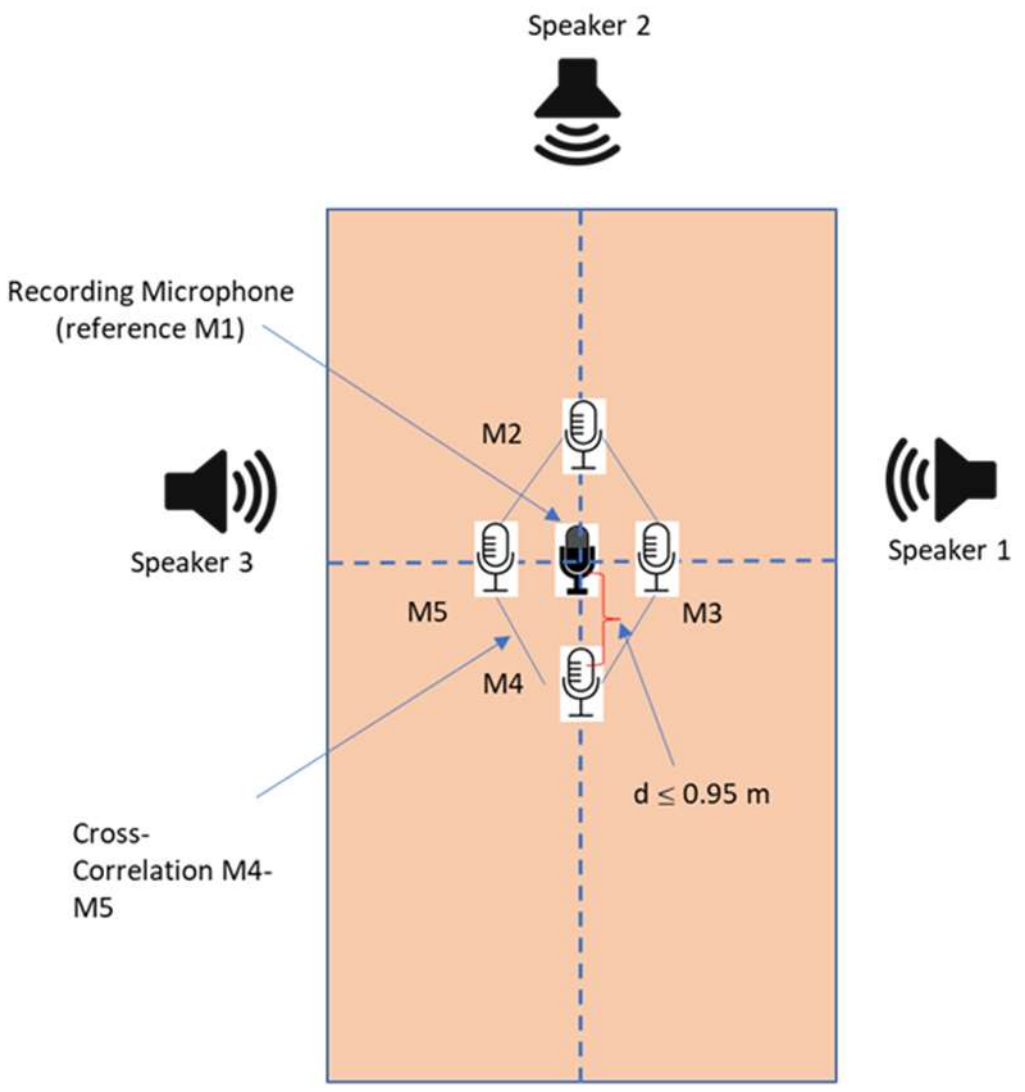

**Figure 18: 2D Model of Fig. 12(b) with Microphone Array.**

If Fig. 18 is an ideal representation, where there are no reflections or room absorption, then for each unique active speaker there will be a set of pairs of microphones with unique TDOAs that correspond to the active speaker. For example, if speaker 3 is active, then the TDOA between M5 and M3 and the TDOA between M2 and M3 will be unique for speaker 3. Having this concept in mind, we recall from Chapter 2 that the cross-correlation from any pair of microphones represents the signal delay between them. To uniquely identify each of the speakers, we are interested in the location of the peak of the cross-correlation function defined by:

$$T_{i,j} = \operatorname{argmax} R_{i,j}(t) \qquad (3.0),$$

where $R_{i,j}(t)$ denotes the cross-correlation between two microphone signals $x_i(t), x_j(t)$.

If a source signal propagates to microphones $i, j$, $T_{i,j}$ represents the time lag that it takes for the signal to reach $j$ after reaching $i$. Thus, $T_{i,j} > 0$ implies that the signal arrived at microphone $i$ before $j$. On the other hand, $T_{i,j} < 0$ implies that the signal arrived at microphone $j$ before $i$. The cross-correlation matrix of all possible values $T_{i,j}$ will be used for determining the locations of the speakers.

Now we move to the problem of simulating the virtual microphones. From equation (2.0) from Chapter 2, it is possible to extend this model for the case of multiple sources and microphones. Suppose that we have $J$ possible sources: $s_1(t), \dots, s_J(t)$ and $N$ possible microphone signals: $x_1(t), \dots, x_N(t)$. Next, let $h_{j,k}(t)$ denote the RIR that describes the propagation from the $j$-th source to the $k$-th microphone. At the $k$-th microphone, we receive signals from all sources as expressed by:

$$x_k(t) = \sum_{j=1}^{J} s_j(t) * h_{j,k}(t) + n(t) \qquad (3.1),$$

where $n(t)$ represents additive white noise. For the model in (3.1), we need to estimate $h_{j,k}(t)$. If $h_{j,k}(t)$ is known, it is possible, at least in theory, to estimate the signal source by deconvolving the signal $x_k(t)$ with $h_{j,k}(t)$ (i.e., $h_{j,k}^{-1}(t)$). Once the sources have been estimated, each virtual microphone can be emulated by just convolving the emulated source with each of the RIRs of the virtual microphones.

Some important factors need to be considered to develop a model for this approach. First, $h_{j,k}^{-1}(t)$ may not exist [70], and it may be necessary to construct virtual microphone approximations to $h_{j,k}(t)$. Second, the speaker feature correlation matrix defined by $T^m$ is estimated under the assumption that speaker $m$ is talking while all other speakers remain quiet: $s_k(t) = 0, k \neq m.$ Third, for each audio segment, we need to compute $T$, the cross-correlation matrix of the actual signal. Finally, we need to estimate the active speaker by solving

$$\max_{m} \ \text{match}(T, T^m) \qquad (3.2),$$

where $\text{match}(.,.)$ is a function of the similarity between $T, T^m$.

We now can turn our attention to estimating the RIRs. As it was presented in Chapter 2, the RIR is a function of the geometry of the room, the relative location of the sources and microphones, and the physics of the room (i.e., the absorption of the room, which characterizes the reverberation). This information will be very difficult if not impossible to obtain from just the audio from the recording microphone, but we could use information from the video recording to estimate some of the parameters needed to calculate the RIR. From the video recording, it would be possible to approximate the location of the speakers and the virtual microphones to each other. This information, along with an empirical approximation of the absorption of the room, is all that is necessary to calculate the RIR.

Calculating the RIR can be a very tedious task. The number of calculations required is a factor in the degree of accuracy desired in the model. If we recall the concept of images from Chapter 2, the reception at a microphone is the result of the sum of the images; therefore, the fidelity will depend on the number of images added as part of the RIR function. The next chapter will present an open-source software package that performs these calculations thus saving some programming time.

So far, this dissertation has presented the fundamentals of the simulation on which the proposed method is based. By using the approximate geometry of the room to calculate the RIR and to simulate the microphones, we should be able to calculate the cross-correlation between microphones and determine the active speaker. The proposed method then can be summarized in 5 steps: 1) Evaluating the room geometry and location of the speakers of the acoustic scene, 2) Estimating a generic RIR model for this geometry, 3) Training the model with known speaker samples, 4) Estimation of the sources that will fit the model for each of the possible active speakers given an unknown audio sample, and 5) Conducting a Cross-Correlation Analysis and classification. The following section explains each of the steps in more detail.

*1) Evaluation of room geometry and location of speakers and microphones*

As it was described before, the RIR is a unique transfer function between two points in space. To calculate the RIR between a source and a microphone, it is necessary to know their spatial locations inside a physical room of known acoustic characteristics. In Fig. 17, it is possible to appreciate the relative location of the three speakers and the recording microphone. This video frame can be used as a reference for the location of the sources and virtual microphones in the model, e.g., from this image we can approximate that the table is about 1.5 meters long by 1 meter wide, that the speakers are separated about 0.7 meters from each other, and the speaker's mouths are about 0.24 to 0.25 m from the table. It is also possible to locate the reference microphone in coordinates that are relative to each of the speakers. These are just approximations to create a generic model from where to calculate the RIRs. Fig. 19 shows a possible 2D model for these approximations.

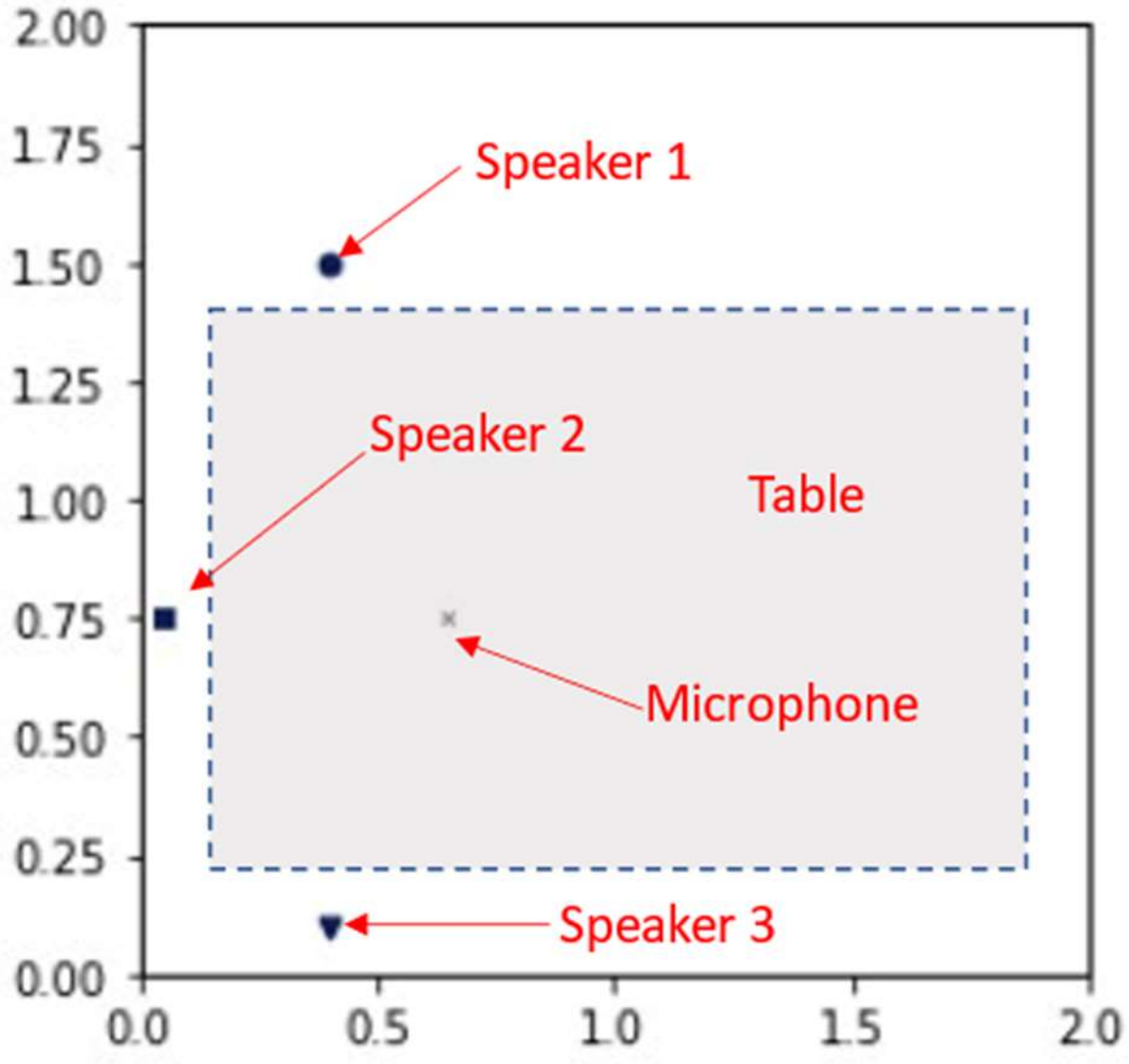

**Figure 19: Possible 2D Model for Fig. 17.**

The location and separation of the virtual microphones can be arbitrary for as long as they do not violate the rules of spatial anti-aliasing. As presented in Chapter 2, the fundamental frequency of human speech varies from 85 Hz to 180 Hz approximately, with some extreme cases going up to 255-300 Hz (children). If it is assumed a max frequency average of 180 Hz using (2.6) and (2.7), the maximum separation $d$ for each microphone would be $\leq \frac{343\frac{m}{s}}{2(180\ Hz)} = 0.95\ m$.

*2) Estimation of the generic RIR model*

The approximation of the geometry of the room provides the basis to implement a generic model to calculate a set of RIRs to estimate the virtual microphone array. This model, as it was mentioned before, is based on an approximate geometry of the room, the location of the speakers, and the number of reflections. It is desirable to reduce the influence of reflections and reverberation in the simulation as they add complexity to the RIR. One way this can be achieved is by an overall reduction of the length of the $T_{60}$. Recalling equation (2.2), we can minimize the volume of the room and maximize its absorption as a means of reducing the length of $T_{60}$. These two parameters are easy to control and implement in the simulation. The number of images to calculate can be set to an acceptable value that compromises the simulation fidelity and the computational burden. Some trial and error may be necessary to optimize the number of reflections.

Another point to consider is the directionality of the human voice. The human voice propagates mostly in one direction to the front of the head; therefore, our model must take this propagation inequality when simulating the audio reception at any point of the room.

One solution implemented in this research consisted of locating the speakers close to the end of the virtual room, so the reflections from the back of the speaker are minimized.

With the approximate physical and acoustical characteristics of the room, it is then possible to calculate the RIRs between the virtual microphones of the array and the speakers. It was indicated in the previous section that it could be possible to implement any arbitrary array of microphones for as long as we follow the rules of spatial anti-aliasing. The calculated value of the distance $d$ is well fitted between the boundaries of the proposed model, but it would be beneficial for the performance of the model to optimize the microphone array for maximum cross-correlation information. This can be accomplished by asymmetric microphone arrays, i.e., arranging the microphones at locations that are offset from equidistant points to the speakers. Also, the microphone arrays should have as many microphones as possible, for as long as the required computational resources remain manageable.

*3) Estimation of sources and virtual microphones*

The next step is to apply our generic model to estimate the signal at the virtual microphone array based on the recorded signal at the reference microphone. To do so, it is necessary first to estimate the sources that would fit the model, i.e., estimate a set of sources that, when convolved with the model's RIRs, will represent the signal at each microphone of the array, including the reference microphone. One way to estimate the sources is to deconvolve the reference signal with the RIR that corresponds to the source we want to estimate. For example, assume that our model has three sources $s_1(t)$, $s_2(t)$ and $s_3(t)$, three microphones $M_1$, $M_{2,}$ and $M_3$, with $M_3$ as the reference microphone. If $x_{3,j}(t)$ is the signal received at $M_3$ with $j = 1,2,3$ for the respective sources $s_{1,}$ $s_{2,}$ and $s_{3,}$ then to estimate $s_1(t)$, $s_2(t)$, and $s_3(t)$ given $x_{3,n}(t)$

$$\tilde{s}_1(t) = x_{3,1}(t) * h_{3,1}^{-1}(t) \qquad (3.3),$$

$$\tilde{s}_2(t) = x_{3,2}(t) * h_{3,2}^{-1}(t) \qquad (3.4),$$

$$\tilde{s}_3(t) = x_{3,3}(t) * h_{3,3}^{-1}(t) \qquad (3.5),$$

with $h_{3,1}^{-1}(t)$ the inverse of $h_{3,1}(t)$ (RIR from $M_3$ to $s_1$), $h_{3,2}^{-1}(t)$ the inverse of $h_{3,1}(t)$ (RIR from $M_3$ to $s_2$), and $h_{3,3}^{-1}(t)$ the inverse of $h_{3,3}(t)$ (RIR from $M_3$ to $s_3$). Fig. 20 shows how the sources can be estimated for the example of Fig. 17 and the model of Fig. 19. Once the sources have been estimated, they can be convolved with the remaining RIRs to obtain the simulated reception on each of the microphones of the array. Fig. 21 shows an example of how microphones $M_1$, $M_{2,}$ and $M_3$ are estimated using $s_2(t)$. The process is extensive to the other sources as well. We can use an estimation or the ground truth for microphone $M_3$.

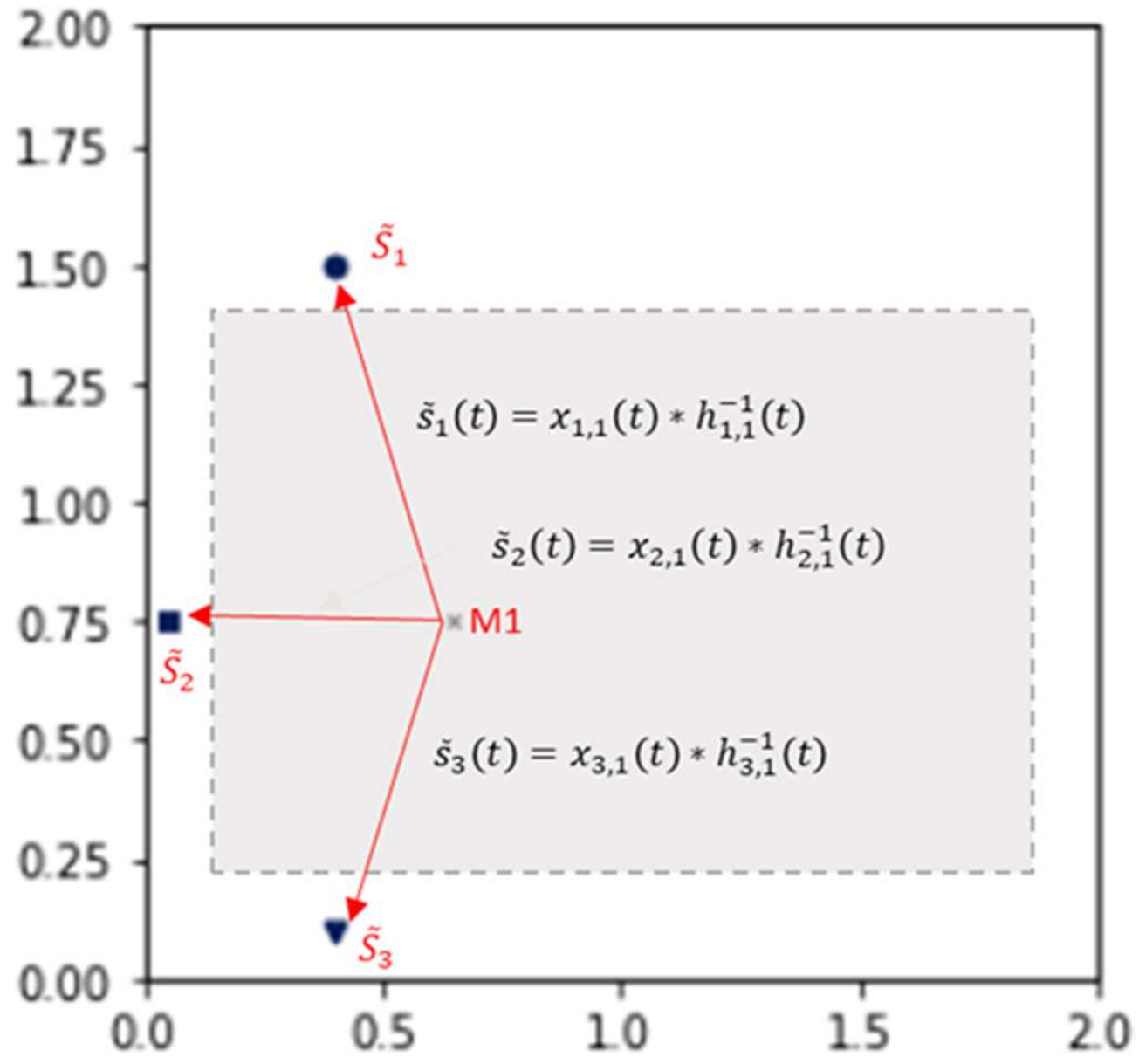

**Figure 20: Estimation of the Sources.**

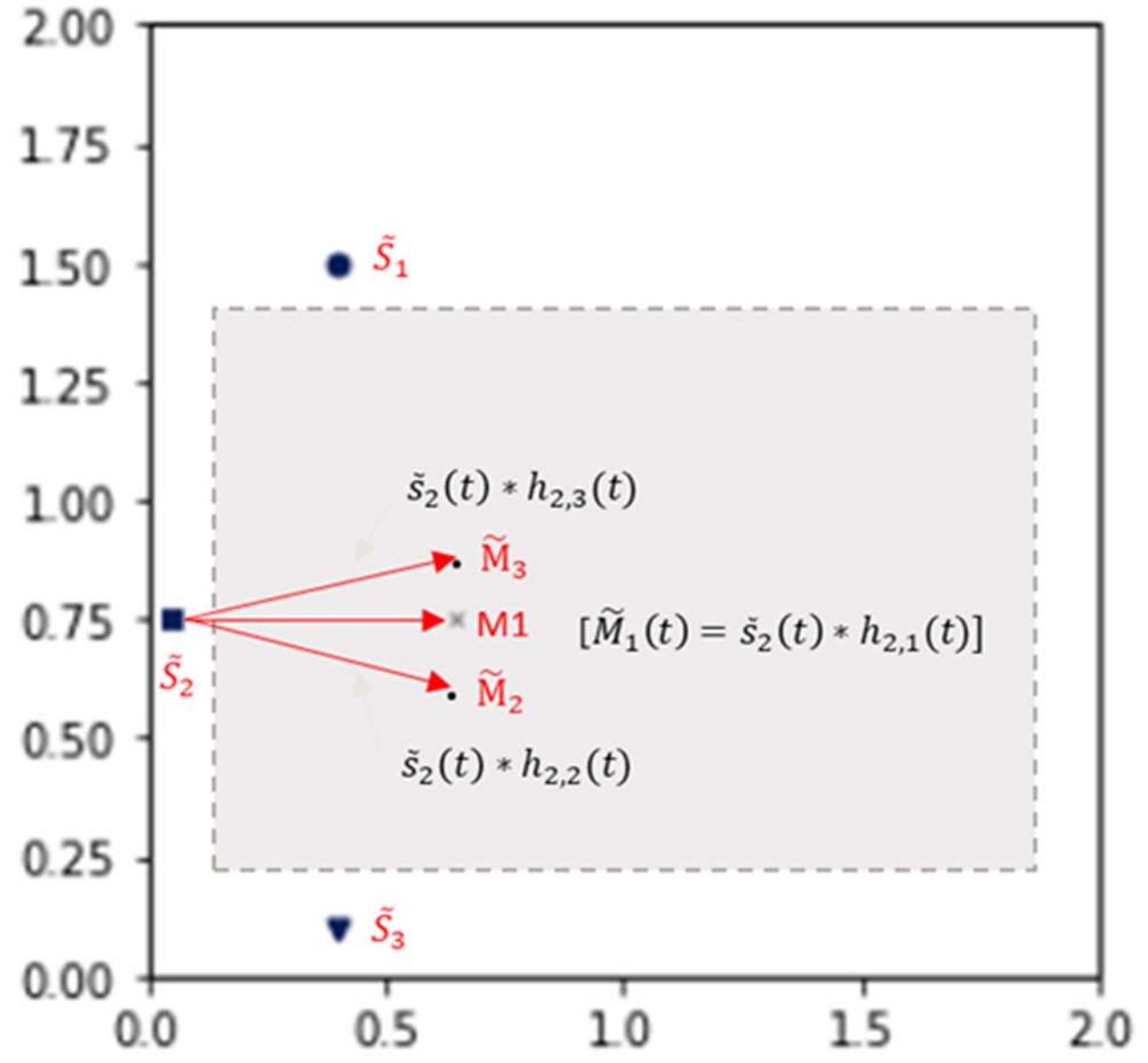

**Figure 21: Estimation of Virtual Microphones.**

It may be noticed at this point that the solution presented above will only work if we know which $n$ source is active at $x_{3,n}(t)$, and which $h_{3,n}(t)$ we need to deconvolve with. To solve this problem, our method simulates each possible source by deconvolving the signal at $M_3$ with each RIR of the model and then simulates the signal at each of the virtual microphones. The result is a set of virtual arrays that correspond to each of the possible active sources. In the three source examples, if $x_3(t)$ is defined as the *unknown* signal at $M_3$, the estimate of both possible sources $\tilde{s}_1(t)$, $\tilde{s}_2(t)$ and $\tilde{s}_3(t)$ is obtained by deconvolving $x_3(t)$ with $h_{3,1}(t), h_{3,2}(t)$ and $h_{3,3}(t)$:

$$\tilde{s}_1(t) = x_3(t) * h_{3,1}^{-1}(t) \qquad (3.6),$$

$$\tilde{s}_2(t) = x_3(t) * h_{3,2}^{-1}(t) \qquad (3.7),$$

$$\tilde{s}_3(t) = x_3(t) * h_{3,3}^{-1}(t) \qquad (3.8),$$

and then emulating two sets of virtual microphones. Each set of the microphones is represented as $x_{a,b}$, where a is the index of the virtual set, $b$ is the index of the virtual microphone of the set, and $h_{c,d}$ is the RIR from source $c$ to microphone $d$.

For set 1:

$$x_{1,1}(t) = \tilde{s}_1(t) * h_{1,1}(t), \qquad (3.9),$$

$$x_{1,2}(t) = \tilde{s}_1(t) * h_{1,2}(t), \qquad (3.10),$$

$$x_{1,3}(t) = \tilde{s}_1(t) * h_{1,3}(t). \qquad (3.11).$$

For set 2:

$$x_{2,1}(t) = \tilde{s}_2(t) * h_{2,1}(t), \qquad (3.12),$$

$$x_{2,2}(t) = \tilde{s}_2(t) * h_{2,2}(t), \qquad (3.13),$$

$$x_{2,3}(t) = \tilde{s}_2(t) * h_{2,3}(t). \qquad (3.14).$$

For set 3:

$$x_{3,1}(t) = \tilde{s}_3(t) * h_{3,1}(t), \qquad (3.9),$$

$$x_{3,2}(t) = \tilde{s}_3(t) * h_{3,2}(t), \qquad (3.10),$$

$$x_{3,3}(t) = \tilde{s}_3(t) * h_{3,3}(t). \qquad (3.11).$$

*4) Cross-Correlation and Model training*

The three sets of virtual microphones give us enough information for cross-correlation analysis and model training. To train the model, we take a small audio sample that contains only one active source (e.g., see Fig. 22), and we then use this information to generate a cross-correlation table that contains all combinations of possible sources and microphone pairs for that source. For the training, we include the background noise, as is shown in Fig. 22. Training the model for noise is explained later in the implementation section. These tables are templates for the classification of each of the active sources. For our example of three sources and three microphones, we calculate three cross-correlation tables for each known sample processed with filters for $s_1$, $s_2$, and $s_3$, as shown in Table I, Table II, and Table III.

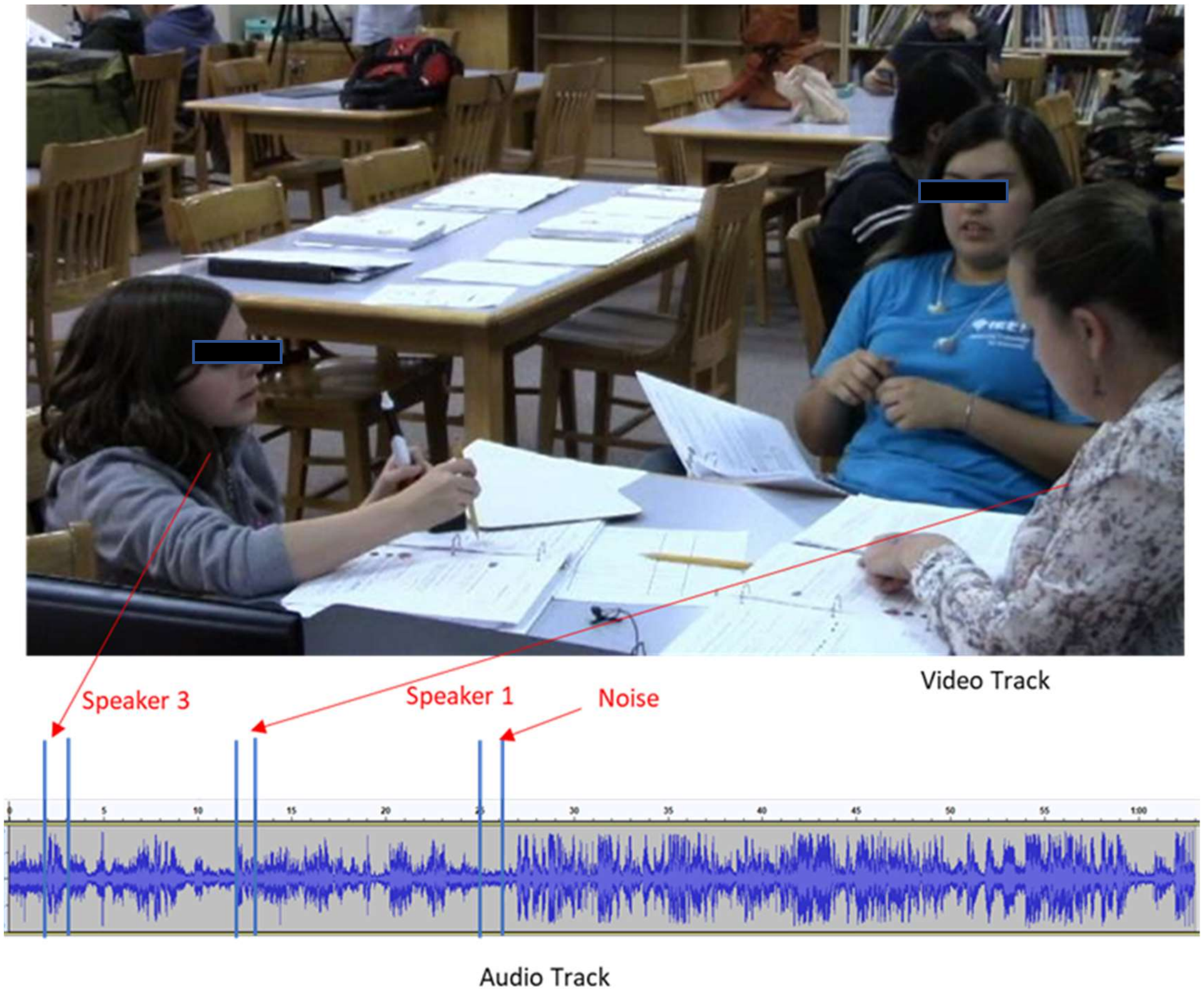

**Figure 22: Training Samples from Each Speaker and Noise.**

***Table I: Template Cross-Correlation Table for Source 1.***

| | $r_{1-2}$ | $r_{1-3}$ | $r_{2-3}$ |
|---|---|---|---|
| $\tilde{s}_{1,1}$ | $V_1$ | $V_2$ | $V_3$ |
| $\tilde{s}_{1,2}$ | $V_4$ | $V_5$ | $V_6$ |
| $\tilde{s}_{1,3}$ | $V_7$ | $V_8$ | $V_9$ |

*Table II: Template Cross-Correlation Table for Source 2.*

| | $r_{1\text{-}2}$ | $r_{1\text{-}3}$ | $r_{2\text{-}3}$ |
|---|---|---|---|
| $\tilde{s}_{2,1}$ | $V_{10}$ | $V_{11}$ | $V_{12}$ |
| $\tilde{s}_{2,2}$ | $V_{13}$ | $V_{14}$ | $V_{15}$ |
| $\tilde{s}_{2,3}$ | $V_{16}$ | $V_{17}$ | $V_{18}$ |

*Table III: Template Cross-Correlation Table for Source 2.*

| | $r_{1\text{-}2}$ | $r_{1\text{-}3}$ | $r_{2\text{-}3}$ |
|---|---|---|---|
| $\tilde{s}_{3,1}$ | $V_{19}$ | $V_{20}$ | $V_{21}$ |
| $\tilde{s}_{3,2}$ | $V_{22}$ | $V_{23}$ | $V_{24}$ |
| $\tilde{s}_{3,3}$ | $V_{25}$ | $V_{26}$ | $V_{27}$ |

where $V_1 \ldots V_n$ are the values of the cross-correlations $r_{1,2}$, $r_{1,3}$, and $r_{2,3}$ corresponding to the microphones $M_1$-$M_2$, $M_1$-$M_3$, and $M_2$-$M_3$ respectively, for each known sample source of $s_1$, $s_2$ and $s_3$. Table I will contain the results for the sample $s_1$, Table II for the sample $s_2$, and Table III for the sample $s_3$. Training needs to be done just once.

*5) Analysis and Classification*

Cross-correlation analysis of multi-speaker audio is not possible unless this is divided into segments. Proper segmentation of the audio is important to the performance of the proposed method. Because the audio from collaborative environments contains multiple speakers, it is possible at any time to have more than one simultaneous active speaker. Also, it is possible to have periods of noise (there are no periods of silence) or overlapping speech when one speaker finishes and another one begins speaking. For optimal cross-correlation location identification, each segment should only contain one active speaker at a time. If the audio contains overlapping speech or mixes of speakers, the location of the peak value of the cross-correlation between microphones will depend on the amount of information from each of the speakers that are contained in the audio segment, making the classification more difficult.

One solution to maximize the probabilities of having only one active speaker in a segment is by minimizing its length: the shorter the segment is, the most likely it is to have content from only one speaker. There is, however, a limit on the minimum length of the segments. The minimum length of the segments is subject to the performance of the cross-correlation algorithm. This means the segments need to be long enough to contain sufficient information for the algorithm to calculate a meaningful cross-correlation. In addition, the total analysis time is affected by the number of segments that need to be cross-correlated and analyzed, hence the desire of reducing the number of segments. There is then a need for optimizing the length of the segments for a balance between the maximum information content and the minimum overlapping between speakers. Recalling from Chapter 2, the best way to segment the audio is to incorporate a VAD. Ideally, a VAD will detect speech content and the change of speakers based on a certain pre-determined energy threshold, as shown in Fig. 23. If the energy threshold is properly adjusted, a VAD can be effective in producing segments of audio that contain only one speaker at the time, and segments that contain mixes of speakers or noise, maximizing their information content.

Because VADs are not perfect, there will be always segments that could contain overlapping, mixes of speakers, or simply being misclassified. To minimize the number of misclassifications, the length of the segments can be limited to a maximum that provides an acceptable number of misclassifications. It was found during this research that segments that are more than 1.5 s long are prone to misclassifications, while segments of less than 500 ms are difficult to cross-correlate.

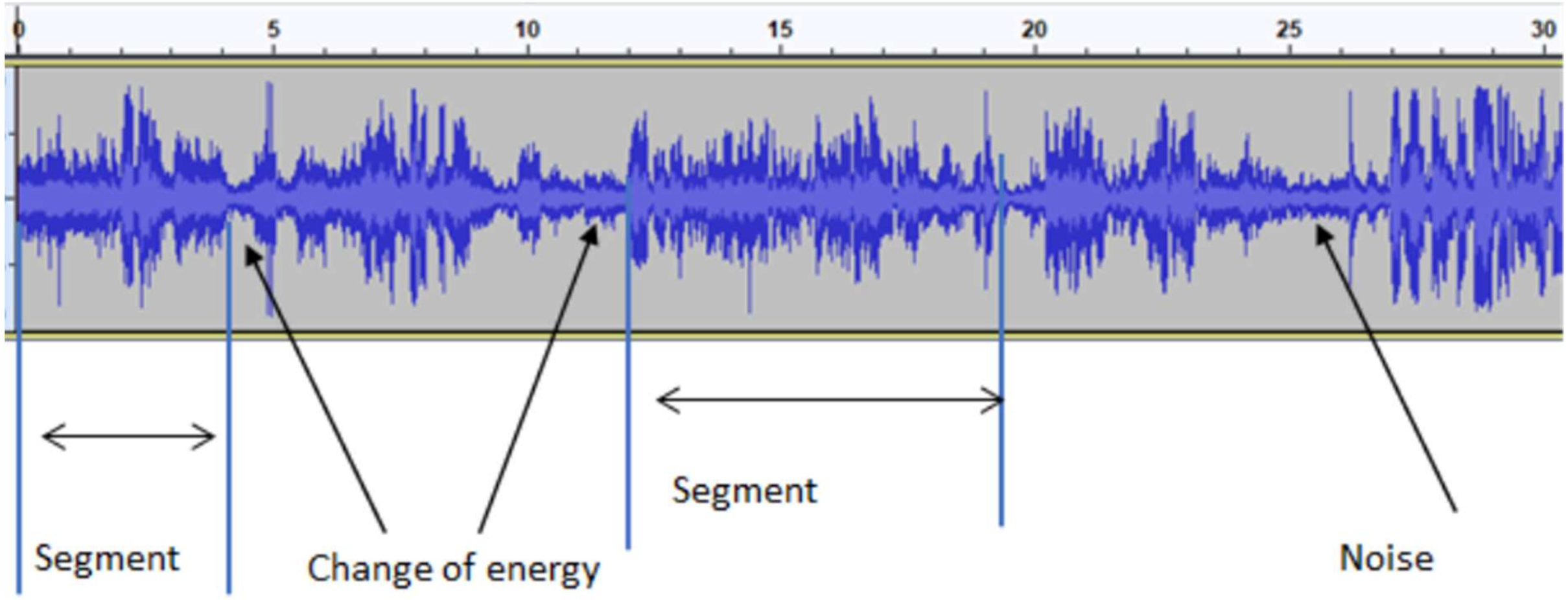

**Figure 23: Audio Segmentation using a VAD.**

Each audio segment will generate a single cross-correlation table that corresponds to each of the possible locations of the speakers, as it is done for training. The classifier will then use the cross-correlation template tables from training to compare with the analysis results and determine the most probable source match. Alternatively, it is possible to cluster the cross-correlation results for later classification. Chapter 4 covers the analysis and classification methods of this research in more detail.

## 3.2 Block Diagram of the Proposed System

Fig. 24 presents the block diagram of the proposed method. Chapter 4 covers the experimental implementation of each of the modules in this diagram, except for the clustering module, which was not implemented for this research.

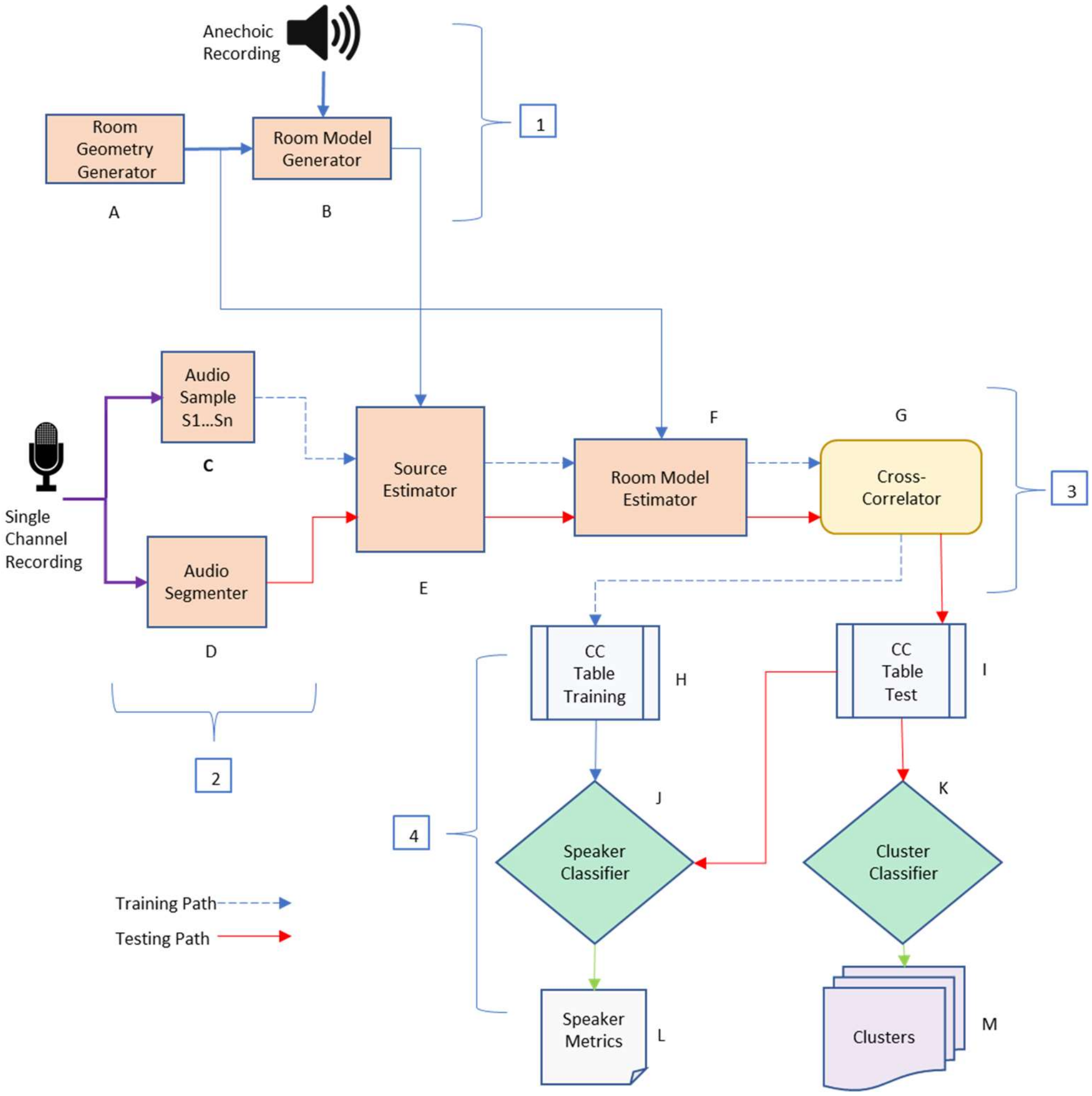

**Figure 24: Block Diagram of the Proposed System.**

The proposed system is divided into 4 subsystems described below:

*1) Room Geometry and RIR Generator:*

The first subsystem of the block diagram is The Room Geometry and RIR Generator (RGRG). This subsystem accepts the room geometry parameters (geometry of the room, absorption, location of the speakers and microphones) and calculates the RIRs using synthetic speech sources. The RGRG consists of two modules: The Room Parameter

Generator (A) and the Room Model Generator (B). The Room Parameter Generator creates the vectors that contain the geometry of the room and the location of the speakers and microphones (virtual and real). The Room Model Generator gets the room geometry vectors and calculates the RIRs between the sources and the microphones.

*2) Audio Pre-Processor:*

The second subsystem is the Audio Pre-Processor (APP). The function of the APP is to prepare the single-channel raw audio for analysis. It consists of two modules: The Training Sample Audio (C), and the Segmenter (D). The Training Sample Audio module contains the training samples of each of the speakers participating in the audio to be analyzed plus a sample of ambient noise. These samples are saved as .wav files and labeled independently.

The Segmenter module uses a VAD to create segments of the audio to be analyzed. Each segment of audio is saved as a .wav file of variable duration, with a minimum and a maximum length threshold. The segments that are less than a predetermined length are discarded.

*3) Analysis Subsystem:*

The function of the Analysis Subsystem (AS) is to calculate the cross-correlation between the emulated microphones. The AS consists of three modules: The Source Estimator (E), the Room Model Estimator (F), and the Cross-Correlator (G). The source Estimator gets the audio from the APP and deconvolves it with the RIR from the RGRG to estimate each possible source. The deconvolution is done to both the audio training samples and the segments for analysis. The Room Model Estimator then emulates each of the microphones by convolving the estimated sources with each of the corresponding RIR calculated using the model room geometry. Again, this is done for both the training samples and the analysis segments. Finally, The Cross-Correlator module calculates the cross-correlation between the microphones for each of the possible source combinations, for both training and testing.

*4) Classifier:*

The output of the AS subsystem is then handled to the Classifier. The classifier creates the cross-correlation sets of tables for training and testing. During training, only one table is created for each of the training audio samples. For testing, there is a cross-correlation table for each possible source for each of the segments, as was described in section 3.1. There are two possible paths of action once the cross-correlation tables are available. One path is to just run a clustering algorithm to group each segment in similar clusters or run a classifier that selects the best source that matches the training cross-correlation table. This research follows the classifier option, which will be discussed in the next section.

The final component of the Classifier subsystem is the Speaker Metric module. The function of this module is to calculate the statistics of each of the participants, e.g., for how long they have been active, and when they have been active. In this research, our metrics only focus on measuring the total time each participant has been active.

# Chapter 4. Experimental Implementation

This chapter presents the software and hardware implementation of the proposed method presented in Chapter 3. It begins by presenting the software tools for simulation, deconvolution, and data handling, and continue with the software implementation based on the AOLME environment. This implementation will be used in the experiments of Chapter 5 to evaluate the performance of the proposed method.

## 4.1 Software and Hardware Tools

The block diagram of Fig. 24 shows the need of developing several software modules to simulate the acoustics characteristics of the room, including RIRs and source image calculations, deconvolution for source estimation, cross-correlation, and classification. In summary, it is necessary to have code that performs the following operations:

1) Simulation of the geometry of a room
2) Calculation of all the RIRs based on the geometry of the room
3) Extraction of audio track from video recording
4) Segmentation of audio recording
5) Deconvolution of audio for source simulation
6) Simulation of microphone array
7) Microphone cross-correlation calculation
8) Analysis of microphone cross-correlation to identify the active speaker

9) Calculation of the metrics for each participating speaker.

Developing code for the above modules is a time-consuming task due to the large number of mathematical algorithms and calculations needed. Fortunately, there are software packages available and code libraries that simplify the implementation of these modules into a software framework for the experimental analysis. This dissertation have combined open-source code and commercial software, saving a considerable amount of time to the alternative of writing code from scratch.

### 4.1.1 Open-Source Code for Room Geometry, RIR Calculation, and Microphone Simulation

The open-source community of code developers offers an extensive variety of software libraries that cover a wide range of topics, from machine learning to financial market analysis, including acoustic simulations. Several acoustic simulation packages are available on GitHub for download. These packages are mainly designed to simulate the acoustics of environments for performing arts, such as theaters, stadiums, and recording studios. From these available packages, Pyroomacoustics was selected for RIR calculation and microphone simulation.

Pyroomacoustics [71] is an open-source acoustic simulation package that calculates the RIRs and simulates the reception of the audio at a set of virtual microphones located inside a virtual room. Pyroomacoustics uses the Image Source Method (ISM) to calculate the RIR between a source and any point inside the virtual room. The location of the images can be visualized with 2-D and 3-D representations of the geometry of the virtual room and the location of the sources and the virtual microphones. After simulating the location of the images, Pyroomacoustics calculates the RIRs to the target microphones and convolves the sample audio to simulate the reception at the microphones. Pyroomacoustics libraries' inputs are the geometry of the room, the location of the sources, the location of the virtual microphones, the absorption of the room, the sampling frequency, and the number of images to calculate. The outputs for the libraries include a set of arrays containing the RIR to each of the microphones and the emulated reception at each of the microphones. Figs. 25(a) and 25(b) show examples of 2-D and 3-D visualizations from Pyroomacoustics of a non-rectangular room 3 x 5 x 2 meters, with a circular microphone array with 6 microphones and one source. Fig 25(c) shows the same room with the simulated images. These types of representations will be used to approximate the AOLME models discussed later. Complete documentation on Pyroomacoustics functions and code can be found in [72].

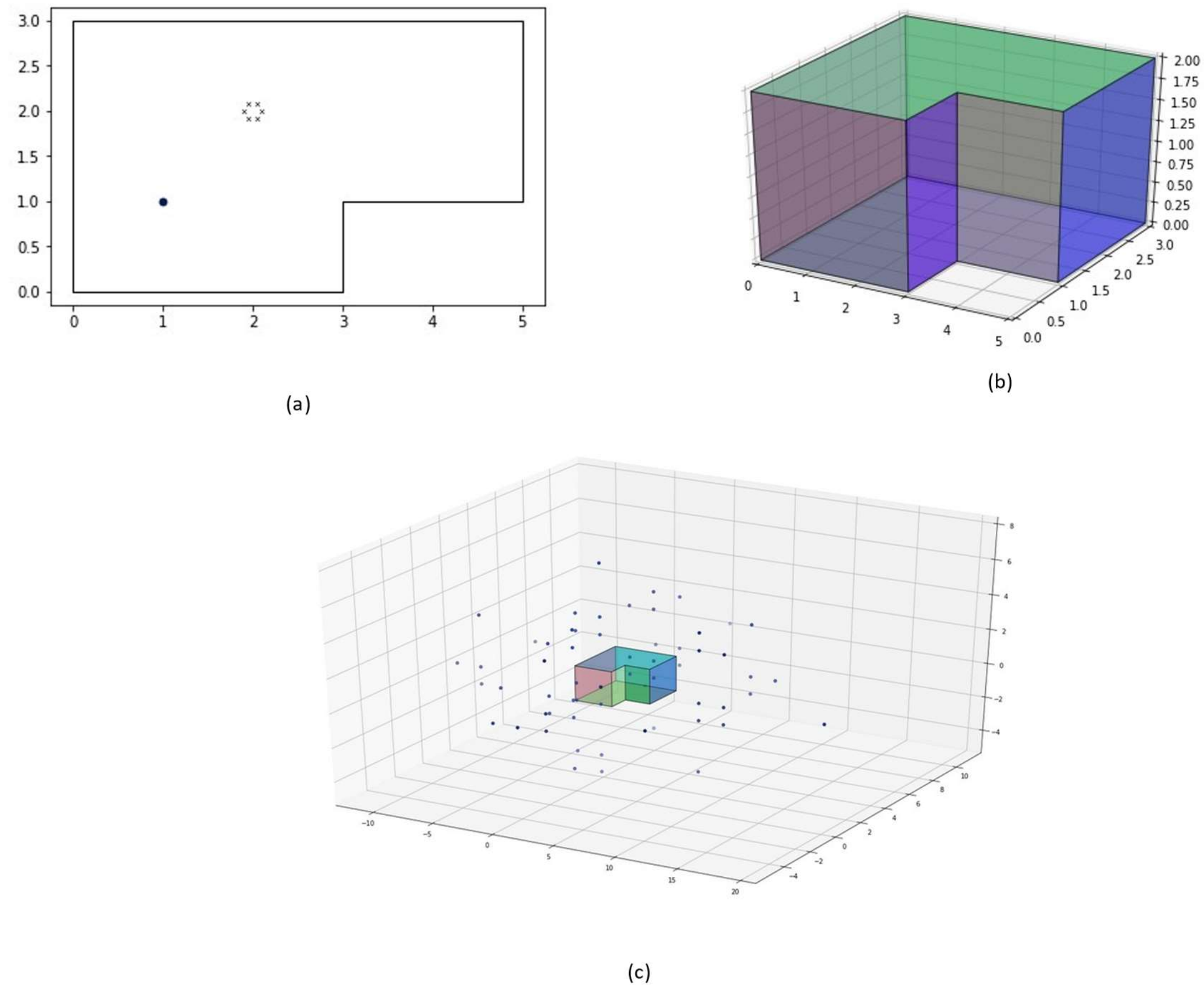

**Figure 25: Pyroomacoustics Models. (a) 2D. (b) 3D. (c) 3D With Images.**

The version of Pyroomacoustics used for this research (0.4.1) has some limitations that needed to be considered when developing the models: 1) Microphones and sources are always modeled as omnidirectional. There are no options to add unidirectional sources or other types of microphones (e.g., cardioids); 2) All rooms are square, with no round corners; 3) There is no option to add objects such as tables inside the room, and 4) Absorption is an empirical parameter that needs to be estimated by other means outside the software.

All experiments in this research were conducted using Pyroomacoustics version 0.4.1. For the room geometry calculations, Pyroomacoustics libraries are called from a Jupyter Notebook under Anaconda 3. The libraries were also called from within LabVIEW using scripts under Python 3.

#### 4.1.1.1 Pyroomacoustics Implementation

Implementation of Pyroomacoustics for virtual microphone simulation is accomplished by four steps that include simulation of the room, placement of the sources and microphones, calculation of the RIR from the sources to the microphones, and convolution of the sources with the RIR for microphone simulation. This is done by calling the classes in the Pyroomacoustics libraries as follows:

a) *Room Simulation:*

The room simulation contains the parameters of the room, such as its dimensions, the absorption, the number of images to be calculated, and the sampling frequency. For example, the following script will generate a room of dimensions 9 x 7 x 3 meters, with a total of 9 images, at a sampling frequency of 9600 Hz:

```
pyroomacoustics.room.Room([9.0,7.0,3.0], fs=9600, max-order=9)
```

b) *Sources and Microphone Placement:*

This script adds the sources and microphones to the model. Pyroomacoustics need a valid audio file for source location. The following script will locate a source at X= 2 m, Y = 3 m, and Z = 1 m from the origin, and two microphones at $X_1$ = 6 m, $Y_1$ = 4, and $Z_1$ = 1 m for microphone 1, and $X_2$ = 6 m, $Y_2$ = 4.5, and $Z_2$ = 1 m for microphone 2:

```
room.add_source([2.0, 3.0, 1.0], signal=audio)
mic_locs = np.c_[
    [6.0, 4.0, 1.0],  # mic 1
    [6.0, 4.5, 1.0],  # mic 2
]
room.add_microphone_array(mic_locs)
```

*c) RIR calculation:*

By calling room.compute_rir() the RIRs are calculated to each of the microphones, and the results are saved in the form of a list of lists at the **rir** attribute of **room**.

*d) Microphone Simulation:*

The final microphone simulation is obtained by calling simulate(). This convolves the sources with each of the RIRs and emulates the signals in each of the microphones. The results of the convolutions are stored in the signals attribute of room.mic_array.

Appendix A includes some of the scripts used in the actual experimental implementation of Pyroomacoustics. Refer to Pyroomacoustics documentation found at [72] for full description of the libraries and their algorithms.

### 4.1.2 Audio Segmentation

Recalling from previous discussions, the practical analysis of long audio recordings is not possible unless they are segmented into smaller frames. Audio segmentation plays a critical role in the overall performance of the proposed method; therefore, careful consideration should be made with the algorithms for audio segmentation. As previously indicated, the audio segments need to comply with two main requirements: 1) Contain audio from only one active speaker with minimum overlapping or mixing between speakers, and 2) are of a length that provides enough signal information for cross-correlation analysis. Both requirements are difficult to achieve, and in chapter 3 it was introduced the concept of VADs as a segmentation method that maximizes the information content of a single speaker. Two segmentation methods were considered during this research: Fixed Segmentation and Voice Activity Detection. In the end, it was opted for VADs due to their better performance results.

#### 4.1.2.1 Fixed Length Segmentation

The simplest way to segment audio is to divide it into fixed-length segments. Fixed length segmentation is a relatively simple and computationally inexpensive method where each audio segment has the same length, independently of their content. Because there is no intelligence in this method, there is a probability of some of the segments containing overlapping speech. Also, it is very unlikely that the audio can be divided into exactly equal parts making the last segment of shorter duration than the others.

A solution to minimize overlapping is to make the segments as short as possible. As it was discussed before, if the segments are too short, they may not contain enough information for calculating the cross-correlation. There is therefore a balance between the optimum length of the segments and the desired classification error.

One empirical way to find the length of the segments is by assessing the audio. If the audio contains well-separated speakers with little overlapping, the length of the segments can be longer than in acoustic scenes with noise or disorganized speech. This research conducted experiments with segment lengths varying from 250 ms to 1.2 s, obtaining different degrees of success. At the end, fixed-length segmentation was abandoned due to an undesirable number of errors and a lower performance when compared with voice activity detection segmentation.

#### 4.1.2.2 Voice Activity Detection

The VAD used in these experiments was programmed using MATLAB by a fellow graduate student at the University of New Mexico [73]. He used MATLAB Fast Fourier Transform (FFT) and Inverse Fast Fourier Transform (IFFT) functions to convert the audio from the time domain to the frequency domain and vice versa. First, the audio is converted into the frequency domain by applying the FFT, and then a 3000 Hz low pass filter and a 1000 Hz high pass filter are applied to remove some of the noise. The filtered audio is brought back to the time domain using the IFFT, and it is normalized afterward. An Amplitude Trigger (AT) with a threshold of 0.1 is used to determine the presence of speech or noise. If the amplitude is exceeded at any time, this will be the beginning of a speech, and this time is marked as $T_1$. The level is checked 300 ms after $T_1$. If the AT is exceeded again, we mark that time as $T_2$, and check the AT again after another 300 ms. If the audio does not exceed the AT, then it is marked as the end of the audio with a time $T_2 = T_1$ +300 ms; otherwise, the end of the audio will be $T_1 + T_2$. When the audio does not exceed the AT and is not in the time range of speech, it is classified as noise.

Using the information obtained from the filtered audio and adding a time offset (250 ms) to compensate for information missing in the filtered audio, we split the audio giving a maximum and a minimum time. If a noise segment is too small (under 250 ms), they will be combined with the audio segments since this is a pause of a person speaking. If there is noise or the audio is too long, it gets split into batches of a maximum time of 1.2s, with small exceptions that can go up to 1.449s.

It is important to notice that audio segmentation produces artifacts at the beginning and the end of the segment [74]. These artifacts are the familiar "clicks" we hear when listening to a sequence of segments. In audio processing, it is common practice to apply a window, a filter, and overlap of the segments [74] to allow for a smooth transition between them. Our method did not apply windowing or overlapping due to the possibility of altering the spatial content of the segments; therefore, it was better to accept a reasonable error instead.

### 4.1.3 Implementation Using the LabVIEW Graphical Programming

This research work used LabVIEW for deconvolution, array manipulation, classification, metrics, and user interface. LabVIEW is very popular in engineering due to its wide variety of built-in functions and its simplicity to create graphical user interfaces. It is not an open-source language, requiring the purchasing of a license.

LabVIEW requires additional toolkits for some advanced digital signal processing, statistics, and software integration. This research applied the Advanced Signal Processing toolkit for cross-correlation analysis, convolution, and deconvolution calculations. Because Pyroomacoustics was used for all room simulations, it was necessary to install the Python Integration Toolkit provided by Enthought. This toolkit provides LabVIEW with the capability of calling Python code directly. LabVIEW is then used as a wrapper to call Pyroomacoustics Python libraries from within LabVIEW. In this way, LabVIEW provides the user interface for the Pyroomacoustics inputs (i.e., room geometry, number of images, audio files), and processes the outputs (i.e., RIRs, microphone simulations), saves the data, and displays the results. All documents and detailed description of LabVIEW can be found at the NI website [75].

Instead of scripting code, LabVIEW uses a graphical interface that contains functional modules called VIs (short of Virtual Instruments). The VIs perform basic functions such as adding, subtracting, array manipulation, and logical operations, among others. There are more advanced VIs to calculate more complex operations such as convolution and inverse convolution, correlation and cross-correlation, and file manipulation, for example. Each VI transfers data using a wired connection, and there is a mechanism to handle and display execution errors. Fig. 26(a) and Fig. 26(b) show a screenshot of the internal block diagram and user interface, respectively, of the LabVIEW implementation used during this research for convolution, deconvolution, correlation, and cross-correlation operations between two files. The user can select between any of the operations using a drop-down selector. The implementation reads two text files that correspond to the audio files to be analyzed or convolved and a third file that corresponds to the RIR for convolution operations only. There are four graphics that represent the input files, the RIR, and the output of the cross-correlation calculation. The results can be saved as text files for later conversion into audio or any other format.

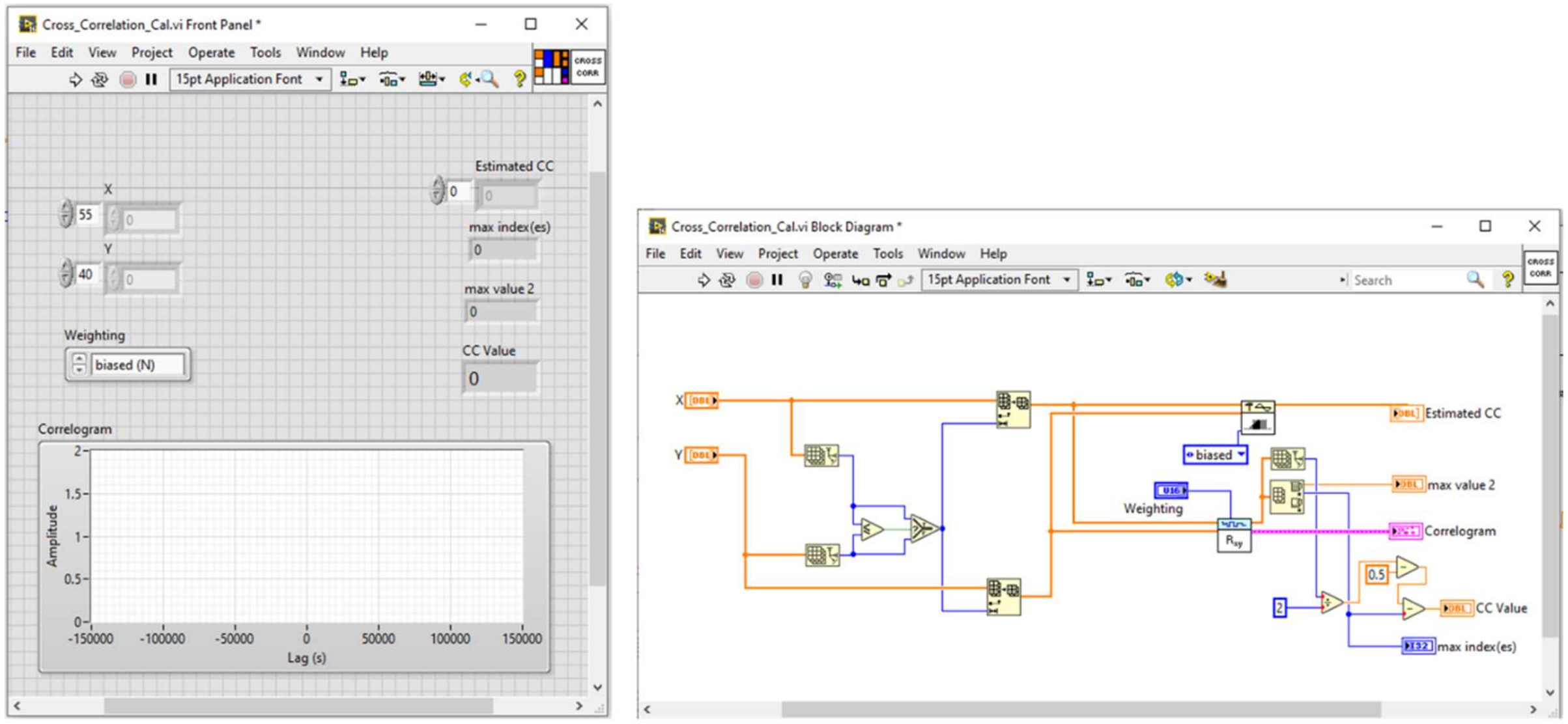

(a) (b)

**Figure 26: LabVIEW Sub VI for Cross-Correlation Calculation.**

Despite the popularity of LabVIEW among the engineering community, LabVIEW is many times regarded by hard-core coders as a language for those who do not know how to code. Its major deficiencies lie in the fact that its built-in functions are rarely modifiable, the block diagrams can get confusing if they are not divided into smaller VIs, and it is difficult to document and comment. The decision was to use LabVIEW because of the time-savings advantages it has over scripted languages.

#### 4.1.3.1 LabVIEW Implementation

The code written for this research used several built-in VIs available in LabVIEW version 2016, 32 bits. These VIs were implemented into more complex sub-VIs to run the calculations, data handling, user interface, file manipulation, and display of results. Although the code required the use of dozens of different VIs for simple mathematical operations and data flow, important calculations such as convolution and deconvolution were handled with LabVIEW built-in functions.

##### 4.1.3.1.1 Function VIs

The three functions VIs in this section were used to calculate convolution, deconvolution, and cross-correlation. They are part of LabVIEW's built-in library for signal processing. The algorithms for these functions are explained next.

###### 4.1.3.1.1.1 Convolution VI

This VI computes the convolution of two vectors **x** and **y.** The convolution can be computed by selecting either a direct method or a frequency domain algorithm that uses the FFT, being the latter the one used for this research. The VI that represents the convolution is shown in Fig. 27. Documentation on this VI can be found at [76].

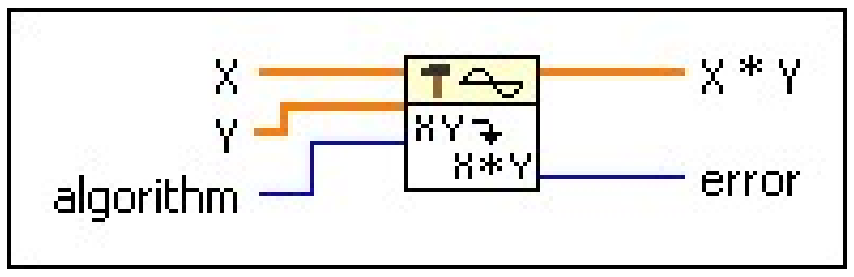

**Figure 27: LabVIEW Convolution VI.**

The algorithm works by padding the ends of **x** and **y** with zeros to make their lengths $M + N - 1$, as shown in (4.0) and (4.1):

$$x'_i = \begin{cases} x_i, & i = 0,1,\dots,N-1 \\ 0, & i = N,\dots,M+N-2 \end{cases} \qquad (4.0),$$

$$y'_i = \begin{cases} y_i, & i = 0,1\dots,M-1 \\ 0, & i = M,\dots,M+N-2 \end{cases} \qquad (4.1),$$

The convolution is computed by calculating the inverse FFT of the product of the FFTs of $x'$ and $y'$

$$\boldsymbol{x}'(f) = FFT(x') \qquad (4.2),$$

$$\boldsymbol{y}'(f) = FFT(y') \quad (4.3),$$

$$\boldsymbol{x} * \boldsymbol{y} = IFFT\big(\boldsymbol{x}'(f) \cdot \boldsymbol{y}'(f)\big) \quad (4.4),$$

where *IFFT* is the inverse *FFT*.

#### 4.1.3.1.1.2 Deconvolution VI

The deconvolution VI computes the inverse convolution of two vectors **x*y** and **y**. It returns the value of vector **x**. Fig. 17 shows the symbol for this VI. Documentation on this VI can be found at [77].

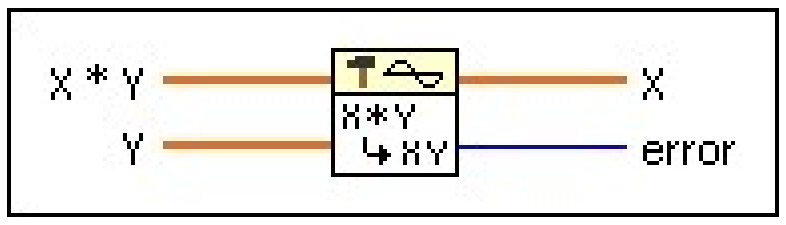

**Figure 28: LabVIEW Deconvolution VI.**

This VI implements the deconvolution by computing the Fourier Transform of the input **x*y** and **y**, then dividing them to create a new vector *h*. The vector **x** is computed by applying the IFFT to the sequence *h*.

#### 4.1.3.1.1.3 Correlation VI

The Correlation VI calculates the correlation coefficient *r* between two vectors **x** and **y**. Fig. 18 shows the icon for this VI. Documentation on this VI can be found at [78].

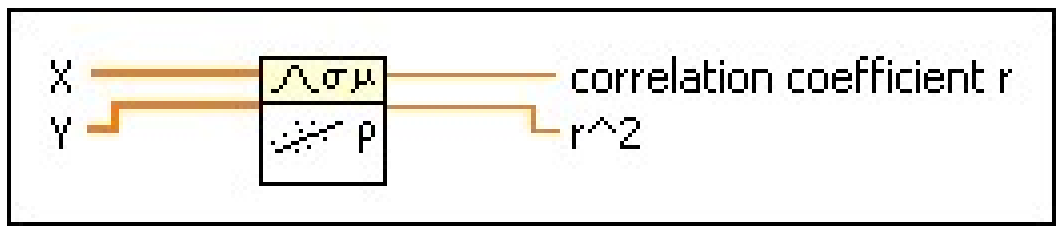

**Figure 29: LabVIEW Correlation VI.**

This VI calculates the linear correlation coefficient, also known as Pierson's correlation by (eq. number)

$$r = \frac{\sum z_x z_y}{n} \qquad (4.5),$$

where $z_x$and $z_y$are the standardized z-values of **x** and **y**. The standardized z-values indicate how many standard deviations **x** and **y** are above or below the mean.

#### 4.1.3.1.1.4 Cross-Correlation VI

The cross-correlation VI computes the cross-correlation between two vectors **x** and **y**. The inputs for this VI are the vectors $\boldsymbol{x_t}$ and $\boldsymbol{y_t}$, the weighting specifies the use of a biased or unbiased weighting in the cross-correlation calculation, being the former the one used in all the calculations. The maximum lag specifies the maximum value of the lag this VI uses to compute the cross-correlation. The maximum lag used equals max (*M*, *N*) – 1, where *M* and *N* are the lengths of $\boldsymbol{x_t}$ and $\boldsymbol{y_t}$, respectively. Fig. 30 shows the icon for this VI. Documentation on this VI can be found at [79].

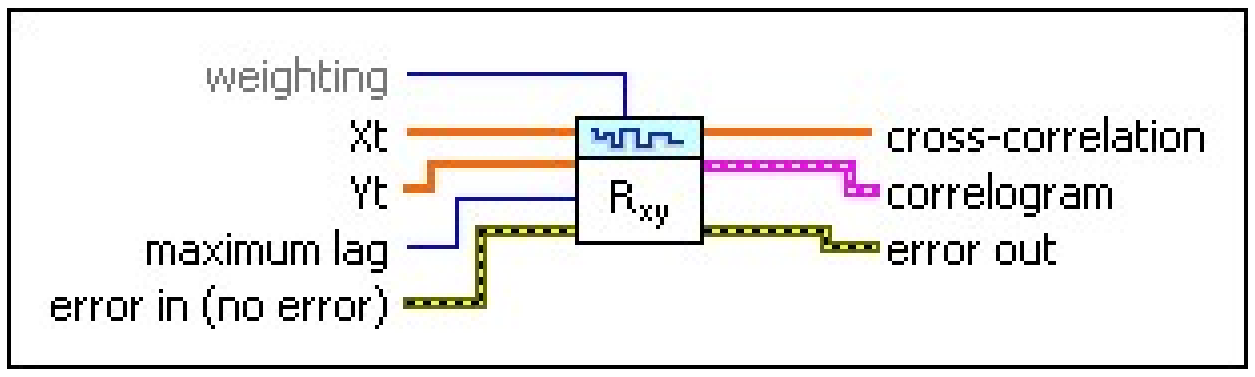

**Figure 30: LabVIEW Cross-Correlation VI.**

This VI computes the cross-correlation values between two univariate time series $\boldsymbol{X_t}$ and $\boldsymbol{Y_t}$ according to the following equation:

$$r_{xy}(k+N-1) = \frac{1}{a \cdot b \cdot w(k)} \sum_{n=0}^{N-1} X_t(n) Y_t(n+k), \quad 1-N < k < M \qquad (4.6),$$

where $= \sqrt{\sum_{n=0}^{w-1} X_t^2(n)}$ , $a = \sqrt{\sum_{n=0}^{M-1} Y_t^2(n)}$, $\boldsymbol{X_t}$ has length *N* and $\boldsymbol{Y_t}$ has length *M*. The length of the output is *N*+*M*–1. *w* is the weighting factor which in our case, *w*(*k*) = 1.

#### 4.1.3.1.2 Operational Sub-VIs

This section will cover the Sub-VIs that form the core of the code that performs the computations needed for the analysis. These Sub-VIs contain the function VIs covered in the previous section. Appendix B contains the front panels and block diagrams of these sub-VIs.

#### 4.1.3.1.2.1 Room Parameters Reader

The Room Parameters Reader Sub-VI reads the source locations, microphone locations, and 2D room dimension files created by the Pyroomacoustics Room Geometry Generator and formats them for the Room Model Generator Sub-VI. The room absorption, the room extrusion, and the number of images to calculate are just a pass thru. Appendix B section (a) shows the front panel and blocks diagrams for this Sub-VI.

#### 4.1.3.1.2.2 Room Model Generator

The Room Model Generator Sub-VI reads the room geometry parameters formatted by the Room Parameters Reader Sub-VI and runs the Python scripts that call the Pyroomacoustics libraries that compute the RIRs for the room model. This Sub-VI also reads the synthetic speech or noise .wav files used by Pyroomacoustics for the RIR calculations. The calculated RIRs are saved in .txt files for later retrieval by the Source Estimator Sub-VI. The Room Model Generator is used twice, first to calculate the room model RIRs for the source estimation, and again to emulate the virtual microphones using the estimated sources. Section Appendix B section (b) shows the front panel, block diagram, and inputs and outputs with more detail.

#### 4.1.3.1.2.3 Source Estimator

The Source Estimator takes the model RIR and estimates all the sources that will correspond to the audio segment that is being analyzed. For this estimation, this Sub-VI takes the segment of audio under analysis (corresponding to the real recording microphone) and deconvolves it with the RIRs for each of the source locations. The emulated sources are saved under .txt files for virtual microphone simulation using another instance of the Room Model Generator. Appendix B section (c) shows the details of this Sub-VI and a simplified block diagram.

#### 4.1.3.1.2.4 Cross-Correlation Model Calculator

This Sub-VI takes the results of the virtual microphone simulation from the second run of the Room Model Generator and calculates all the cross-correlations between the virtual microphones. The results are saved as cross-correlation tables and used for training and classification. Appendix B section d shows the details of this Sub-VI.

The output of this Sub-VI is a table that contains all possible cross-correlations between microphones for each of the possible sources. For the three speakers and three microphones example, the cross-correlation table would look like the one represented in Table IV. The first row is the cross-correlation microphone combinations, and the first column is the speakers. The numbers represent the array index where the max occurs.

**Table IV: Example Cross-Correlation Table Output from Model Calculator**

| | 1-2 | 1-3 | 2-3 |
|---|---|---|---|
| 1 | 96 | 5 | -5 |
| 2 | -32 | 0 | -83 |
| 3 | 5 | 5 | -130 |

#### 4.1.3.1.2.5 Model Classifier

The Model Classifier Sub-VI takes all the correlation tables, from training and testing, and performs the classification by comparing the testing results against the training templates. This is a very simple classifier that works by comparing each CC table for best similarity. For example, assume that the CC table IV corresponds to the training of speaker S1, and the analysis of an unknown audio segment produces the three CC tables shown in Table V(a), (b), and (c). The classifier simply counts the number of matches between each CC table and the training CC table. In this example, table V(a) has the greatest number of matches, indicating that the unknown segment corresponds to speaker 1. Appendix B section (e) shows the icon and front panel.

**Table V: Cross-Correlation Tables for Classification**

| | 1-2 | 1-3 | 2-3 | Match |
|---|---|---|---|---|
| 1 | 96 | 1 | -5 | 2 |
| 2 | -15 | 2 | -3 | 0 |
| 3 | 5 | 1 | -130 | 2 |
| Total for S1 | | | | 4 |

(a)

| | 1-2 | 1-3 | 2-3 | Match |
|---|---|---|---|---|
| 1 | 9 | 1 | -1 | 0 |
| 2 | -32 | 6 | -8 | 1 |
| 3 | 2 | 9 | -13 | 0 |
| Total for S2 | | | | 1 |

(b)

| | 1-2 | 1-3 | 2-3 | Match |
|---|---|---|---|---|
| 1 | 6 | 8 | -50 | 0 |
| 2 | -2 | 2 | -8 | 1 |
| 3 | 5 | 5 | -15 | 1 |
| Total for S3 | | | | 2 |

(c)

###### 4.1.3.1.2.6 Multi-Function Convolution and Correlator Visualizer

The Multi-Function Convolution and Correlator Visualizer is a full stand-alone Sub-VI used to manually convolve and deconvolve audio files and for correlation and cross-correlation analysis of files. Appendix B section (f) provides more information about this Sub-VI.

#### 4.1.3.2 Audio Laboratory

The purpose of the Audio Laboratory was to capture real audio in a controlled environment. This laboratory allowed to conduct experiments knowing the position of the speakers and microphones and control the content, duration, and characteristics of the analyzed speech. The results from the experiments performed at the audio lab were compared against the results obtained from our proposed method and the simulation software.

The audio laboratory consisted of a set of microphones, an audio processing device, an audio amplifier, loudspeakers, and the computer running the software that captures the recordings. The audio laboratory was physically configured to follow the common acoustic scene found on the videos analyzed in this research. This configuration used a set of loudspeakers located at the approximated position of the speakers sitting around a table. A set of microphones captured the audio at different locations of the lab, and one microphone was located at the same relative position as the recording microphone at the videos.

Fig. 31 shows a block diagram of the lab components. The set of microphones were the same type used in the recording of AOLME video. These microphones were connected to the Tascam® Audio Processor. This processor can capture simultaneous audio from all six microphones and send it digitally to the computer via USB. The computer processes the audio using Tracktion Waveform® audio processing software [80]. This software processes the audio from the microphones and saves it in separate .wav files that correspond to each of the microphones.

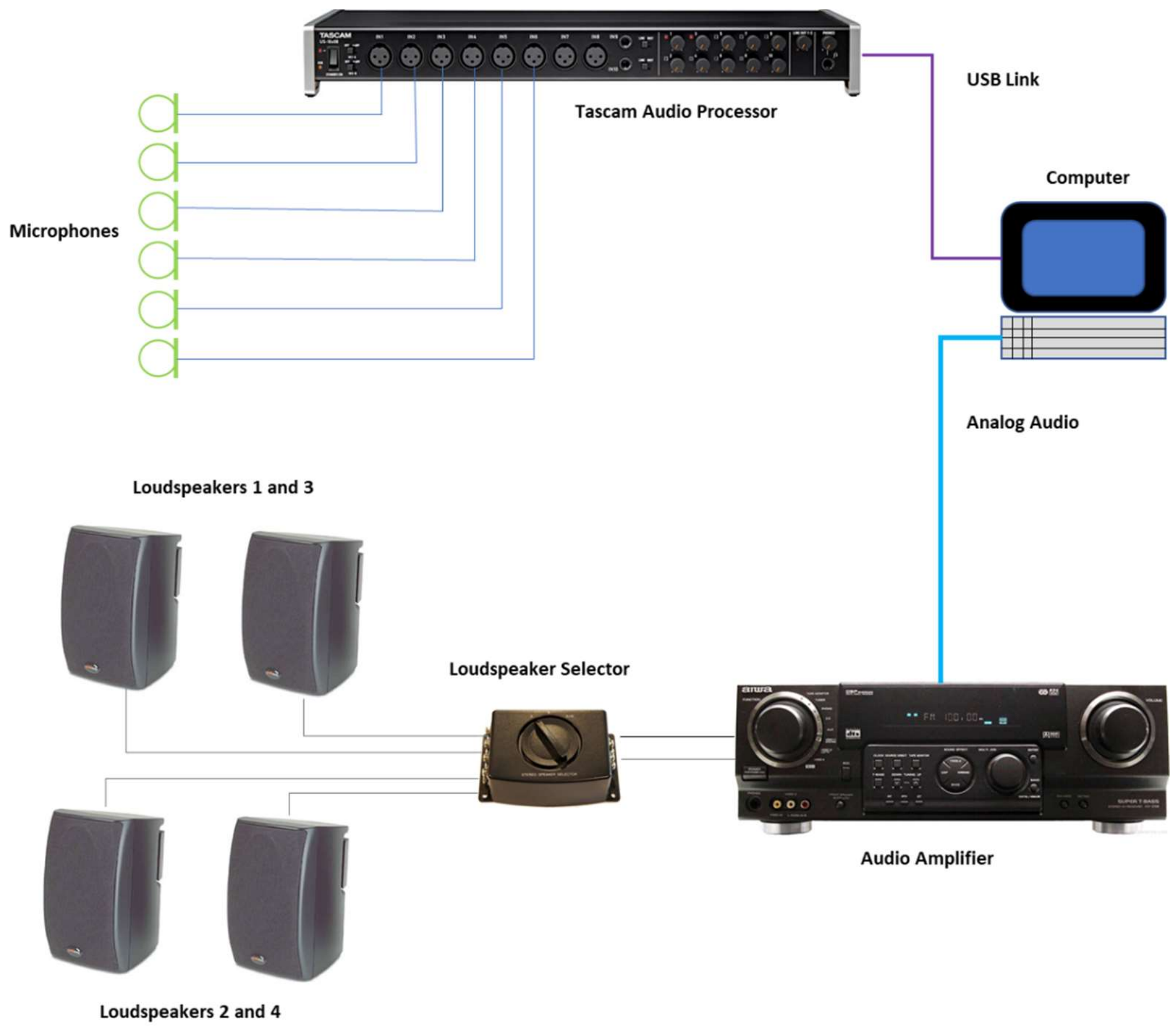

**Figure 31: Audio Lab Components.**

The simulation of the speakers is accomplished using a set of four loudspeakers connected to a stereo audio amplifier. Speakers 1 and 2 were simulated with the left stereo channel, while speakers 3 and 4 were simulated with the right stereo channel. A switch allows selecting between loudspeakers 1 and 3, and 2 and 4. The lab also included a Compact Disk (CD) player located at a certain distance from the table. This CD player was used to inject background noise during the experiments. Fig. 32 shows the actual audio laboratory setup where we can appreciate the location of its components.

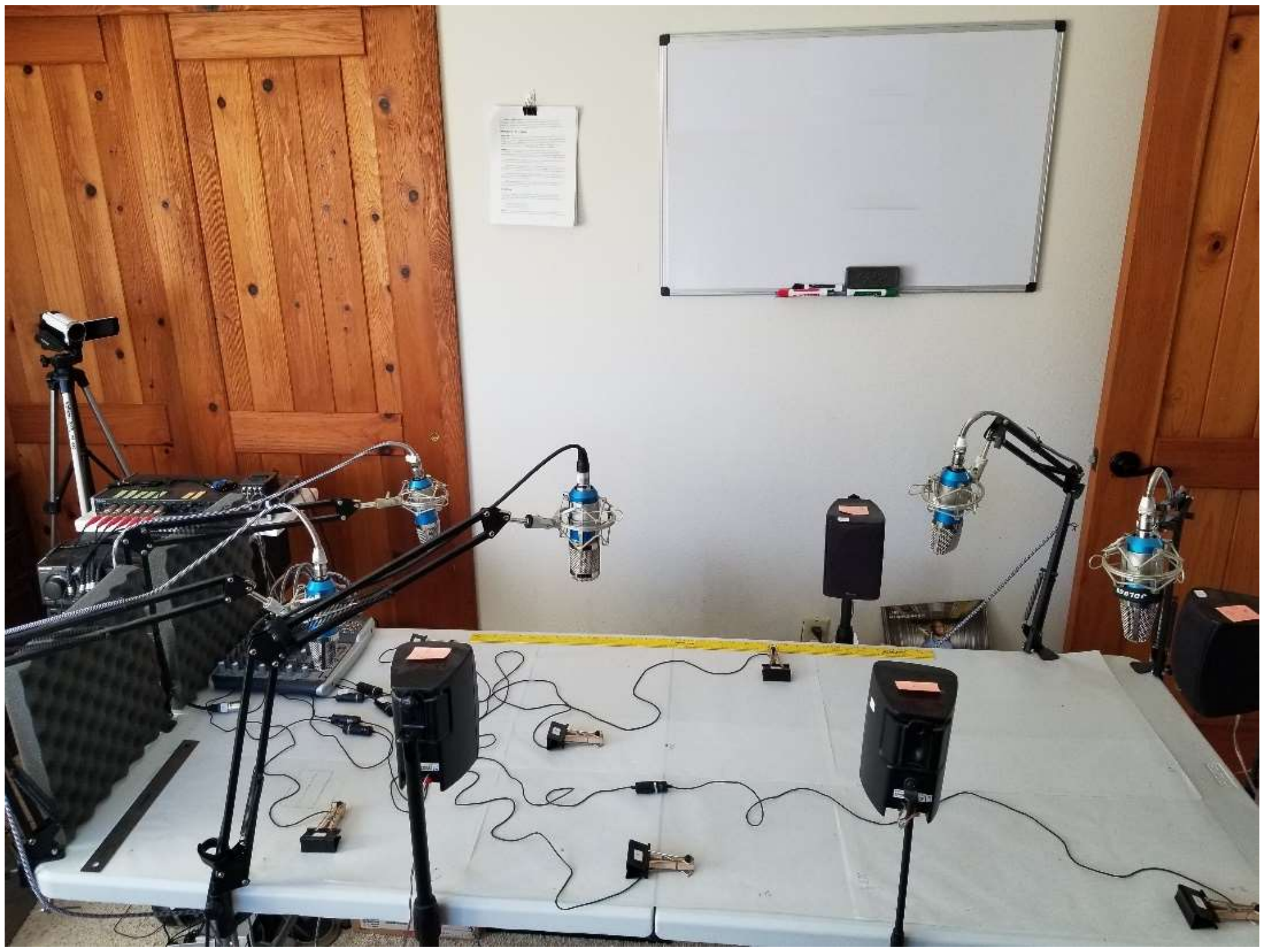

**Figure 32: Audio Lab Setup**

## 4.2 The AOLME Environment

The dissertation focuses on the analysis of audio from AOLME videos to assess the level of engagement of the participants. The AOLME environment is characterized by the presence of background noise, crosstalk, and other interferences that make it challenging for speaker identification tasks; therefore, to improve the identification rate, the simulation models must be optimized to fit this environment. This section studies the AOLME environment to find out how to best adapt the models to the acoustic characteristics of this environment and implement these models for the experimental section.

### 4.2.1 Characteristics of the AOLME Environment

Fig. 33 shows a screen capture from one of the AOLME videos analyzed in this research. The scene shows a typical collaboration table with four students and one instructor. It is common to have 5 to 10 of these tables, with three to six participants each, distributed in a room of approximated dimensions of 9 x 14 x 2.5 m. The camera is recording the audio via a single omnidirectional microphone that is resting on the top of the table. In addition to normal room noise, this environment presents other elements that make its dynamics more complex. For example, it is typical to have the participants shuffling papers, leaning over the table, eating, speaking simultaneously, and accidentally covering the microphone with books or other utensils. Furthermore, there are occasions when another staff member walks in and joins the group for a conversation.

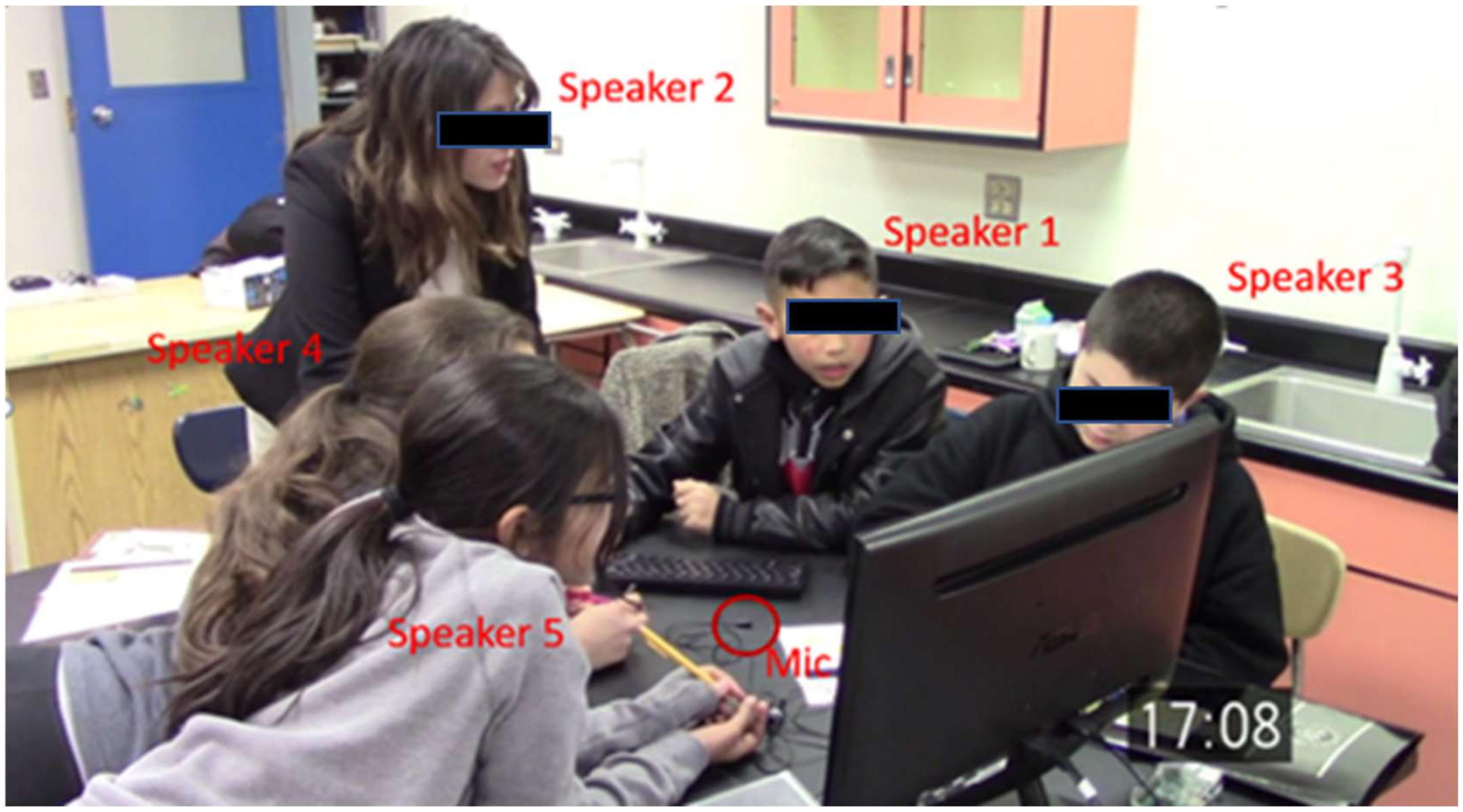

**Figure 33: Common AOLME Environment Setup.**

The first step in building the models is to approximate the location of the speakers and the recording microphone. By analyzing the scene in Fig. 33, it is possible to get some clues that can be used to approximate these locations. From Fig. 20, it is possible to estimate the relative locations of each of the speakers with respect to each other and the recording microphone. It is noticeable also that the position of the speakers forms a rectangle that can be translated into a 3D figure whose bottom area is the table and its height is defined by the tallest speaker.

The second step is to approximate the geometry of the room. From Fig. 33, it is possible to recognize that there is a nearby wall behind speakers 1 and 3. The second wall is located behind speaker 2 at a farther distance from speaker 2 than the first wall is located from speakers 1 and 3. There is no indication of any other wall or the presence of the celling, which we are assuming exists. It is also assumed that there are other tables nearby, but these cannot be seen in Fig. 33.

### 4.2.2 Preparation of the Experimental Models

As mentioned earlier, the models are based in part on the geometry of the room and the location of the speakers. Because this exact information is not available, the models need approximations based on the observations made from the video shot. Also, recalling from section 4.1.1, our version of Pyroomacoustics does not allow us to simulate complex environments like the one shown in Fig. 33, where we have the participants sitting around a table. Fortunately, the models do not need to be perfect, and we can make assumptions that will reduce their complexity.

#### 4.2.2.1 Approximating the Models Using Video Observations

We are ready to make some assumptions and approximations based on observations from the video. Fig. 34 shows another frame from the same AOLME video recording, where it is easy to estimate the relative distances between the participants. In Fig. 34, $\boldsymbol{H_1}$ represents the height of speaker 2, while and $\boldsymbol{H_2}$ represents the relative height of speakers 1, 3, 5, and 4. ***S*** represents the separation between speakers, and ***D*** represents the width of the table. We are assuming also in this observation that speakers 1 and 4 are separated by the same distance ***D.***

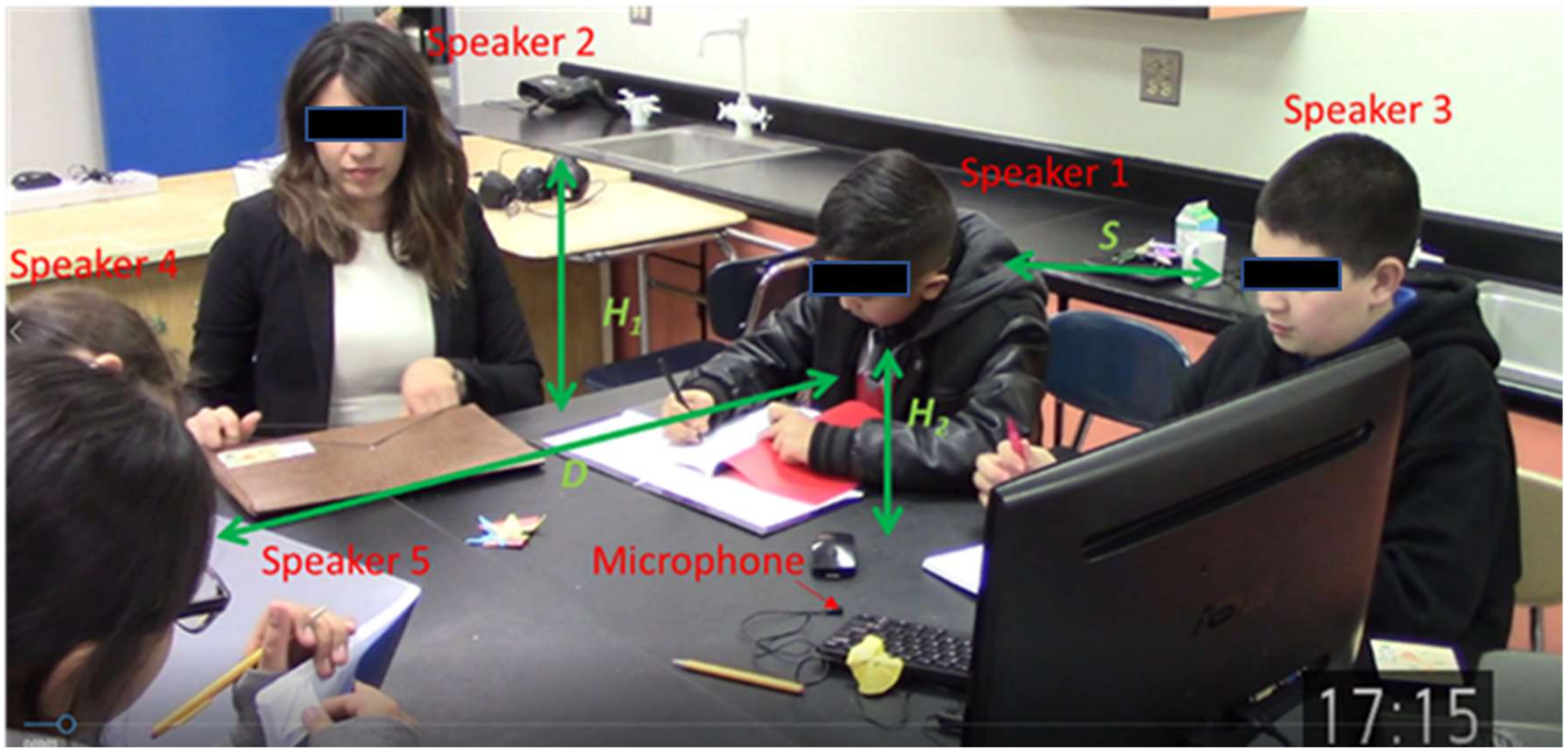

**Figure 34: Relative Positions of AOLME Participants.**

In Fig. 34, ***D*** can be approximated to the width of two standard commercial tables, which we can assume is 0.8 m x 2 = 1.6 m total. Speaker 2 is sitting about half of this distance, about 0.8 m from each edge of the table combination. Speaker 1 is close to one of the corners of the table, as it is speaker 4. The separation ***S*** between speakers can be approximated to 0.3 m, and the recording microphone can be located at half of this distance at the center of the table. Finally, $\boldsymbol{H_1}$ can be approximated using as reference the average waist to head distance of a young female, to about 0.5m, and $\boldsymbol{H_2}$ to the average waist to head distance of kids 11 years old, to approximately 0.4 m. These values are just examples to illustrate the principle on which we are basing the approximations. The actual model will not necessarily use these values.

There is no prior knowledge of the dimensions of the room that can be used to approximate its geometry. Observations about the location of the walls and the ceiling only provide a reference for the location of two walls. Nevertheless, it is possible to recognize, given the appearance of the scene, that the remaining walls are at a greater distance than the visible ones. This assumption does not provide a numeric value to the location of the walls or the ceiling, but it gives a clue of the behavior of the sound in the room.

Recalling Section 2.4, the human voice propagates mostly unidirectionally to the front of the speaker. Speakers 1 and 3 will project their voices toward speakers 4 and 5 and vice versa. Most of the sound energy from speakers 1 and 3 is absorbed by speakers 4 and 5, with some energy reflected by the table, some traveling to the ceiling of the room, and some other amount propagating to the walls behind. The walls reflect the residual sound energy to speakers 4 and 5, and the process repeats until all the energy is absorbed, following the $T_{60}$ rule. The same process applies when speakers 4 and 5 are active. In the case of speaker 2, there are no reflecting surfaces directly located in front of her, and the computer screen is located at a distance where the sound reflections from it can be considered of minimal influence, making the table the only reflecting surface. Under this model, it is possible to conclude that the sound energy of the participants is mainly contained within the boundaries of the table, and the contributions of the reflections due to the walls can be considered in practice as negligible, given the directionality of speech, the absorption of the speakers, and the separation of the speakers to the wall and the ceiling of the room.

The previous analysis indicated that it is not critical that the models take into consideration the reflections from the walls, suggesting that the rooms can be modeled as to be of infinite dimensions or to have an absorbance that is close to 1. Unfortunately, having a room of infinite dimensions will lead to a problem when modeling the sources. As discussed previously, the simulation software only allows for omnidirectional sources and microphones. In a wall-less room, Pyroomacoustics will create images from speech that equally propagates in all directions from the speaker, which we know is not accurate. The solution is to place the sources at very close proximity from the walls of the model and make the virtual room of the size of the table, thus reducing the propagation behind each speaker to negligible levels.

The analysis described above gives the basis for a first model representing the location of the speakers and the recording microphone. Recalling the 2D model of Fig. 19, we can set up a 2D model based on the acoustic scene of Fig. 34, representing the location with respect to the table of the 5 speakers and the real (recording) microphone. Note that this model includes a 6$^{th}$ "speaker" that represents the room noise. Representing the noise as a separate speaker allows for better discrimination between audio segments containing noise and those containing speech.

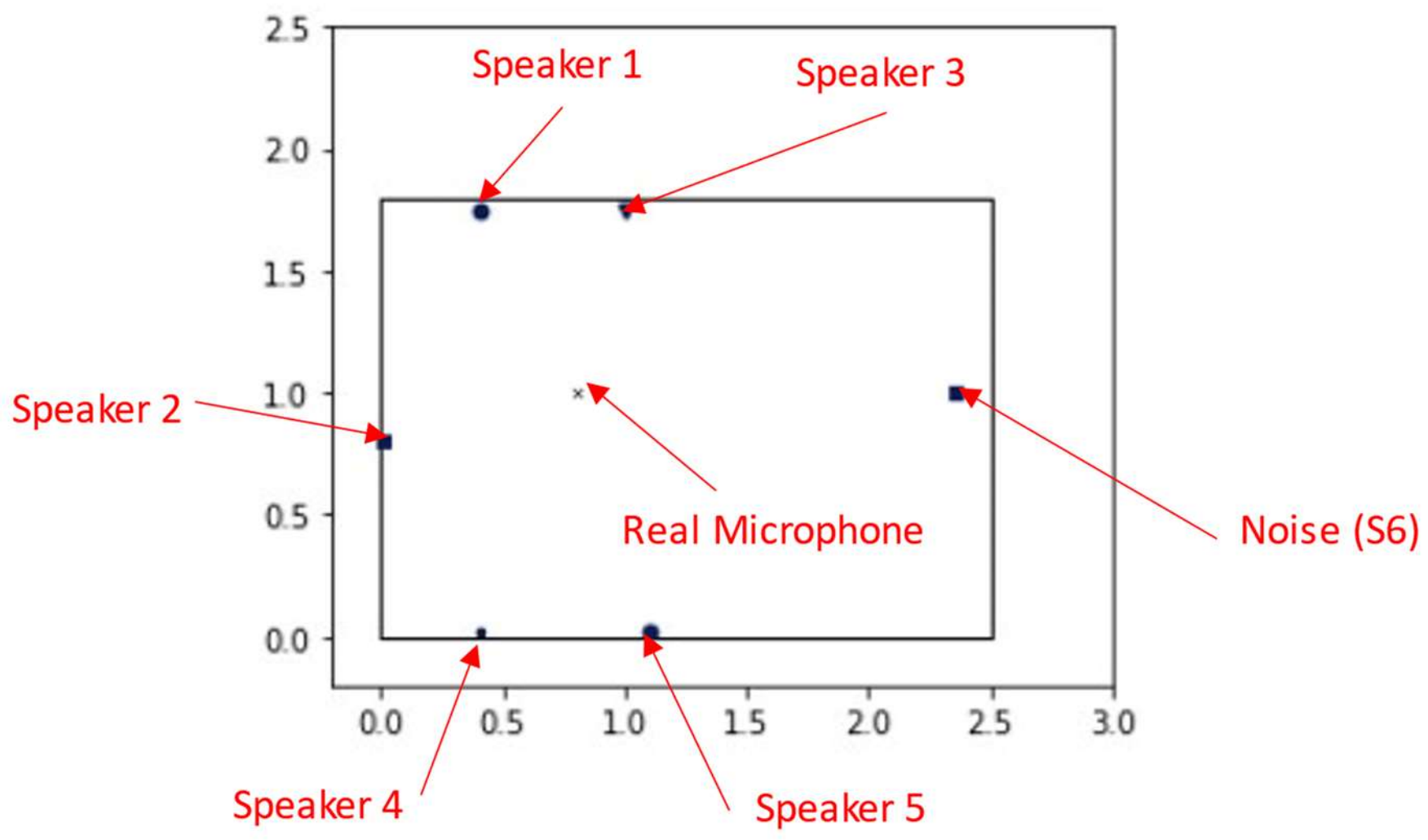

**Figure 35: Location of Speakers and Real Microphone.**

The Z dimension (room height) needs to be added to convert the 2D model into a 3D model. Because the perimeter of the room is limited to the size of the table, the table itself can be modeled as the floor of the room. With this approach, all locations will be zero-referenced with respect to the table.

The total height of the room can be approximated in a similar manner as it was done for the perimeter of the room. Because of the directionality of the human voice, it is expected that there will be a little transmission of voice energy to the ceiling; therefore, the reflections coming from above can be neglected. The ceiling can then be located at any height for as long it is above the maximum height of the taller speaker. Empirically, this value can be set, for example, at 1 m above the table. The 3D model for the room dimensions and the speakers is shown in Fig. 36.

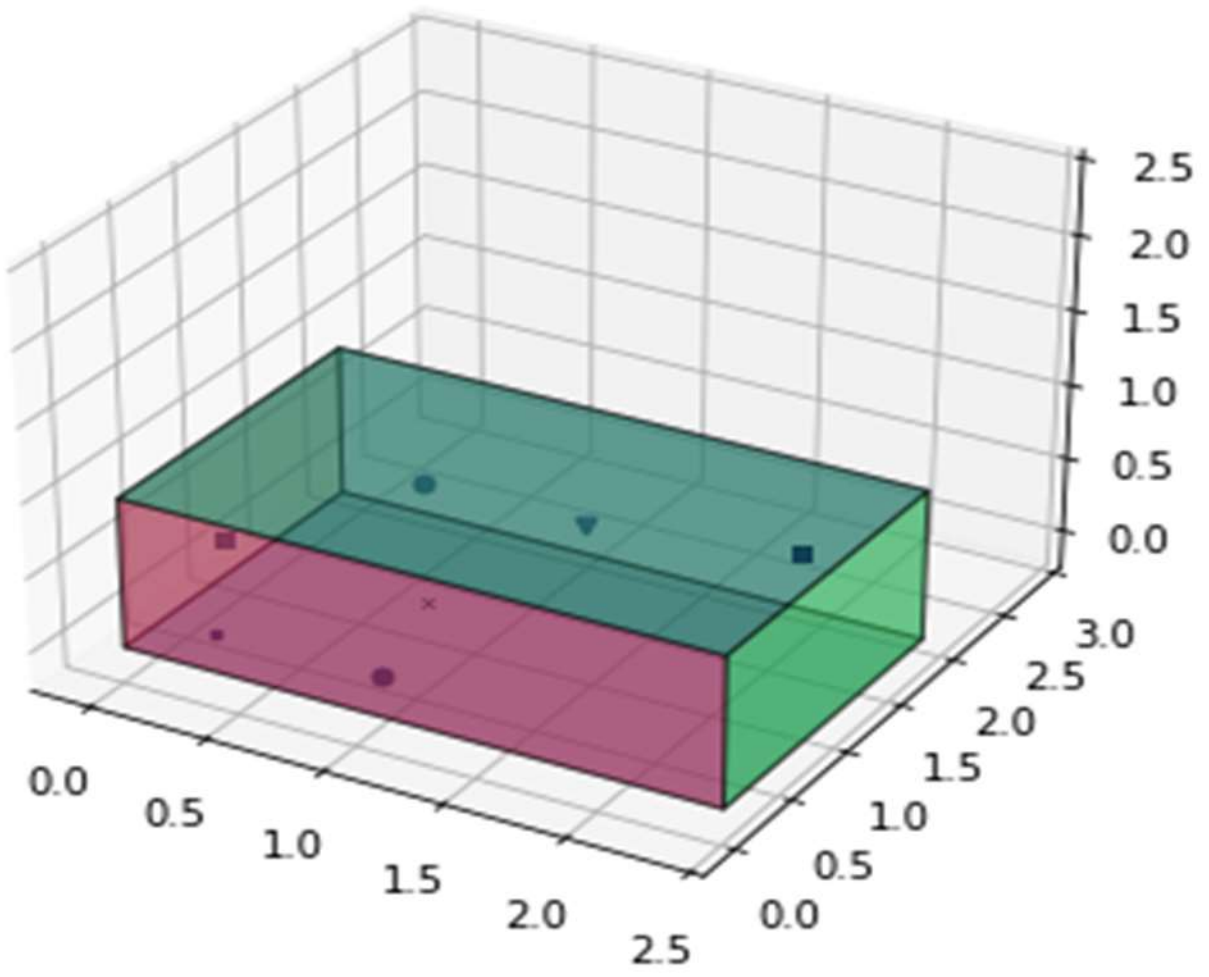

**Figure 36: 3D Model of the Virtual Room**

The last element needed to complete the model is the location of the virtual microphones. Their location is constrained by the dimensions of the virtual room and the maximum anti-aliasing distance between them. Also, it is necessary to consider that the array of microphones consists of a set of virtual microphones plus a real microphone, which is resting at the top of the table. At this location, the real microphone receives no sound reflections from the bottom; therefore, it can be assigned a Z value of zero. Because the real microphone is resting on the table, there are mechanical vibrations transmitted from the table. To simulate these vibrations, all models in this research include some value for the Z component of the real microphone.

The location of the virtual microphone array can be arbitrary, and the separation between microphones is not critical because the distance between two adjacent microphones will never exceed the maximum for anti-aliasing. However, it is of interest to have unique cross-correlation values between microphones. For this, the array should be in an asymmetric position with respect to the speakers in such a way that the value of the magnitude of the cross-correlation between microphones is different for each speaker. The Z value of the virtual microphones can be arbitrary, but because Pyroomacoustics can only simulate omnidirectional microphones, it is of advantage to locate them a certain height above the reference microphone. All the models in this research have microphones located at approximately the height of the speakers, allowing for simulation from all directions. Fig. 37 shows the complete 2D model derived from the five-speaker AOLME environment example. This type of model is used in all experiments in this dissertation, with the variations needed to fit the objective of the experiment.

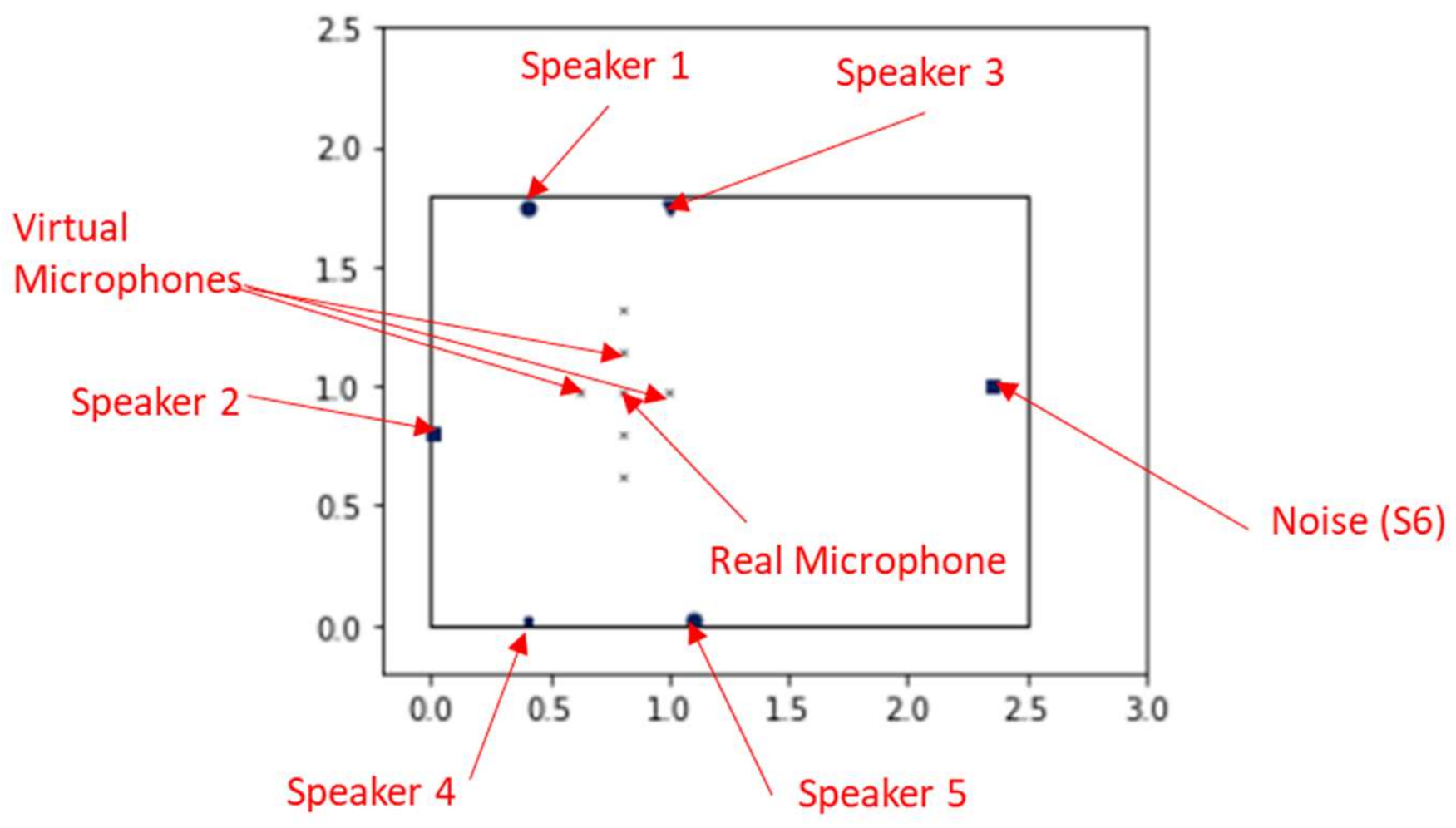

**Figure 37: Final 2D Model for AOLME Example.**

# Chapter 5. Results

This chapter presents the experiments conducted to evaluate the capability of the proposed method to identify speakers in audio segments. The experiments focused on three objectives: 1) To determine the suitability of Pyroomacoustics as a simulation package; 2) to evaluate the performance of the proposed method for diarizing and identifying speakers; and 3) to compare the performance of the proposed method against Amazon AWS and Google Cloud. These experiments included both real audio recordings from the audio lab and AOLME videos.

## 5.1 Evaluation of Pyroomacoustics

The objective of this experiment was to evaluate Pyroomacoustics as simulation software. This experiment compared the cross-correlation measured between real microphones and the cross-correlation between emulated microphones using Pyroomacoustics. This experiment was performed using the audio lab, with a Pyroomacoustics simulation based on the geometry discussed in Chapter 4.

### 5.1.1 Microphone Calibration

All audio recording devices have an electronic delay that varies from equipment to equipment. To measure the real cross-correlation between physical microphones, it is necessary to measure this electronic delay for each of the microphones and apply a calibration factor if necessary. Because Pyroomacoustics version 0.4.0 simulates all microphones as ideal and does not consider any delays, it is necessary to calibrate the real microphones to compensate for their delays before comparing them against any simulation.

One way to calibrate the microphones is to place them in an array configuration and locate this array in the proximity to an audio source. Fig. 38 shows a block diagram of the components needed to calibrate the microphones. This calibration setup consists of an audio source, speaker, sound processor, and microphone array. The audio source is driven by a signal generator, and the sound processor can acquire six audio channels simultaneously.

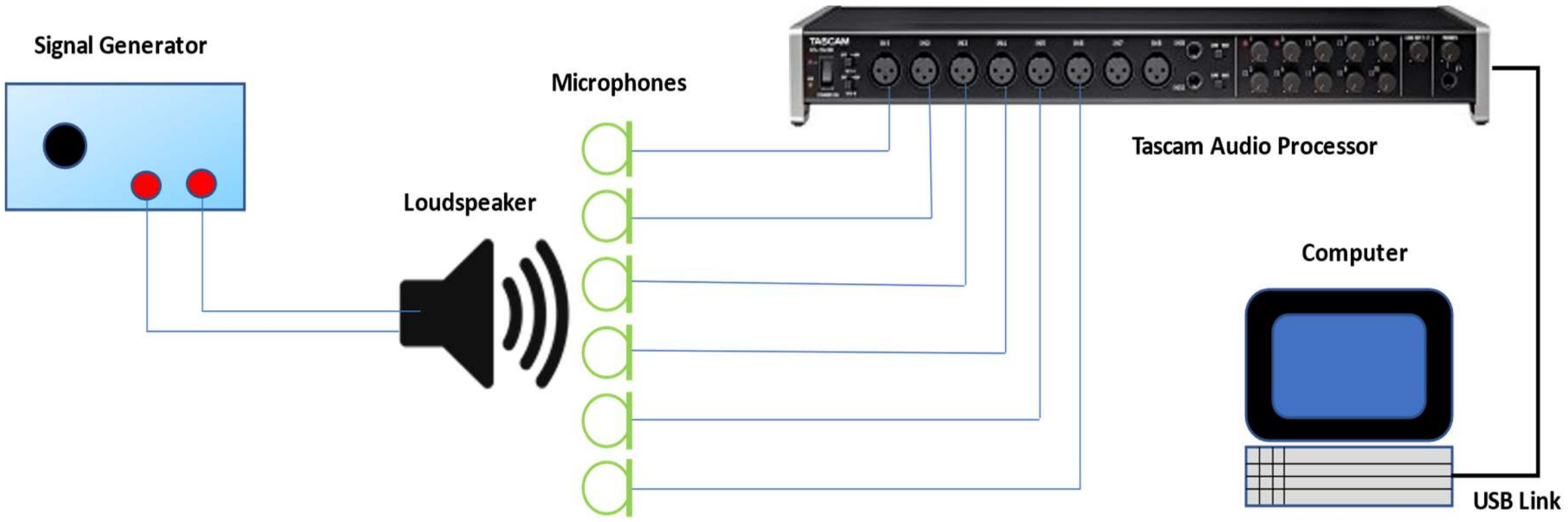

**Figure 38: Block Diagram of a Microphone Calibration Setup.**

*a) Calibration Preparation*

A homemade jig made of cloth pins was used to hold the six microphones for calibration. The configuration and separation of the microphones are shown in Fig. 39(a). The array of microphones was located next to one of the loudspeakers, as shown in Fig 39(b). With this configuration, the distance of each microphone to the sound source is about the same for all microphones, making the time differential of arrival between them neglectable.

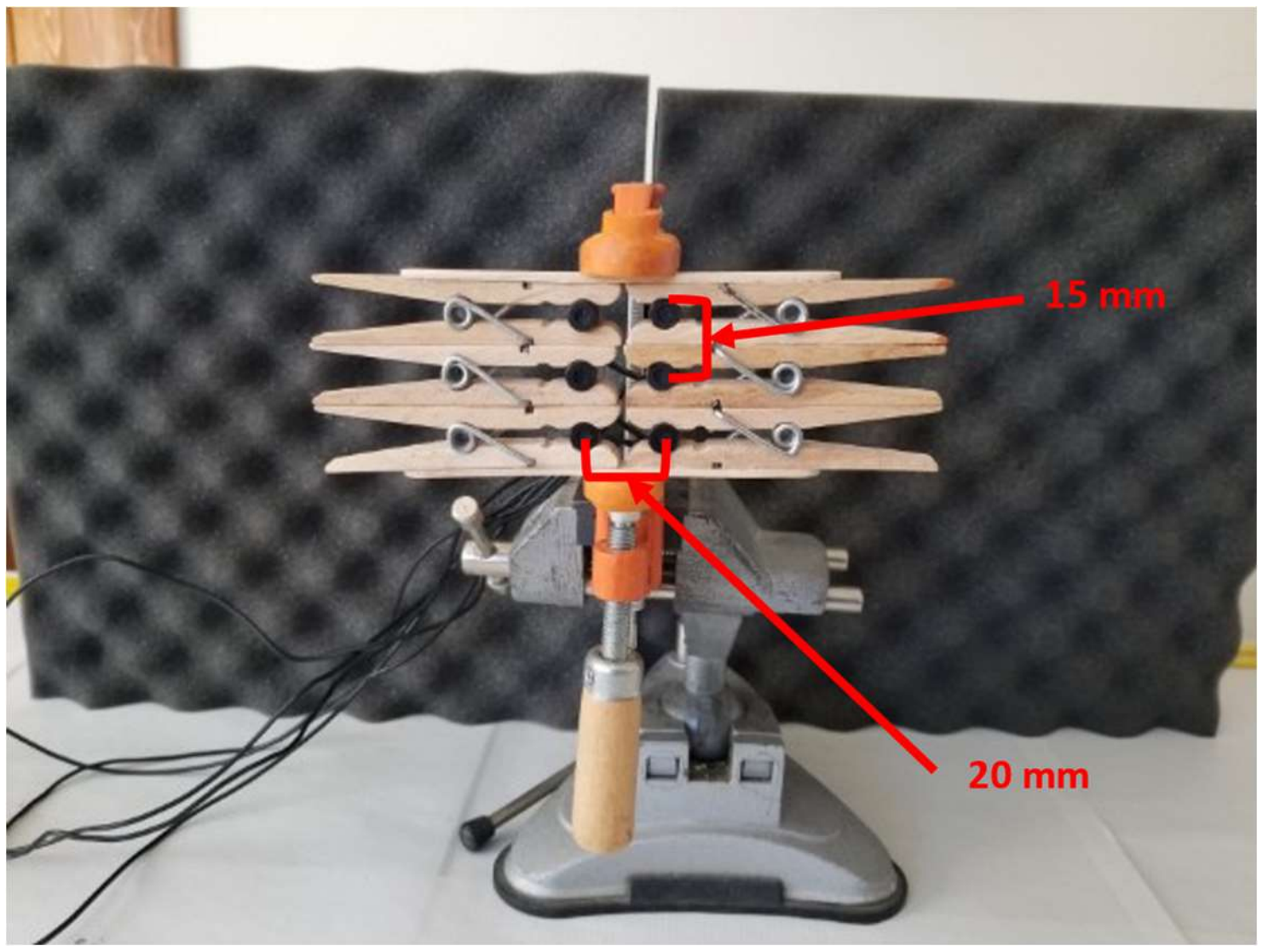

(a)

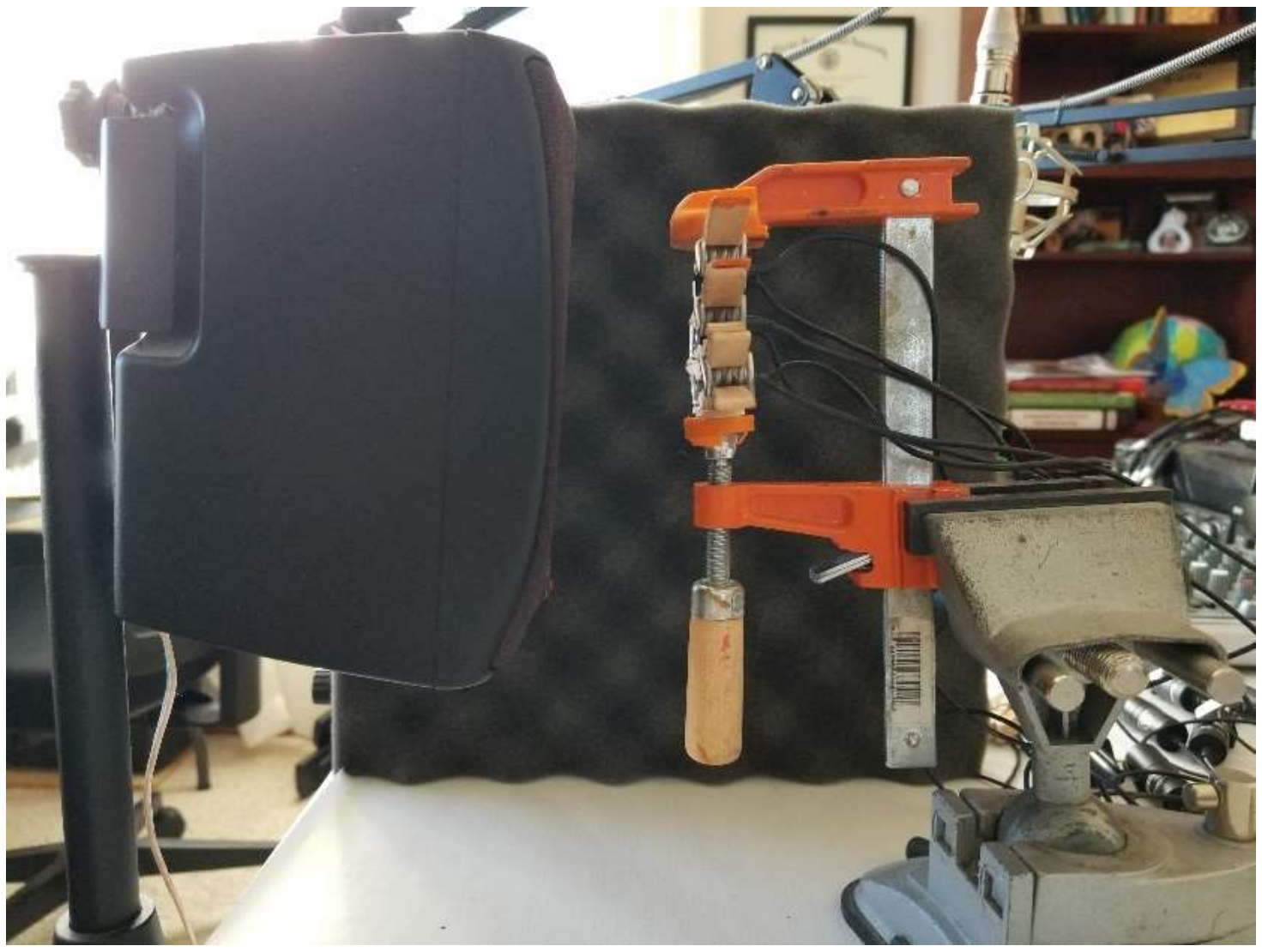

(b)

**Figure 39: (a) Microphone Calibration Jig. (b) Location to Loudspeaker.**

### b) *Calibration Execution*

A 450 Hz signal was applied to the loudspeaker using a signal generator, to the loudspeaker, and the output of the six microphones was collected simultaneously using the sound processor and the computer running Tracktion Waveform® software. Each channel recording was saved as a separated .wav file of 2 s duration, sampled at 48 kHz.

To measure the delay between microphones, each of the .wav files was converted into .txt files for cross-correlation analysis using the Multi-Function Convolution and Correlator Visualizer Sub-VI. Each combination of microphones was cross-correlated as shown in Table VI. The results in Table VI show that Microphones 1, 3, and 6 had zero cross-correlation between them. The same was observed between microphones 2, 4, and 5. Rather than apply a calibration factor, it is more convenient to segregate the microphones into groups and measure the cross-correlation between pairs that belong to the same group. Note that the results shown by Table VI correspond to the index of the array where the max cross-correlation occurs.

**Table VI: Cross-Correlation Table for Microphone Calibration.**

| | | Microphones | | | | | |
|---|---|---|---|---|---|---|---|
| | | 1 | 2 | 3 | 4 | 5 | 6 |
| Microphones | 1 | - | -50 | 0 | -49 | -50 | 0 |
| | 2 | 50 | - | 50 | 0 | 0 | 50 |
| | 3 | 0 | -50 | - | -49 | -50 | 0 |
| | 4 | 49 | 0 | 49 | - | 0 | 49 |
| | 5 | 50 | 0 | 50 | 0 | - | 50 |
| | 6 | 0 | -50 | 0 | -49 | -50 | - |

### 5.1.2 Audio Lab Setup and Model Configuration

Fig. 40 shows the laboratory setup for this experiment. The setup follows the general model configuration described in Chapter 4, but the microphones were distributed between the loudspeakers to maximize the cross-correlation value differences between microphones. The dimension of the lab setup allows for the microphones to be within the anti-aliasing distance already calculated of 0.95 m. Microphone 3 was kept in the same location as the recording microphone of the draft model.

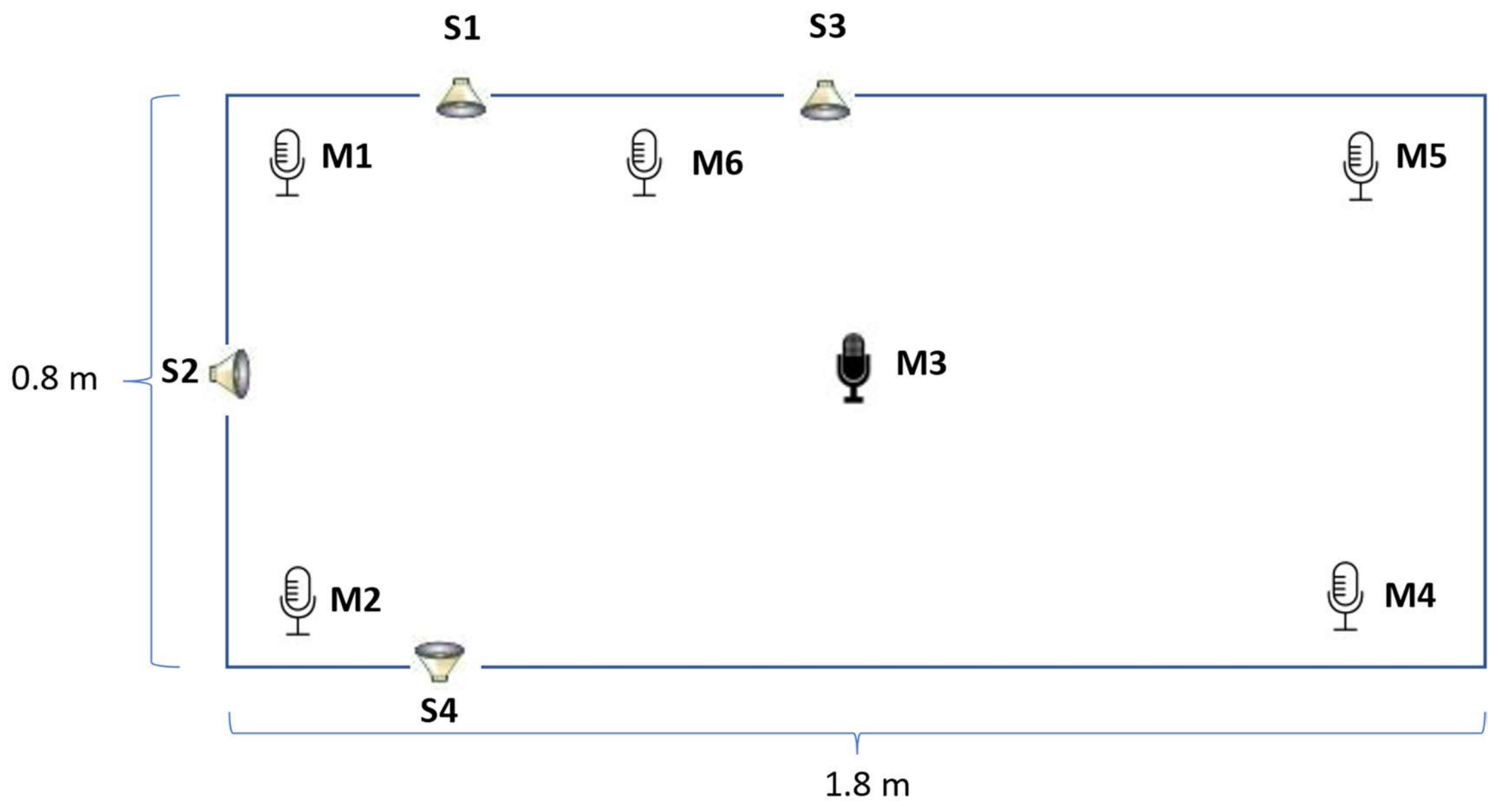

**Figure 40: Audio Lab Set-Up for Pyroomacoustics Evaluation.**

The Pyroomacoustics model was set up following the configuration of the audio lab. Because the audio lab has only 4 loudspeakers, speakers 5 and 6 were not included in the model. The virtual room perimeter was set to the size of the lab table, and the height of the room was set to a value of 1 m. The absorption of the model was set empirically to 0.95 and the number of images at 8. The microphone height was set to 0.025 m for all microphones, following the observations made in Chapter 3. Table VII shows the final dimensions of the virtual room and the location of the sources (loudspeakers) and microphones used to create the Pyroomacoustics model. The final 2D model geometry generated by Pyroomacoustics is shown in Fig. 41.

**Table VII: Dimensions of Virtual Room and Location of Sources (in m).**

| | | X | Y | Z |
|---|---|---|---|---|
| **Sources** | S1 | 0.4 | 0.79 | 0.25 |
| | S2 | 0.01 | 0.4 | 0.25 |
| | S3 | 1 | 0.79 | 0.25 |
| | S4 | 0.4 | 0.01 | 0.25 |
| | S5 | --- | --- | --- |
| | S6 | --- | --- | --- |
| **Mics** | M1 | 0.015 | 0.79 | 0.025 |
| | M2 | 0.015 | 0.01 | 0.025 |
| | M3 | 0.9 | 0.35 | 0.025 |
| | M4 | 1.39 | 0.1 | 0.025 |
| | M5 | 1.39 | 0.7 | 0.025 |
| | M6 | 0.7 | 0.79 | 0.025 |
| **Room** | CORNER 1 | 0 | 0 | -- |
| | CORNER 2 | 0 | 0.8 | -- |
| | CORNER 3 | 1.4 | 0.8 | -- |
| | CORNER 4 | 1.4 | 0 | -- |
| | EXTRUDE | 1 | | |

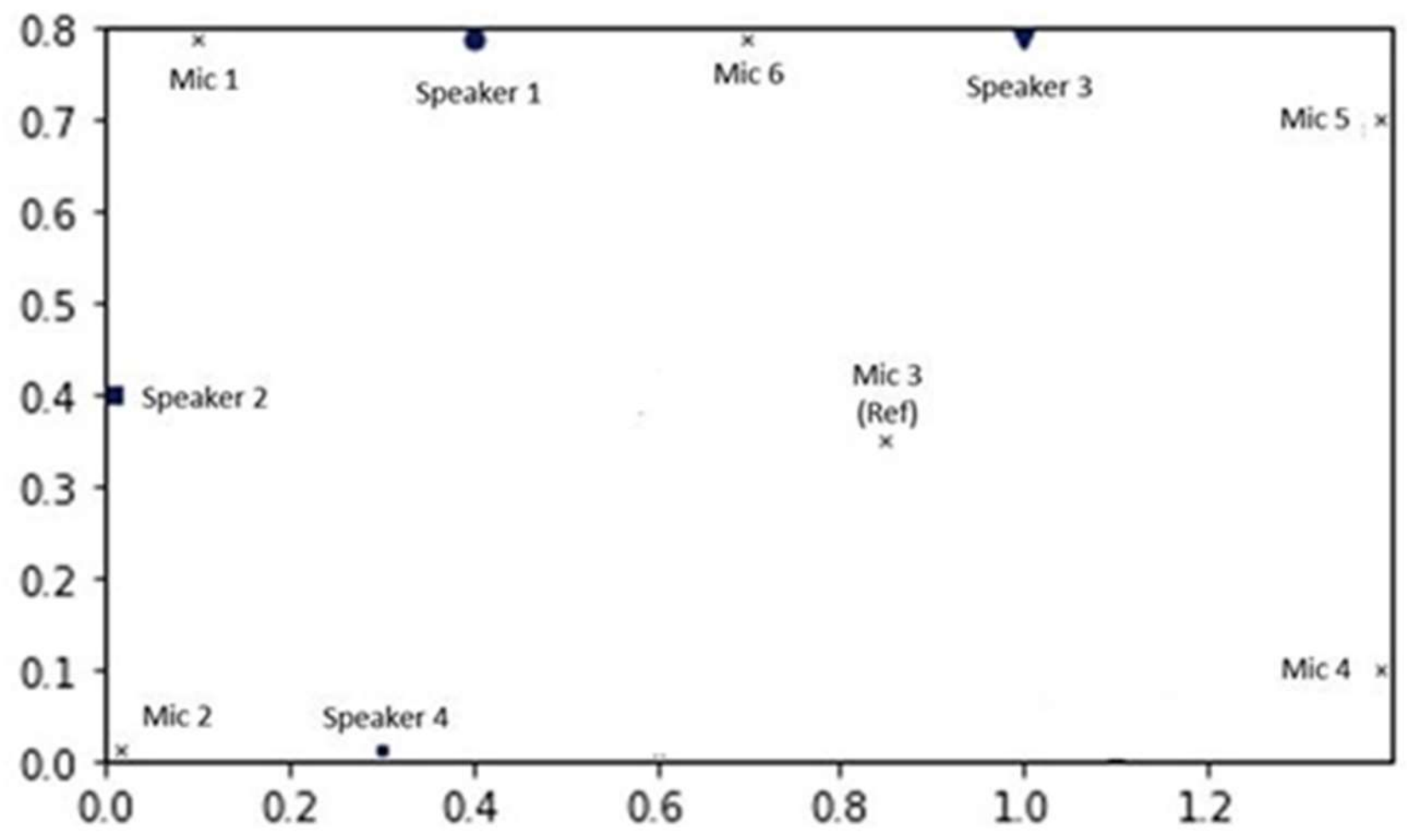

**Figure 41: Final 2D Model of Audio Lab Setup.**

### 5.1.3 Experimental Execution

Both audio lab and simulation sections of this experiment used as a source one anechoic male voice of 2 s of duration. The source was played sequentially on each of the loudspeakers corresponding to S1, S2, S3, and S4, and it was captured simultaneously into the six-channel audio processor, corresponding to each of the microphones. The six-channel audio then was saved as six independent audio files using Tracktion Waveform®.

The simulation with Pyroomacoustics used the geometric model of Fig. 41. Because there was no need to estimate the sources, the simulation of the reception at microphones M1, M2, M3, M4, M5, and M6 was accomplished by only running the Room Model Generator Sub-VI with the geometric model and playing the source at the location of speakers S1 to S4. The Sub-VI saved the results of each microphone simulation as a separate .txt file.

### 5.1.4 Results

The final analysis consisted of running the Multi-Function Convolution and Correlator Visualizer Sub-VI to calculate the cross-correlation for each of the real microphone audio files (ground truth) and the simulated microphone audio files. The cross-correlation was calculated between microphones of the same group as it was determined during calibration. There was no need for audio segmentation due to the short duration of the sample audio. Table VIII shows the results in ms of the offset between the ground truth and the simulated signals, corresponding to a sampling rate of 48 kHz.

**Table VIII: Experimental Results for Simulation Software Evaluation.**

| | S1 | | | S2 | | | S3 | | | S4 | | |
|---|---|---|---|---|---|---|---|---|---|---|---|---|
| | **Sim.** | **G.T.** | **Diff** | **Sim.** | **G.T.** | **Diff** | **Sim.** | **G.T.** | **Diff** | **Sim.** | **G.T.** | **Diff** |
| **1-3** | -0.58 | -0.34 | 0.24 | -1.24 | -1.62 | 0.38 | 1.56 | 1.88 | 0.32 | 0.72 | 0.38 | 0.34 |
| **1-6** | 0.20 | 0.30 | 0.10 | -1.06 | -1.30 | 0.24 | 1.82 | 1.96 | 0.14 | 0.08 | 0.06 | 0.02 |
| **3-6** | 0.80 | 0.68 | 0.12 | 0.18 | 0.30 | 0.12 | 0.26 | 0.06 | 0.20 | -0.60 | -0.32 | 0.28 |
| **2-4** | -1.08 | -0.88 | 0.20 | -1.90 | -1.56 | 0.34 | 0.48 | 0.50 | 0.02 | -1.98 | -1.56 | 0.42 |
| **2-5** | -0.48 | -0.50 | 0.02 | -1.90 | -1.62 | 0.28 | 1.62 | 1.46 | 0.16 | -2.58 | -2.08 | 0.50 |
| **4-5** | 0.58 | 0.40 | 0.18 | 0.00 | -0.02 | 0.02 | 1.12 | 0.96 | 0.16 | -0.56 | -0.50 | 0.06 |

Table VIII shows that the simulation correctly predicts the sign of the cross-correlation for each of the microphone pairs. The maximun offset difference is 0.5 ms which corresponds to a difference of 20%, and the average difference is 0.2 m, hence the simulation model appears to be sufficiently accurate for differentiating speakers based on their positions.

## 5.2 Controlled Environment Experiments

The objective of this next set of experiments is to evaluate the performance of our method to identify speakers in single-channel audio segments that were recorded under controlled conditions at the audio lab. There were two controlled experiments: The first experiment demonstrated the capability of the proposed method to identify two speakers based only on their location. The second experiment demonstrated the capability of the proposed method to identify multiple speakers independently of their spoken words.

### 5.2.1 Methodology

The approach for these experiments is to physically emulate an open collaborative environment such as AOLME in which we record audio containing speech with a single microphone. Because the geometry of the acoustic scene is known, we can create a model that numerically follows this real scene, and then evaluate the performance of the proposed method using this model. Conversely, by having control over some of the parameters, such as the location of the sources, it is possible to experiment with different microphone arrays and absorptions values to evaluate the performance of different models.

The controlled experiments used the same audio lab configuration and the same Pyroomacoustics model from the previous experiment. A change was made to the location of the speakers to better fit the distance that will be used for the AOLME experiments.

#### 5.2.1.1 Audio Lab and Model Preparation

Table IX represents the audio lab configuration for this experiment, with the location of the loudspeakers and the recording microphone (MIC3). The audio was recorded using the Canon video camera connected with MIC3, and the video recording was saved in the internal SD card of the camera, the same way it is done with AOLME recordings. Ambient noise was injected using the CD player with background noise from one of the AOLME video sessions.

**Table IX: Distribution of Microphones and Sources for Controlled Experiments.**

| | | X | Y | Z |
|---|---|---|---|---|
| **Sources** | S1 | 0.4 | 0.79 | 0.25 |
| | S2 | 0.16 | 0.5 | 0.25 |
| | S3 | 0.65 | 0.79 | 0.25 |
| | S4 | 0.3 | 0.2 | 0.25 |
| | S5 | --- | --- | --- |
| | S6 | 0.98 | 0.4 | 1.5 |
| Mics | M1 | 0.6 | 0.6 | 0.025 |
| | M2 | 0.65 | 0.55 | 0.025 |
| | M3 | 0.6 | 0.55 | 0.025 |
| | M4 | 0.6 | 0.45 | 0.025 |
| | M5 | 0.6 | 0.5 | 0.025 |
| | M6 | 0.55 | 0.55 | 0.025 |
| | M7 | 0.6 | 0.65 | 0.025 |
| **Room** | CORNER 1 | 0 | 0 | -- |
| | CORNER 2 | 0 | 0.8 | -- |
| | CORNER 3 | 1 | 0.8 | -- |
| | CORNER 4 | 1 | 0 | -- |
| | EXTRUDE | 2 | | |

The lab setup was translated into the Pyroomacoustics 2D model shown in Fig. 42. Noise is represented as "speaker" S6 and placed it in a relative location that resembles the location of the CD player. All sources and microphones kept the same Z coordinate value as in the previous experiment (0.25 m), except for the noise S6, which is located at Z= 1.5 m, to better represent the location of the CD player.

This experiment (and for all subsequent experiments in this research), used a linear cross-type virtual microphone array with 7 elements, with the recording microphone located at the center of the array. This type of microphone configuration is flexible and compact and allows its implementation in other models with different geometries. The separation between microphones in the array was set to 0.05 m, which is a distance commonly found in commercial microphone arrays, which is around 0.025 m to 0.040 m. The virtual microphone array is located at an offset position to the loudspeakers, avoiding any symmetry with them. This location should provide more distinctive cross-correlation results between microphones for better differentiation.

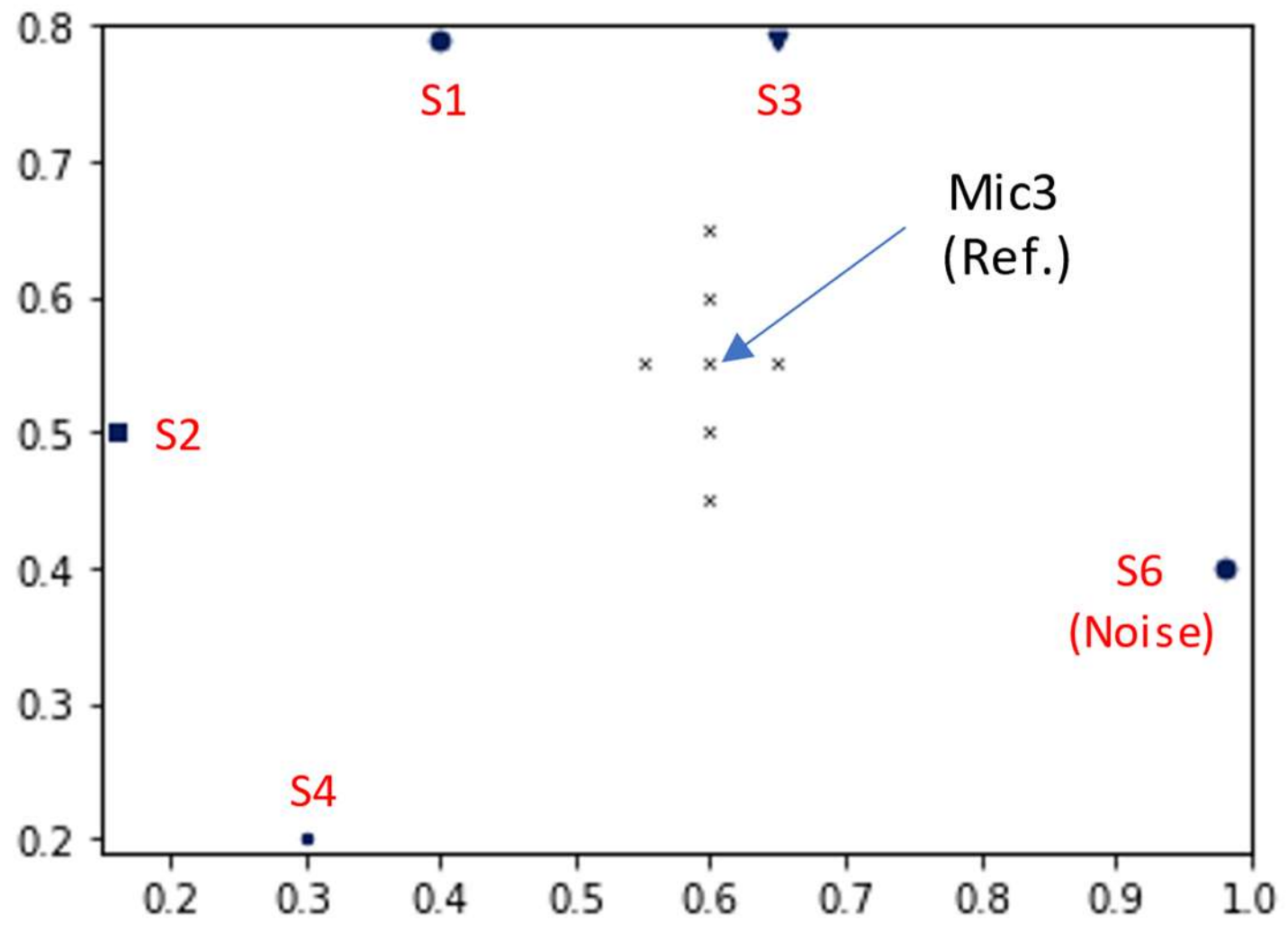

**Figure 42: 2D Model for Controlled Experiments.**

### 5.2.1.2 Evaluation Criteria

A common method to measure the performance of diarization systems is the Diarization Error Rate (DER) [81], [82]. The DER is defined as the fraction of the time that is not attributed correctly to a speaker or non-speech [38]. It can be calculated as the summation of all errors as follows:

$$DER = \frac{FA+Miss+Overlap+Confusion}{Reference\ Lengh} \quad (5.0),$$

where *FA* is the length of False Alarms, *Miss* is the length missed speech segments, *Overlap* is the total length of overlapped speech, *Confusion* is the total length of misclassified segments, and the *Reference Length* is the total length of the audio reference. Overlap was not used in any of the tests.

### 5.2.2 "HAL 9000" Experiments

The objective of this experiment is to demonstrate that the proposed method can identify speakers solely on the location of the speaker and independently of their speech characteristics. This was accomplished by using non-anechoic audio as the speech source, obtained from a raw video clip of a classic movie.

Many of the software packages for speech processing found during this research provided some sort of test files for evaluation. One of these demos included a phrase from the classical 1967 movie "2001, a Space Odyssey". In this movie, the human crew faced the rebellion of the spaceship's computer, "HAL 9000", which after some malfunction, attempts to kill the crew. The phrase "I'm sorry Dave, I'm afraid I can't do that" is still very well-known nowadays when we discuss the implications of artificial intelligence taking over the control of critical missions.

This experiment used a clip of 128 seconds of duration where this famous phrase is spoken. This clip included the conversation between Dave, who is inside a space pod (Fig. 43, clip1), and HAL at the mothership (Fig. 43, clip2). The video scenes switched between the space pod and the spaceship, with voices coming from radio transmissions, or the inside of the spaceships, depending on the scene. There is also some background noise from the electronic equipment at the space pod. This clip can be downloaded from YouTube at https://www.youtube.com/watch?v=Wy4EfdnMZ5g&t=11s

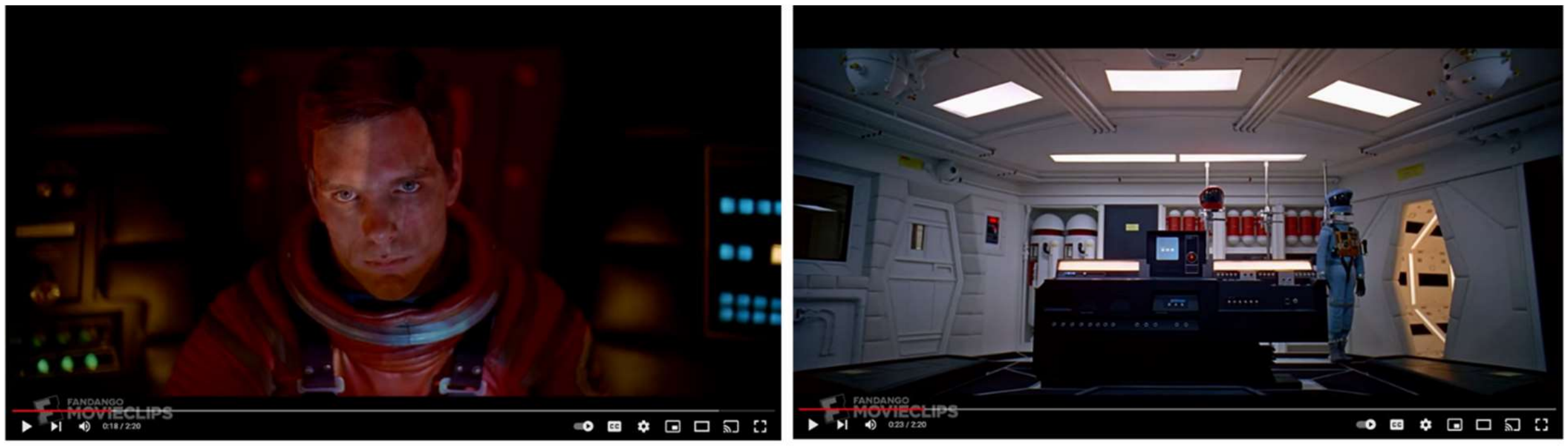

(Clip1) (Clip2)

**Figure 43: Video Clips of Dave (Clip1), and HAL (Clip2) SOURCE: Fandango Movie Clips.**

Two sets of experiments were performed: Experiment 1 was aimed to determine if there was any biasing on the results as the product of the location of the speakers. Experiment 2 evaluated the effects of training in the results.

#### 5.2.2.1 Source Preparation and Editing

This experiment played Dave's and HAL's voices independently at the loudspeakers. To do so, the YouTube video was converted into a single channel using Audacity ® version 2.4.2 [83] and saved as a MP4 48 kHz audio (See Fig 44 (a)). Then, using Audacity, the segments with voices of Dave and HAL were cut and pasted in two separate channels of a new stereo track (Fig. 44 (b)). The intervals with noise were converted into silence to allow the recording noise to come from an external source. Dave was placed on the right track and Hal was placed on the left track. The noise segments were copied and pasted into a separate audio track and burnt into a CD (Fig. 44 (c)).

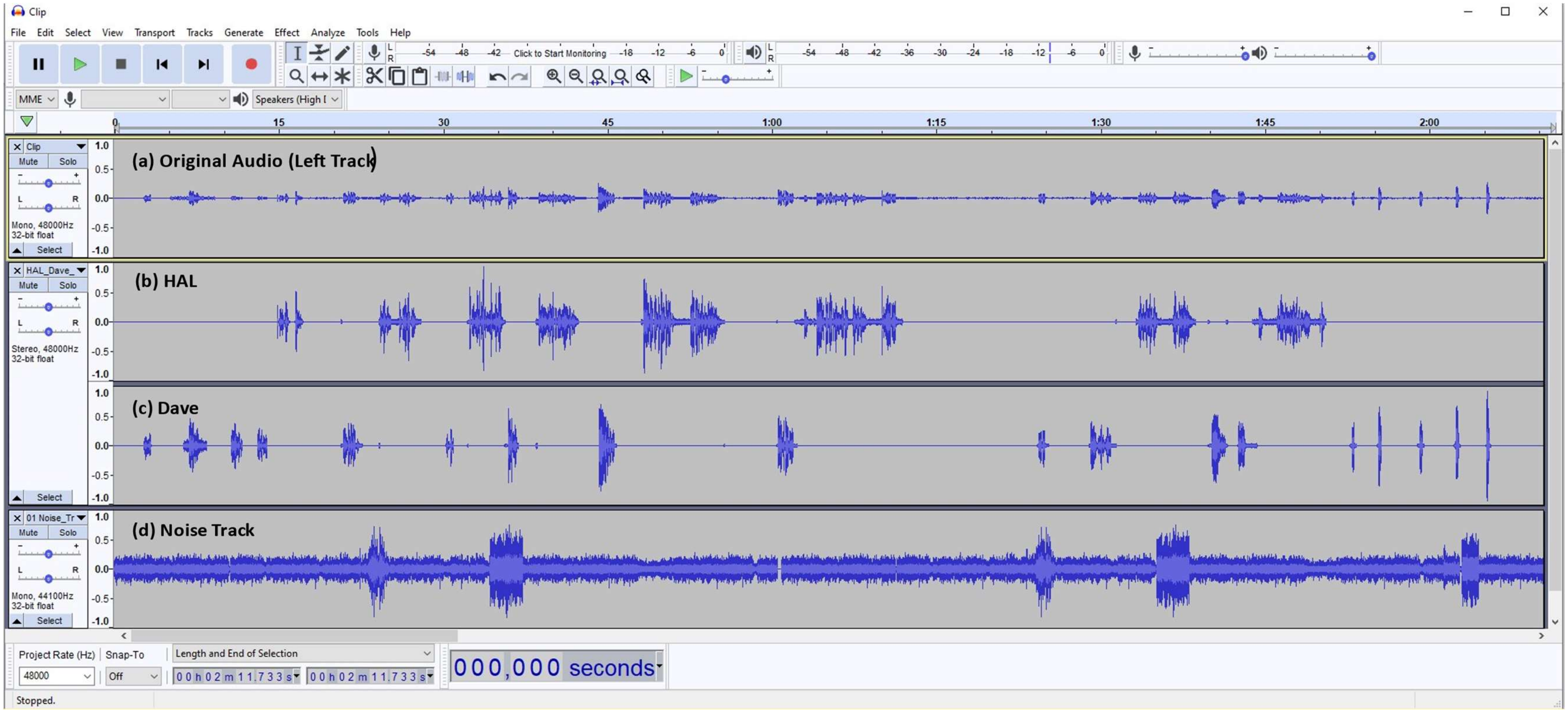

**Figure 44: Sources and Noise for HAL 9000 Experiment**

With this configuration, it was possible to play Dave at loudspeakers 1 (S1) and 2 (S2), and HAL at loudspeakers 3 (S3) and 4 (S4), by using the loudspeaker switch. The noise was played at the CD player in a continuous loop and modeled as S5 or S6.

#### 5.2.2.2 Ground Truth Recording

Two sets of recordings were taken for this experiment. Set 1A consisted of playing the audio track using the loudspeakers 1 (Dave) and 3 (HAL). Set 1B consisted of playing the loudspeakers 2 (Dave) and 4 (HAL). The noise track was played in a continuous loop by a CD player located at the position of Source 6. The audio was recorded using the Canon video camera with the microphone located at the position of microphone 3 in Table IX. The recording was transferred to the computer for segmentation and training. Because the camera records audio in stereo mode and the code can only handle mono audio, the stereo track was converted into mono audio by removing the right channel. This conversion kept intact all the spatial information contained in the left channel. Using Audacity's "convert to mono" feature would have mixed both channels, rendering the spatial information useless. Fig. 45 shows the final sets 1A and 1B of audio captured by the camera.

#### 5.2.2.3 Training and Segmentation

Both models for experiments 1 and 2 were trained with segments of speech from Dave and HAL, and a segment of noise. For experiment 1A, the model was trained with Dave as S1 using a 1.98 s long segment, and HAL as S3 with a 1.76 long segment, as shown in Fig 46.

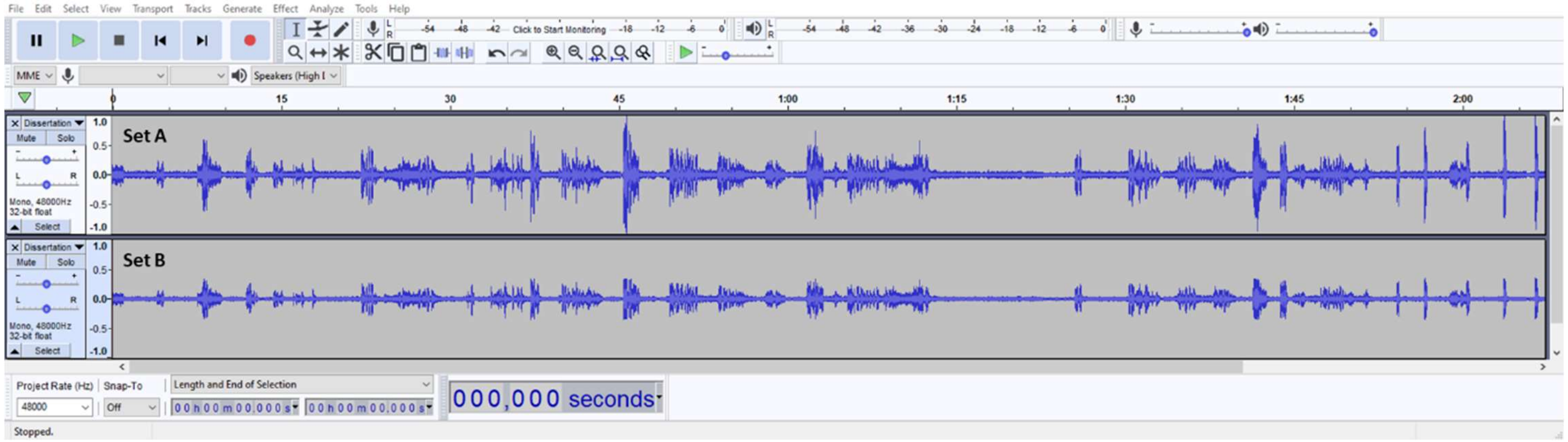

**Figure 45: Ground Truth Sets A and B for HAL 9000 Experiment.**

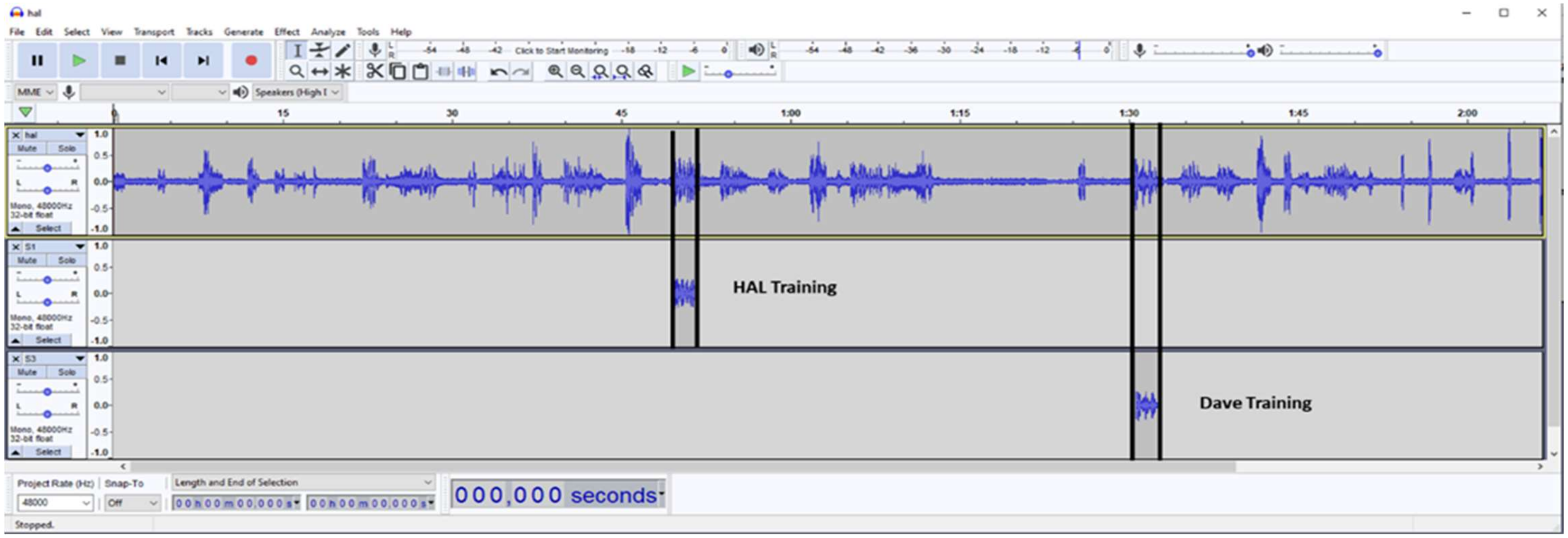

**Figure 46: Training Segments for HAL and Dave.**

Noise was trained as S6 with a 2 s long segment. S2, S4, and S5 were set to silence. For experiment 1B, Dave was trained as S2, and HAL was trained as S4. S1, S3, and S5 were set to silence, and S6 was noise. All training segments were about the same length as in 1A.

For Experiment 2, the model was trained with HAL as speaker S1 and S4, and Dave as speaker S2 and S3. The noise was trained as S6, and S5 was set to silence.

The recorded audio was segmented in two different ways. For experiment 1, the VAD was set with a maximum length segment of 5 seconds, ending with an audio of 120.39 s after subtracting the dropped segments. For experiment 2, the length of the segments was limited to a maximum of 1.5 s. Frames of less than 500 ms were discarded for both experiments.

#### 5.2.2.4 Testing and Results

Table X shows the results of experiments 1 and 2. We can appreciate that the length of the segments has an influence on the DER. In this experiment, the longer the segments, the less the error. These results agree with our previous discussion on the amount of information needed for proper cross-correlation. It is important then to optimize the length of the segments so they can contain as much information as possible and maximize the matching probabilities with the training template.

**Table X: DER Results for HAL 9000 Experiments.**

| Exp | Test | No. Segments | Properly Classified Segments | False Alarms (s) | Miss (s) | Confusion (s) | DER |
|---|---|---|---|---|---|---|---|
| 1 | A | 71 | 58 | 5.68 | 1.46 | 2.88 | 0.083 |
| | B | 71 | 58 | 1.22 | 3.01 | 2.32 | 0.054 |
| 2 | -- | 122 | 99 | 11.31 | 1.2 | 10.98 | 0.195 |

### 5.2.3 Multi-Speaker Identification Experiments

The objective of the experiments in this section is to measure the performance of the proposed method to identify several speakers in single-channel recording, independently of the content of their speech. As with the previous experiments, the geometry of the room and the location of the speakers is known, allowing for models that represent more accurately the actual acoustic scene under analysis. The experiments in this section used the same lab setup and models of the "HAL 9000" experiment.

The experiment was divided into four separate tests, that included two speakers and four speakers. Three of the experiments have two independent speakers repeating the same phrases, at different positions. The last experiment has four separate speakers at four different locations.

#### 5.2.3.1 Source Preparation and Editing

The speech sources for the experiments consisted of four different speakers, two male, and two females, sampled at a rate of 48 kHz. These sources were downloaded from the Telecommunications and Signal Processing Laboratory of McGill University, database version 2 [84]. The lengths of these sources vary between 1.2 to 3 s, approximately.

A total of four audio tracks were prepared for analysis. Tracks A, B, and C had two speakers, while track D had four. The sources were arranged into one stereo track, so they can be played at the loudspeakers LS1 and LS3, and then switched to be played at LS2 and LS4, as it was done with the HAL 9000 experiments. A small pause was inserted to allow for switching between loudspeakers.

Table XI shows the structure of each of the audio sample. Each sequence in the table indicates the label of the active speaker, the loudspeaker playing the speech, and the label of the spoken phrase. For example, audio sample A contains two sequences, 1 and 2. Sequence 1 is played at loudspeaker S1 by speaker 1, speaking phrase “a”. Sequence 2 is played by speaker 2, loudspeaker S3, speaking phrase “b”. In samples B, C, and D, speakers repeat some of the phrases with the objective to demonstrate the ability of the proposed system to differentiate the speakers regardless of their speech content.

**Table XI: Multi-Speaker Experiment Sequence Table**

| | | Duration (s) | Conditions | Sequence | | | | | | | | | |
|---|---|---|---|---|---|---|---|---|---|---|---|---|---|
| | | | | 1 | 2 | 3 | 4 | 5 | 6 | 7 | 8 | 9 | 10 |
| Sample | A | 4.78 | Loudspeaker | S1 | S3 | -- | -- | -- | -- | -- | -- | -- | -- |
| | | | Speaker | 1 | 2 | -- | -- | -- | -- | -- | -- | -- | -- |
| | | | Phrase | a | b | -- | -- | -- | -- | -- | -- | -- | -- |
| | B | 6.05 | Loudspeaker | S2 | S4 | S2 | -- | -- | -- | -- | -- | -- | -- |
| | | | Speaker | 1 | 2 | 1 | -- | -- | -- | -- | -- | -- | -- |
| | | | Phrase | a | a | b | -- | -- | -- | -- | -- | -- | -- |
| | C | 39.58 | Loudspeaker | S1 | S1 | S3 | S1 | S2 | S4 | S4 | S4 | S4 | -- |
| | | | Speaker | 1 | 1 | 2 | 1 | 1 | 2 | 2 | 2 | 2 | -- |
| | | | Phrase | a | b | c | d | e | f | g | h | i | -- |
| | D | 27.73 | Loudspeaker | S1 | S3 | S1 | S3 | S3 | S2 | S2 | S4 | S2 | S4 |
| | | | Speaker | 1 | 2 | 1 | 2 | 2 | 3 | 3 | 4 | 3 | 4 |
| | | | Phrase | a | a | b | b | c | d | e | f | g | h |

### 5.2.3.2 Ground Truth Recording

The audio was captured the same way as in the "HAL 9000" experiments, using the Canon video camera and saving the video recording in the camera's internal SD storage. Ambient noise was injected by paying background noise using the CD player, as it was done for the "HALL 9000" experiments. The background noise was extracted from one of the AOLME video recordings.

As with the "HAL 9000" experiments, the stereo recording from the video camera was converted into single-channel audio by removing the right channel, before the analysis.

#### 5.2.3.3 Training and Segmentation

The training was done with segments that had a maximum length of 1.5 s for each of the speakers, plus 1.5 s segment of noise. The custom VAD was used for segmentation. The number of segments produced for each of the audio tracks varied as is shown in the results table.

#### 5.2.3.4 Testing and Results

Testing was conducted in the same manner as the "HAL 9000" experiments. The results for each of the segments are shown in Table XII.

**Table XII: Controlled Environment Experiments Diarization Error Rate Results**

| Audio Sample | No. Speakers | No. Segments | Properly Classified Segments | False Alarms | Miss | Confusion | DER |
|---|---|---|---|---|---|---|---|
| A | 2 | 9 | 7 | 0 | 2 | 0 | 0.19 |
| B | 2 | 15 | 12 | 0 | 2 | 1 | 0.19 |
| C | 2 | 116 | 98 | 10 | 0 | 8 | 0.12 |
| D | 4 | 37 | 27 | 2 | 0 | 8 | 0.27 |

Table V shows that the DER is not more than 0.27 in the worst case. These results are comparable or better than DER results from methods using databases and neural networks [85]

## 5.3 AOLME Experiments

The controlled environment experiments demonstrated that the proposed method could identify speakers in single-channel recordings. These experiments analyzed audio samples that featured organized speech (one speaker at a time), where the speakers are well separated from each other (no overlapping between speakers). The objective of the AOLME experiments in this section is to evaluate the performance of the proposed method to identify speakers in single-channel audio recordings from videos of noisy multi-speaker collaborative environments.

This section evaluates the process of selection of the AOLME videos for the experimental analysis, and discusses the models employed for the analysis. The analysis of the audios will follow the same approach as the previous experiments.

### 5.3.1 Evaluation and Selection of AOLME Videos

There are several hundred hours of AOLME video recordings available for analysis but, because of the experimental nature of this research work, it was necessary to select videos that met certain characteristics that facilitate the preparation of the models and the setup of the experiments. The models used in the previous experiments proved to perform well, and for this reason, it was necessary to search for AOLME videos with similar geometric characteristics as these models, i.e., the participants were in similar places as the speakers in the model from our previous experiments. The selection consisted of four videos with 2, 3, 4 and 5 participants from the library of videos. The videos were approximately 3 minutes long each. Fig. 47 shows frames from these videos with 2 participants (a), 3 participants (b), 4 participants (c), and 5 participants (d). As was done in the previous experiments, the stereo audio track for each video was extracted and converted into a 48 kHz single channel by removing the right channel.

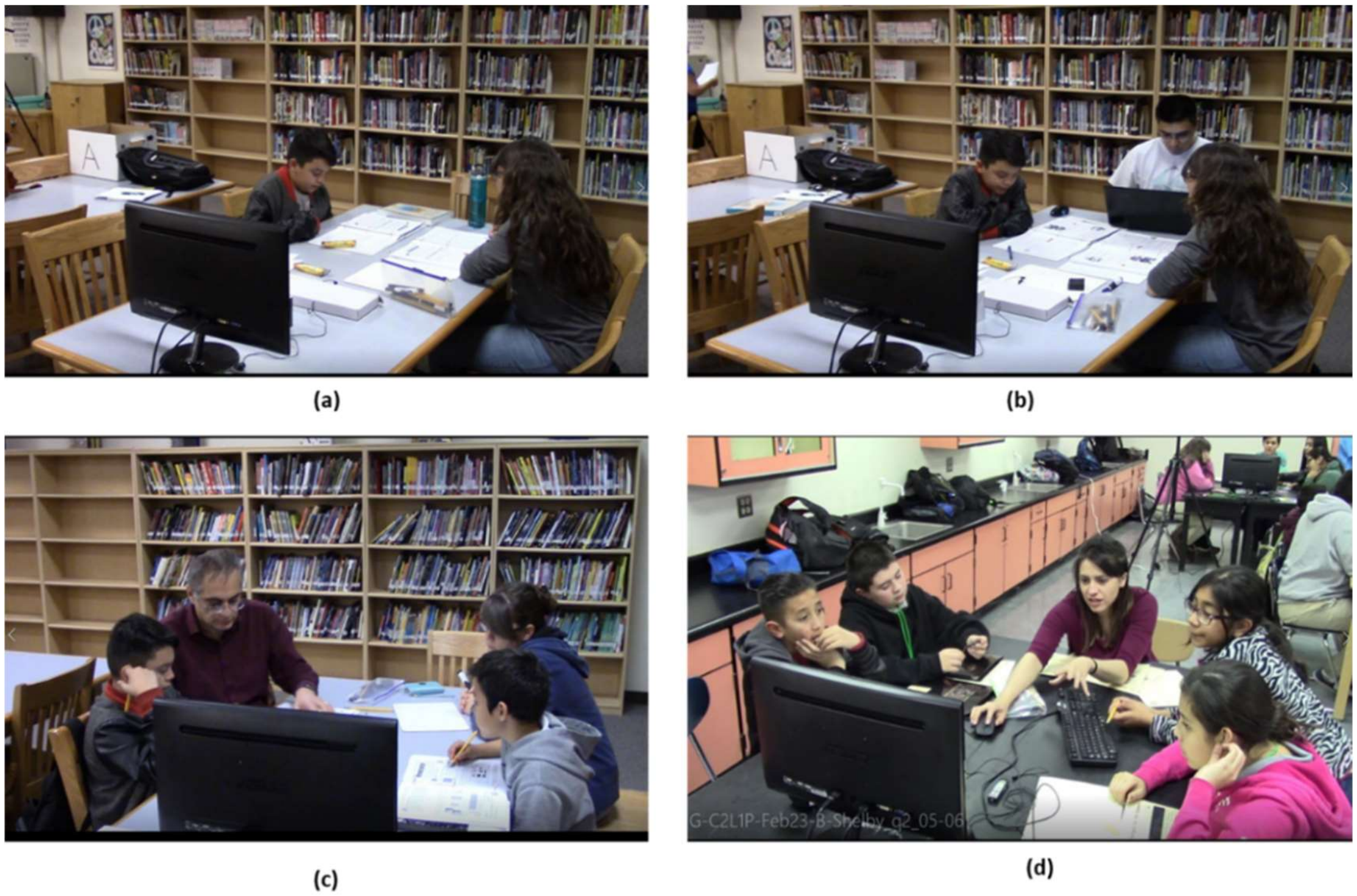

**Figure 47:Video Clips for AOLME Experiments.**

### 5.3.2 Model Preparation

The model used for this experiment followed the same geometry as the previous experiments, with the width of the table adjusted to 1.8 m to fit the AOLME scene more accurately. Instead of generating separate models for each of the videos, the model had all three speakers for all the experiments. As previously done, the locations of the absent speakers were turned off by training with a silence segment. Fig. 48 shows the 2D Pyroomacoustics model and Table XIII shows the locations of the speakers and the microphones.

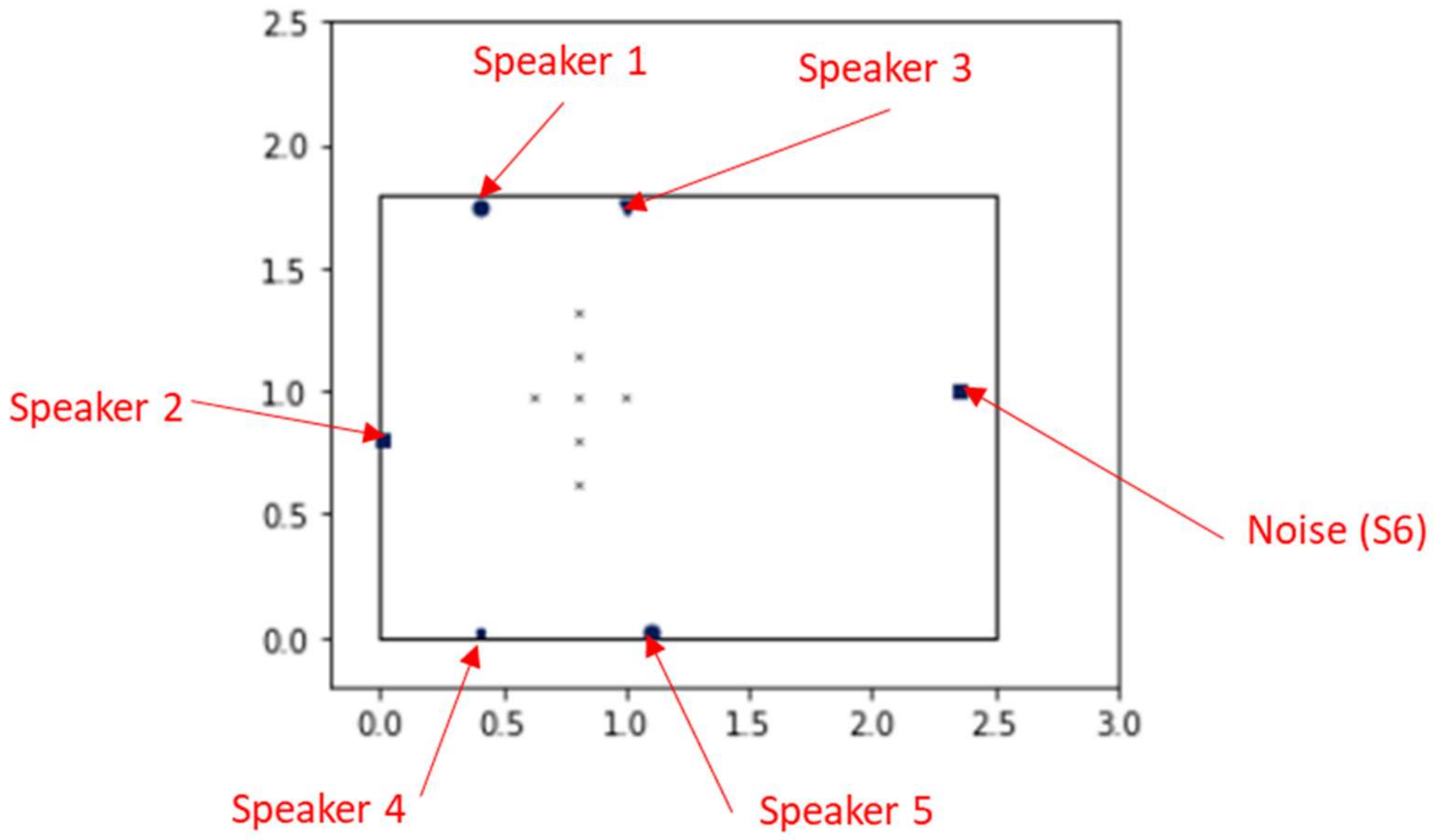

**Figure 48: 2D Model for AOLME Experiments.**

**Table XIII: Location of Speakers and Microphones for AOLME Experiments.**

| | | X | Y | Z |
|---|---|---|---|---|
| Speakers | S1 | 0.40 | 1.79 | 0.25 |
| | S2 | 0.01 | 0.80 | 0.25 |
| | S3 | 1.00 | 1.79 | 0.25 |
| | S4 | 0.40 | 0.01 | 0.25 |
| | S5 | 1.20 | 0.01 | 0.25 |
| | S6 | 2.40 | 1.00 | 1.50 |
| Mics | M1 | 0.75 | 1.00 | 0.03 |
| | M2 | 0.85 | 1.00 | 0.03 |
| | M3 | 0.80 | 1.00 | 0.03 |
| | M4 | 0.80 | 1.05 | 0.03 |
| | M5 | 0.80 | 0.95 | 0.03 |
| | M6 | 0.80 | 1.10 | 0.03 |
| | M7 | 0.80 | 0.90 | 0.03 |
| Room | CORNER 1 | 0.00 | 0.00 | -- |
| | CORNER 2 | 0.00 | 1.80 | -- |
| | CORNER 3 | 2.50 | 1.80 | -- |
| | CORNER 4 | 2.50 | 1.80 | -- |
| | EXTRUDE | 2.00 | | |

### 5.3.3 Training and Segmentation

The same training and segmentation principles were used as in the previous experiments. Training used a 1.5 to 2 s long sample of each of the participants, plus a similar length segment of background noise. Because the same model was used for all participants, non-active speakers were trained with a silence segment of 2 s duration. Table XIV shows the speaker assignment for each of the experiments.

**Table XIV: Speaker Assignment for AOLME Experiments.**

| | | Speaker Assignment | | | | | |
|---|---|---|---|---|---|---|---|
| | | S1 | S2 | S3 | S4 | S5 | S6 |
| **Audio Sample** | A | ✓ | Silence | Silence | ✓ | Silence | Noise |
| | B | ✓ | ✓ | Silence | ✓ | Silence | Noise |
| | C | ✓ | Silence | ✓ | ✓ | ✓ | Noise |
| | D | ✓ | ✓ | ✓ | ✓ | ✓ | Noise |

The Ground Truth for each audio was segmented using the VAD, discarding segments with less than 0.5 s duration, and limiting the length of the segments to 1.5 s maximum. The total number of segments for each sample is shown in the results table.

### 5.3.4 Testing and Results

The same type of analysis was applied as in the previous experiments. Table XV (a) shows an example of the Cross-Correlation results of analyzing one segment of Audio Sample B. Tables VII (b), (c), and (d) show the training CC tables with the score of each possible speaker. Each match is represented by a zero (0). In this case, the segment corresponds to speaker 2.

## Table XV: CC Tables for AOLME Experiment.

(a) Microphone Cross Correlation. Unknown Segment

| Speakers | | 1-2 | 1-3 | 1-4 | 1-5 | 1-6 | 1-7 | 2-3 | 2-4 | 2-5 | 2-6 | 2-7 | 3-4 | 3-5 | 3-6 | 3-7 | 4-5 | 4-6 | 4-7 | 5-6 | 5-7 | 6-7 |
|---|---|---|---|---|---|---|---|---|---|---|---|---|---|---|---|---|---|---|---|---|---|---|
| | 1 | -64 | -45 | -96 | -100 | 12 | -68 | 18 | -74 | -276 | 71 | -9 | -72 | -41 | 57 | -27 | 14 | 285 | -58 | 294 | 3 | -80 |
| | 2 | -65 | -75 | -236 | -236 | -75 | 0 | -30 | -103 | -103 | -11 | 65 | -69 | -69 | 19 | 75 | 0 | 98 | 236 | 98 | 236 | 75 |
| | 3 | 11 | 89 | 137 | 306 | 90 | -28 | 56 | 203 | 92 | 79 | -34 | -41 | 35 | 21 | -123 | 82 | 18 | -94 | -8 | -107 | -125 |
| | 6 | 63 | 72 | 133 | 133 | 97 | 0 | 8 | 57 | 57 | 21 | -63 | 42 | 42 | 1 | -72 | 0 | -36 | -133 | -36 | -133 | -97 |

(b) Microphone Cross Correlation. Training Speaker 1. Score: 25

| Speakers | | 1-2 | 1-3 | 1-4 | 1-5 | 1-6 | 1-7 | 2-3 | 2-4 | 2-5 | 2-6 | 2-7 | 3-4 | 3-5 | 3-6 | 3-7 | 4-5 | 4-6 | 4-7 | 5-6 | 5-7 | 6-7 |
|---|---|---|---|---|---|---|---|---|---|---|---|---|---|---|---|---|---|---|---|---|---|---|
| | 1 | 91 | 0 | 0 | -14 | 5 | 18 | 4 | 4 | -139 | -9 | -5 | 8 | 0 | 0 | -4 | -127 | 56 | -333 | 206 | -1 | 4 |
| | 2 | -53 | 0 | -83 | -83 | 110 | 0 | 0 | 0 | 0 | -1 | 53 | 0 | 0 | 1 | 0 | 0 | 0 | 83 | 0 | 83 | -110 |
| | 3 | 0 | 11 | -145 | 197 | -5 | 6 | 1 | 192 | 5 | -11 | 17 | -11 | 2 | 1 | -2 | 21 | 2 | 0 | 5 | 37 | -15 |
| | 6 | 0 | 4 | 0 | 0 | 12 | 0 | 1 | -1 | -1 | -1 | 0 | 1 | 1 | -1 | -4 | 0 | 1 | 0 | 1 | 0 | -12 |

(c) Microphone Cross Correlation. Training Speaker 2. Score: 32

| Speakers | | 1-2 | 1-3 | 1-4 | 1-5 | 1-6 | 1-7 | 2-3 | 2-4 | 2-5 | 2-6 | 2-7 | 3-4 | 3-5 | 3-6 | 3-7 | 4-5 | 4-6 | 4-7 | 5-6 | 5-7 | 6-7 |
|---|---|---|---|---|---|---|---|---|---|---|---|---|---|---|---|---|---|---|---|---|---|---|
| | 1 | -245 | 0 | 0 | -14 | 5 | 18 | 4 | 204 | -270 | -13 | -5 | -1 | 14 | 0 | -4 | 5 | 394 | -333 | 208 | 0 | 0 |
| | 2 | 0 | -1 | 21 | 21 | 0 | 0 | 0 | 0 | 0 | 0 | 0 | 0 | 0 | 1 | 1 | 0 | 5 | -21 | 5 | -21 | 0 |
| | 3 | 0 | 0 | -145 | 218 | -5 | -9 | 1 | 419 | 5 | 5 | 1 | -11 | 3 | 3 | -11 | 0 | 1 | 0 | 0 | -202 | 0 |
| | 6 | 0 | 4 | 12 | 12 | 0 | 0 | 4 | -1 | -1 | 0 | 0 | 0 | 0 | -1 | -4 | 0 | 1 | -12 | 1 | -12 | 0 |

(d) Microphone Cross Correlation. Training Speaker 3. Score: 21

| Speakers | | 1-2 | 1-3 | 1-4 | 1-5 | 1-6 | 1-7 | 2-3 | 2-4 | 2-5 | 2-6 | 2-7 | 3-4 | 3-5 | 3-6 | 3-7 | 4-5 | 4-6 | 4-7 | 5-6 | 5-7 | 6-7 |
|---|---|---|---|---|---|---|---|---|---|---|---|---|---|---|---|---|---|---|---|---|---|---|
| | 1 | 13 | -1 | 5 | -14 | 5 | 13 | -6 | -198 | -72 | -9 | -5 | 0 | 1 | 1 | 3 | 4 | -9 | -267 | 201 | -5 | 4 |
| | 2 | 0 | -1 | -121 | -121 | 0 | 0 | -194 | 0 | 0 | -2 | 0 | 7 | 7 | 191 | 1 | 0 | 0 | 121 | 0 | 121 | 0 |
| | 3 | -5 | 14 | -77 | 202 | -5 | -5 | 0 | 182 | 5 | 5 | -155 | 0 | 3 | 3 | -3 | 6 | -31 | -12 | 5 | 16 | -12 |
| | 6 | 0 | -209 | 0 | 0 | 24 | 0 | 4 | 11 | 11 | 11 | 0 | 2 | 2 | -3 | 209 | 0 | 1 | 0 | 1 | 0 | -24 |

Table XVI shows the results of the analysis of all Audio Samples, with the respective DER for each experiment.

**Table XVI: Classification Results for AOLME Experiments.**

| Audio Sample | Sample Duration (s) | No. Speakers | No. Segments | Properly Classified Segments | False Alarms | Miss | Confusion | DER |
|---|---|---|---|---|---|---|---|---|
| A | 244 | 2 | 311 | 281 | 5 | 10 | 15 | 0.095 |
| B | 256 | 3 | 328 | 302 | 8 | 10 | 8 | 0.079 |
| C | 381 | 4 | 489 | 426 | 10 | 25 | 28 | 0.12 |
| D | 257 | 5 | 339 | 284 | 12 | 15 | 28 | 0.16 |

## 5.4 Comparison with Other Methods

The final set of experiments focus on comparing our proposed method against Google's and Amazon AWS. Google's and Amazon AWS were two of the cloud-based speech processing services introduced in the background section of this dissertation. Microsoft Diarization service was in the process of being updated by the time this dissertation was written and, therefore, it was not possible to run any experiment with it.

### 5.4.1 Methodology for Comparison

The diarization services provided by Google and Amazon differ from the proposed method in three aspects. First, they do not require a sample of audio for training. Second, the audio samples to diarize need to be of a minimum duration of 4 s, approximately. Third, their output does not provide a label of the active speaker, but rather a set of text transcripts that contain the speech segment, the abstract speaker label (e.g., speaker 0, speaker 2), the active time of the speaker on the transcript segment, and the confidence rate. Given these constraints, the only fair comparison criteria are to manually measure each speaker's ground truth active time manually and compare these times with the results of the analysis by all three methods. It was necessary to add a section of code to the proposed method to measure the length of each of the segments that are already classified and totalize the time for the same speaker plus noise.

### 5.4.2 Selection, Preparation, and Ground Truth Measurements of Videos for Analysis

The analysis consisted of a total of 8 AOLME videos containing 2, 3, 4, and 5 speakers. The duration of each video was limited to a maximum of 3 minutes. The audio from each video was extracted using Audacity and downshifted to 16 kHz for upload to Goggle and Amazon. The audio files for our methods were sampled at a rate of 48 kHz.

Each speaker's active time from the ground truth audio was measured using a stopwatch. In some of the AOLME videos, it was difficult to assess this time due to several speakers being active simultaneously. In these cases, each speaker's time was recorded by listening to his/her voice and watching his/her lip movement on video, even if their speech overlapped at any moment.

### 5.4.3 Training and Segmentation

The system was trained with audio samples of about 1.8 s long from each speaker and noise, using a VAD with a maximum segment length of 1.2 s. All segments with a duration of less than 0.5 s were dropped. There was no need for training on Google or Amazon; these systems trained by using the uploaded audio and their databases.

### 5.4.4 Testing and Analysis

Each of the audio files from the videos was analyzed using the modified code that totalizes each speaker's time, with no other additional steps. For Amazon and Google, the audio was uploaded to the cloud.

Because both Amazon and Google's methods return only abstract labels, the output transcriptions of each of the speakers were used to manually match the identity of the speaker on each segment, noting that both Amazon and Google label the first active speaker they detect as "speaker 0".

### 5.4.5 Results

Table XVII shows the results of this experiment, with the percentage error highlighted in light blue. The error was calculated using (5.1).

$$\text{Percent error} = \frac{\text{estimated time} - \text{true time}}{\text{true time}} * 100 \qquad (5.1).$$

**Table XVII: Experimental Comparison Between Methods.**

| Audio Sample | No. of Speakers | Speaker | Ground Truth Time (s) | Proposed Method | | Amazon AWS | | Google Cloud | |
|---|---|---|---|---|---|---|---|---|---|
| | | | | Time (s) | Error % | Time (s) | Error % | Time (s) | Error % |
| 1 | 2 | S1 | 117.00 | 99.99 | 14.54 | 94.52 | 19.21 | 127.10 | 8.63 |
| | | S2 | 27.52 | 34.62 | 25.80 | 74.47 | 170.60 | 0.00 | 100.00 |
| 2 | 2 | S1 | 107.00 | 113.00 | 5.61 | 120.90 | 12.99 | 73.40 | 31.40 |
| | | S2 | 18.03 | 23.44 | 30.01 | 45.46 | 152.14 | 66.59 | 269.33 |
| 3 | 3 | S1 | 6.00 | 20.69 | 244.83 | 9.88 | 64.67 | 66.59 | 1009.83 |
| | | S2 | 102.52 | 100.52 | 1.95 | 143.74 | 40.21 | 50.80 | 50.45 |
| | | S3 | 9.26 | 13.45 | 45.25 | 0.00 | 100.00 | 10.29 | 11.12 |
| 4 | 3 | S1 | 65.74 | 68.93 | 4.85 | 106.36 | 61.79 | 80.20 | 22.00 |
| | | S2 | 27.66 | 25.38 | 8.24 | 37.67 | 36.19 | 31.39 | 13.49 |
| | | S3 | 10.86 | 15.30 | 40.88 | 0.00 | 100.00 | 0.00 | 100.00 |
| 5 | 4 | S1 | 28.29 | 41.61 | 47.08 | 52.19 | 84.48 | 0.00 | 100.00 |
| | | S2 | 11.17 | 14.69 | 31.51 | 8.93 | 20.05 | 8.30 | 25.69 |
| | | S3 | 42.27 | 68.23 | 61.41 | 0.00 | 100.00 | 35.00 | 17.20 |
| | | S4 | 73.84 | 91.57 | 24.01 | 0.00 | 100.00 | 94.30 | 27.71 |
| 6 | 4 | S1 | 24.48 | 25.39 | 3.72 | 78.70 | 221.49 | 53.59 | 118.91 |
| | | S2 | 22.28 | 13.28 | 40.39 | 36.95 | 65.84 | 29.19 | 31.01 |
| | | S3 | 25.75 | 27.69 | 7.53 | 0.00 | 100.00 | 15.29 | 40.62 |
| | | S4 | 1.20 | 4.20 | 250.00 | 38.05 | 3070.83 | 3.30 | 175.00 |
| 7 | 5 | S1 | 20.25 | 7.99 | 60.54 | 0.00 | 100.00 | 5.09 | 74.86 |
| | | S2 | 69.19 | 64.53 | 6.74 | 88.77 | 28.30 | 24.90 | 64.01 |
| | | S3 | 9.41 | 10.71 | 13.82 | 0.00 | 100.00 | 0.00 | 100.00 |
| | | S4 | 43.12 | 48.86 | 13.31 | 60.04 | 39.24 | 54.70 | 26.86 |
| | | S5 | 12.27 | 10.93 | 10.92 | 0.00 | 100.00 | 46.60 | 279.79 |
| 8 | 5 | S1 | 14.28 | 18.80 | 31.65 | 0.00 | 100.00 | 6.29 | 55.95 |
| | | S2 | 34.56 | 42.05 | 21.67 | 53.13 | 53.73 | 29.59 | 14.38 |
| | | S3 | 2.50 | 3.60 | 44.00 | 0.00 | 100.00 | 7.49 | 199.60 |
| | | S4 | 15.23 | 22.27 | 46.22 | 17.61 | 15.63 | 11.20 | 26.46 |
| | | S5 | 47.67 | 27.54 | 42.23 | 56.02 | 17.52 | 29.59 | 37.93 |

Table XVIII shows the average error for 2, 3, 4, and 5 speakers, as well as the total average error for each method.

**Table XVIII: % Average Error for All Three Methods.**

| No. of Speakers | Proposed Method | Amazon AWS | Google Cloud |
| --- | --- | --- | --- |
| 2 | 18.99 | 88.74 | 102.34 |
| 3 | 57.67 | 67.14 | 201.15 |
| 4 | 58.21 | 470.34 | 67.02 |
| 5 | 29.11 | 65.44 | 87.98 |
| Total | 42.10 | 184.82 | 108.29 |

The results presented in Tables XVII and XVIII show a substantial reduction in the achieved error rate. More specifically, error reduction ranges from 50% to 87%. The color codes used in Table XVII emphasize the results of this experiment. The red highlighting denotes cases of failures where we have a speaker that was completely missed, or the estimated talking time of the speaker had more than a 100% error (e.g., an over-estimating speaker talking time). Out of 28 possible speakers across all examples, Amazon AWS gave failing results for 14 cases (50%), Google cloud gave failing results for 10 cases (36%), while the proposed method gave failing results for 2 cases (7%). It is interesting to notice that the proposed method never failed to detect a speaker (0% error), while Amazon AWS could not detect any talking time for 10 cases (36%). Google cloud failed to detect any talking time for 4 cases (14%). Also, there are failure cases for all 8 samples for Amazon

AWS and Google Cloud. In contrast, for the proposed method, there are 2 samples with examples of over-estimation, with 6 samples being free of dramatic failures.

Teal highlighting denotes cases where the total estimated speaking time gave 20% or less error. Based on this criterion, both AWS and Google Cloud gave satisfactory results in 5 cases (18%) versus 11 cases (39%) for the proposed method.

# Chapter 6. Summary, Conclusions, and Future work

This dissertation presented a method for speaker diarization and identification using virtual microphones and cross-correlation patterns. The proposed method identifies speakers in single-channel recordings taken in noisy collaborative environments, such as classrooms and educational workshops. The method gave an error rate that was over 50% less on average than other available diarization methods when subject to the same testing environments. In contrast with other methods that are considered state-of-the-art, the proposed method requires minimal training and no databases, making it applicable in situations where it is not possible to gather clean speech samples.

The background section of this dissertation presented similar research works on speaker diarization and identification based on microphone arrays. Although some of these works included virtual microphone arrays, none of them approached a full virtual array simulation from a single microphone recording. Given the unprecedented focus on Deep Learning methods, alternative approaches are avoided, limiting the number of researchers interested in pursuing them. Yet, the proposed methodology clearly outperformed commercial Deep Learning methods and demonstrated some of their limitations due to their needs for large training datasets.

The method presented in this dissertation offers an alternative for educational researchers that are involved with collaborative environments and depend mostly on the analysis of data provided by video recordings. The work in this dissertation showed that other available methods perform poorly under these environments when determining who speaks, when, and for how long. The deficiencies presented by these methods are even more prominent when the participants are from underrepresented groups from which large training databases may not exist. The proposed method demonstrated a significant performance improvement by capitalizing on real video information of the environment under analysis, rather than depending on unrelated training data. Also, by no requiring previous speaker enrollment, this method opens the possibility of analysis of a wide variety of video data that may not have been recorded with the known intention of posteriors analysis.

The dissertation method constitutes more of a proof of concept than a fully operational method. The success of the proposed method is due to the possibility of simulating acoustic wave propagation, including speech. Even though this modeling can be complex, we have now powerful personal computers to execute the calculations required by the signal processing algorithms. Furthermore, the code for the simulations is available from large repositories that contain open-source libraries ready for implementation; nevertheless, there is work that needs to be done to address some of the weaknesses observed so far, such as it is the case where participant speakers move and change their original locations, and when they “invade” other’s speakers’ physical location. Under this area, it is possible to eventually adapt the methods from the research work done at the ivPCL lab regarding object and subject tracking. The location of the speakers and the general geometry of the room could be dynamically modified in the models based on the information from video data, thus improving the error rate. Also, the experiments only considered one type of microphone array, leaving open the question of the performance of other types of arrays, such as circular or even volumetric. In addition, the simulation version available during the development of this dissertation had some limitations that impacted the accuracy of the models. Pyroomacoustics released a new version that includes improvements to the models' parameters, such as physical modeling of room absorption, reverberation modeling, and multi-pattern microphone simulation. Finally, the method depends on proper audio segmentation and final classification. Most of the misclassifications in the method were the product of improper pre-segmentation and sub-optimal classification. A more sophisticated classifier using machine learning or neural networks would help improve the overall performance. It is possible also to apply clustering classification for unsupervised identification of the speakers.

Finally, the method could be extended to support other applications of speech processing, as it can be incorporated as a front end or pre-processor. For example, the method can be used to improve the accuracy of spatial filters for speech enhancement or speaker separation from mixtures. The parameters of the spatial filters can be better determined by estimating the location of the speaker and then optimizing the parameters for that location.

# Appendix A: Pyroomacoustics Scripts

This section describes the two Python scripts that call the Pyroomacoustics libraries to generate the room geometry parameters, calculate the RIRs, and emulate the virtual microphones.

### *a) Room Geometry Generator:*

This script accepts the room dimensions and locations of the sources and virtual microphones and generates the 2D and 3D geometric models. The room geometry is saved as a set of .txt files that contains the geometry arrays. This script runs under a Jupyter Notebook.

```
#Location of Sources and Microphones
Source6=[0.98,0.4]
Source6_3D=[0.98,0.4,0.98]

Mic_X = [0.6,0.65,0.6,0.6,0.6,0.55,0.6]
Mic_Y = [0.6,0.55,0.55,0.45,0.5,0.55,0.65]
Mic_Z = [0.25,0.25,0.01,0.25,0.25,0.25,0.25]

#Add room
room = pra.Room.from_corners(corners, fs=fs)

#Location of Microphones Array
R = np.array([Mic_X, Mic_Y])  # [[x], [y], [z]]

#Add source to 2D room

room.add_source(Source1, signal=s1)
.
.
room.add_source(Source6, signal=s6)

room.add_microphone_array(pra.MicrophoneArray(R, room.fs))

#Execute Location
room = pra.Room.from_corners(corners, fs=fs)
room.extrude(1.0)
```

```
R = np.array([Mic_X, Mic_Y, Mic_Z])  # [[x], [y], [z]]
room.add_microphone_array(pra.MicrophoneArray(R, room.fs))
room.add_source(Source1_3D, signal=s1)
.
.
room.add_source(Source4_3D, signal=s4)

#Save Geometry
np.savetxt(r'C:\Users\User\Desktop\PhD
Folder\Dissertation\Experiments\Model_Estimation\Room_parameters\corner
s_array.txt',corners[:,:],delimiter=',', fmt='%f')
.
.
.
np.savetxt(r'C:\Users\User\Desktop\PhD
Folder\Dissertation\Experiments\Model_estimation\Room_parameters\mic_ar
ray.txt',R[:,:],delimiter=',', fmt='%f')
```

### b) *Pyroomacoustics Virtual Microphone Simulation Script*

This script is used twice to first calculate the RIR from the model sources to the virtual microphones, and then again to emulate the signal at the virtual microphones using the estimated sources. This script is called within LabVIEW, and its outputs are saved in .txt files.

```
#Setup Python
import numpy as np
import matplotlib.pyplot as plt
from scipy.io import wavfile
from scipy.signal import fftconvolve
import pyroomacoustics as pra
#Define Variables
Abs = 0
max_o = 0
room_extrude = 0
corners_array = 0
Source1 = 0
.
.
Source6 = 0
mic_array = 0
#Define model
def model_generation():
```

```
#Delimit the corners of the room
corners = np.array(corners_array).T  # [x,y]
room = pra.Room.from_corners(corners)
room.extrude(room_extrude)
#Read Sources
fs, s1 = wavfile.read(r"C:\.....)
.
.
fs, s6 = wavfile.read(r"C:\.....)
room = pra.Room.from_corners(corners, fs=fs)
#Add microphone array
R = np.array(mic_array)  # [[x], [y], [z]]
room.add_microphone_array(pra.MicrophoneArray(R, room.fs))
# set max_order for RIR
room = pra.Room.from_corners(corners, fs=fs, max_order=max_o,
absorption=Abs)
#Set Extrusion
room.extrude(room_extrude)
#Add source arrays and microphones
Source1_3D=np.array(Source1)
#Source 1
room.add_source(Source1_3D, signal=s1)
room.add_microphone_array(pra.MicrophoneArray(R, room.fs))
#Compute image sources
room.image_source_model(use_libroom=True)
room.compute_rir()
#Save Data
np.savetxt(r'C:\....,room.rir[0][0],delimiter=',', fmt='%f')
#Data Mic
data_mic=room.mic_array.signals[0,:]
np.savetxt(r'C:\....,room.rir[0][0],delimiter=',', fmt='%f')
data_mic=room.mic_array.signals[1,:]
np.savetxt(r'C:\....,room.rir[0][0],delimiter=',', fmt='%f')
data_mic=room.mic_array.signals[2,:]
np.savetxt(r'C:\....,room.rir[0][0],delimiter=',', fmt='%f')
data_mic=room.mic_array.signals[3,:]
np.savetxt(r'C:\....,room.rir[0][0],delimiter=',', fmt='%f')    #Data
Mic5
data_mic=room.mic_array.signals[4,:]
np.savetxt(r'C:\....,room.rir[0][0],delimiter=',', fmt='%f')    #Data
Mic6
data_mic=room.mic_array.signals[5,:]
np.savetxt(r'C:\....,room.rir[0][0],delimiter=',', fmt='%f')
data_mic=room.mic_array.signals[6,:]
np.savetxt(r'C:\....,room.rir[0][0],delimiter=',', fmt='%f')
#Reepeat for all sources
.
.
.
return
```

# Appendix B: LabVIEW Sub-Vis

## *a) Room Parameters Reader*

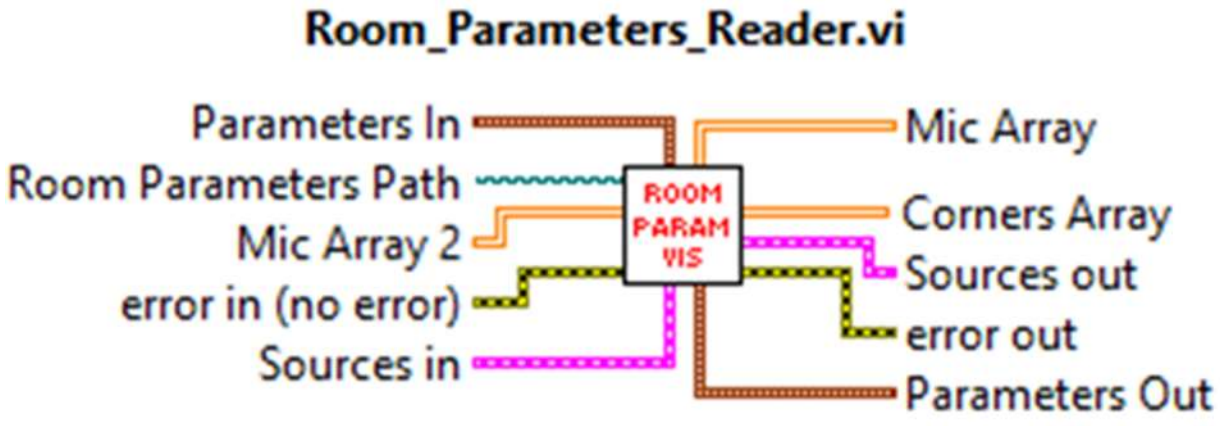

**Figure 49: Room Parameters Reader Inputs and Outputs**

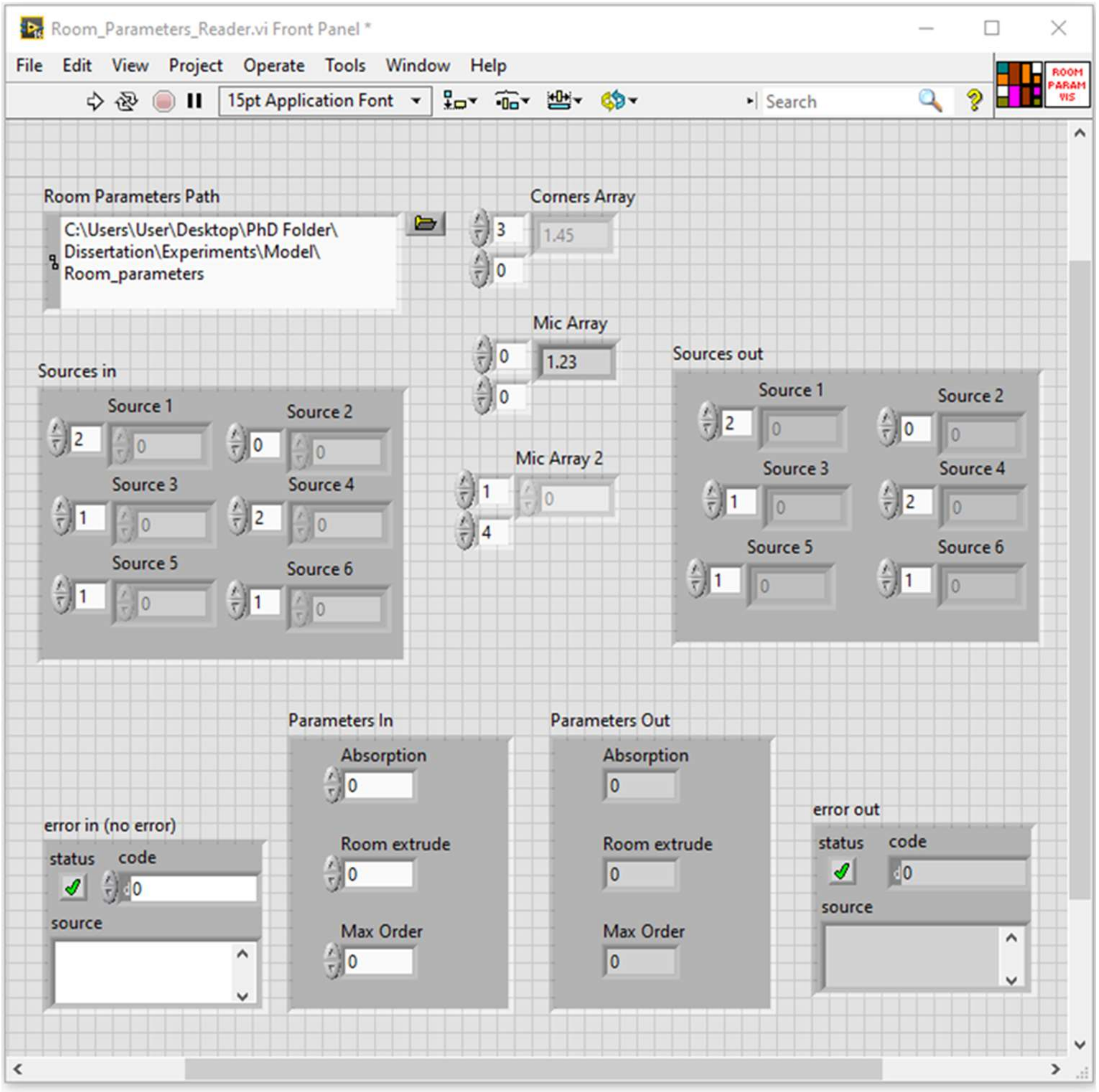

**Figure 50: Room Parameters Reader Front Panel.**

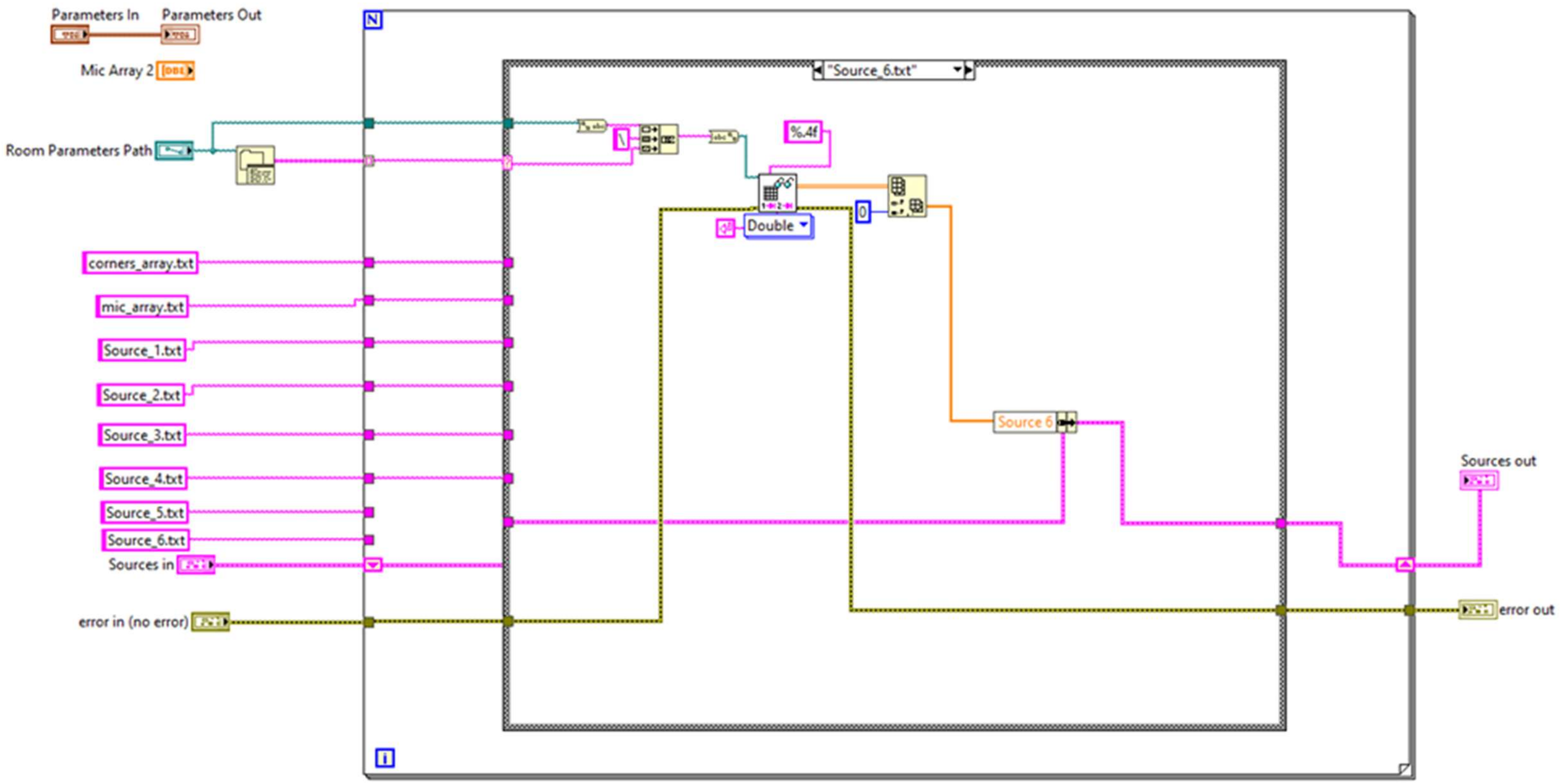

**Figure 51: Room Parameters Reader Block Diagram.**

### *b) Room Model Generator*

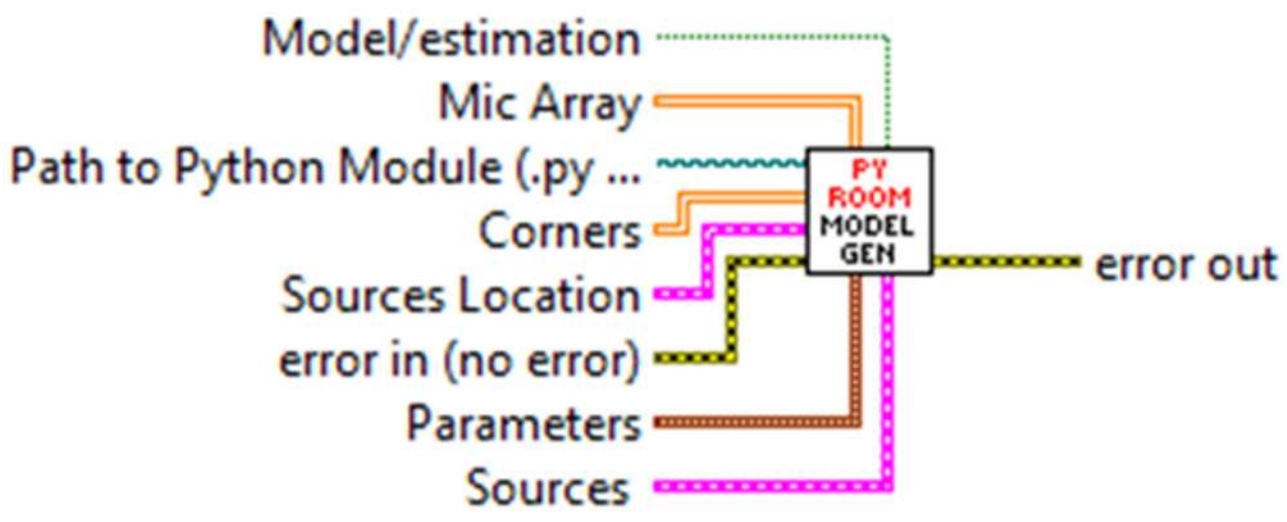

**Figure 52: Room Model Generator Icon.**

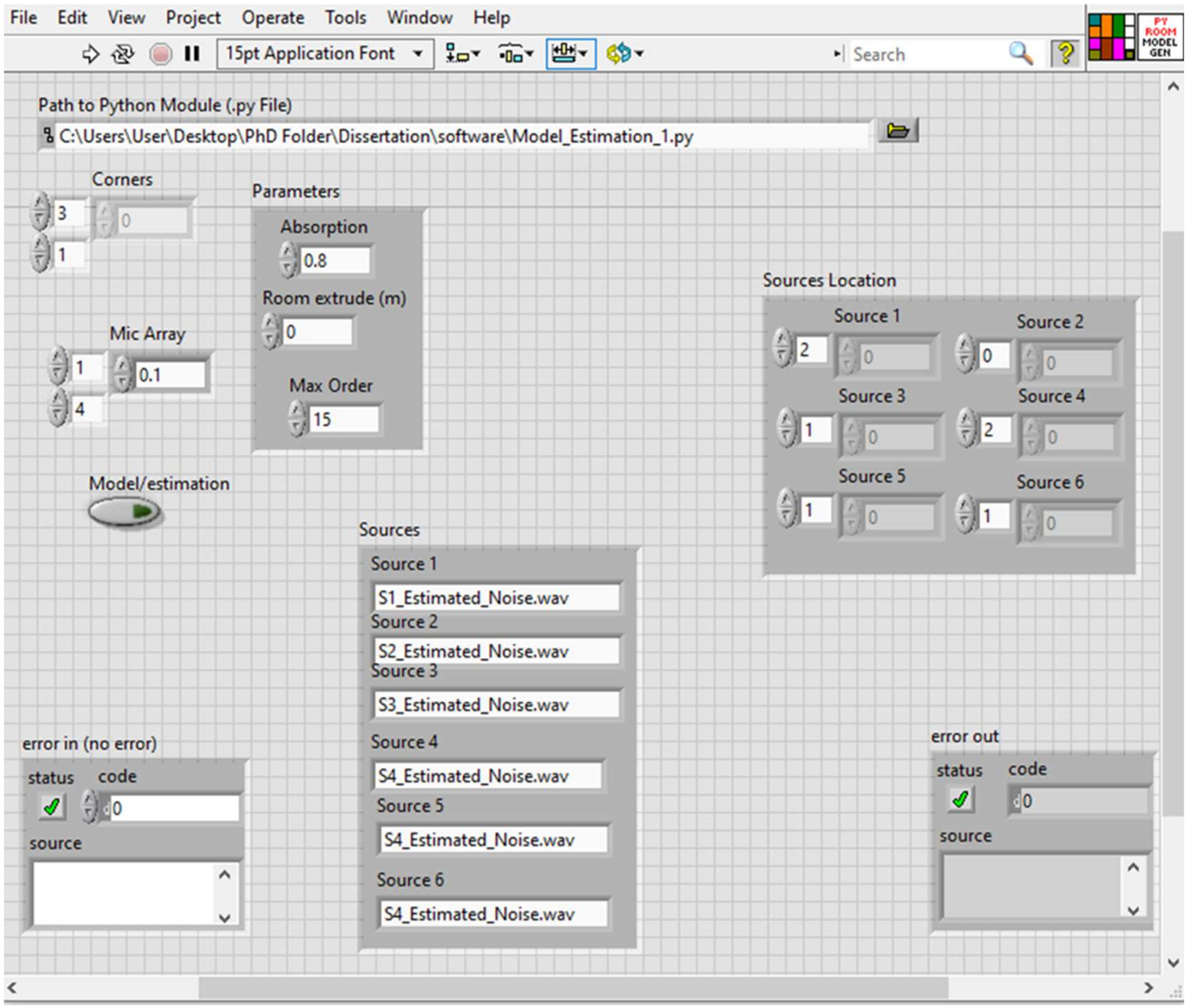

**Figure 53: Room Model Generator Front Panel.**

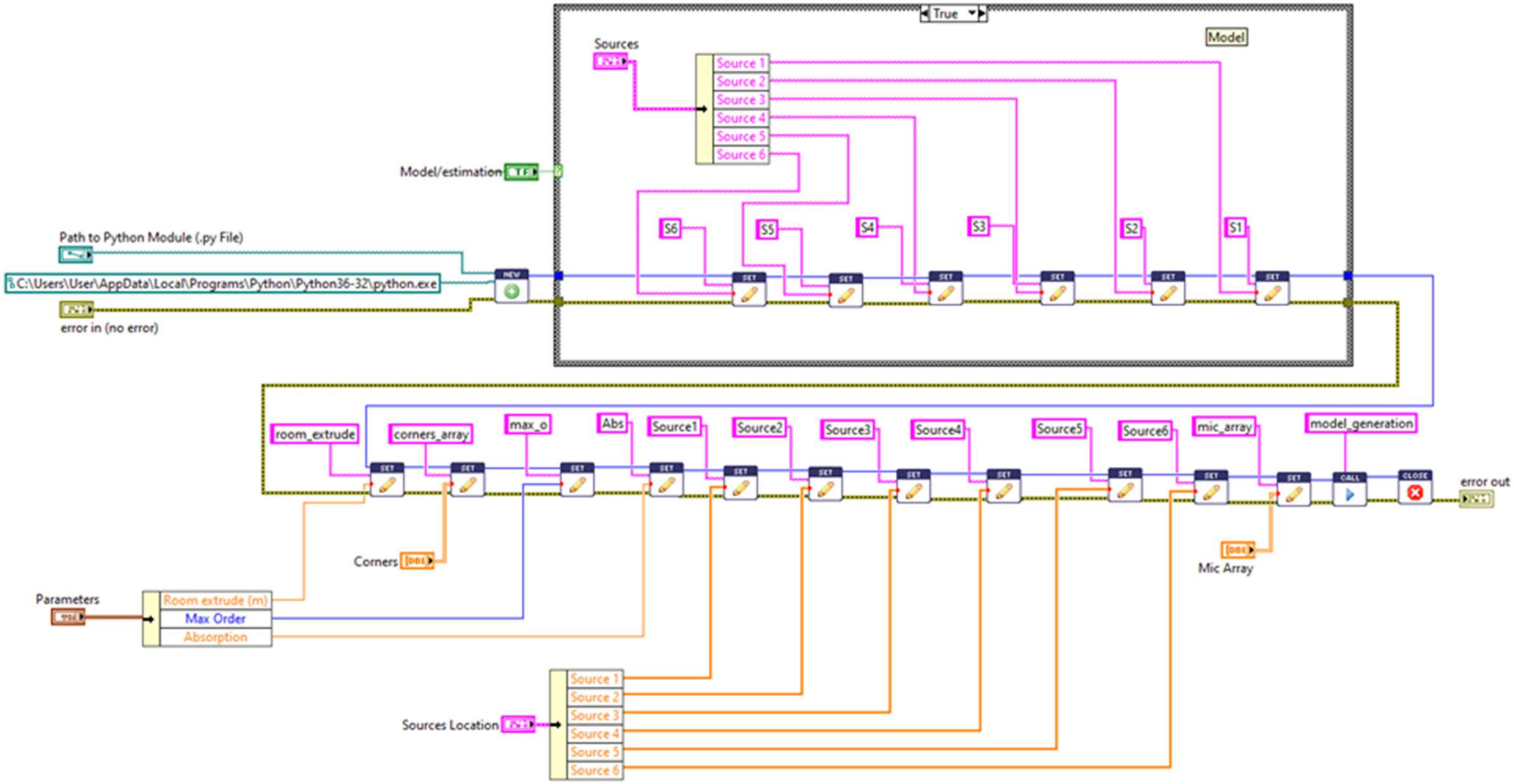

**Figure 54: Room Model Generator Block Diagram.**

## c) *Source Estimator*

This Sub- VI is too complex to display its source block diagram. Instead, a simple functional block diagram is shown.

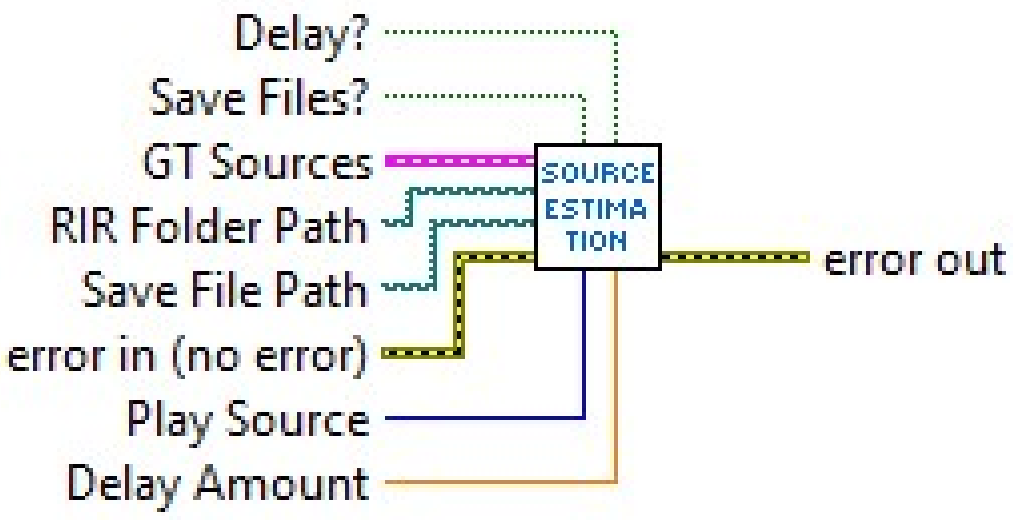

**Figure 55: Source Estimator Icon.**

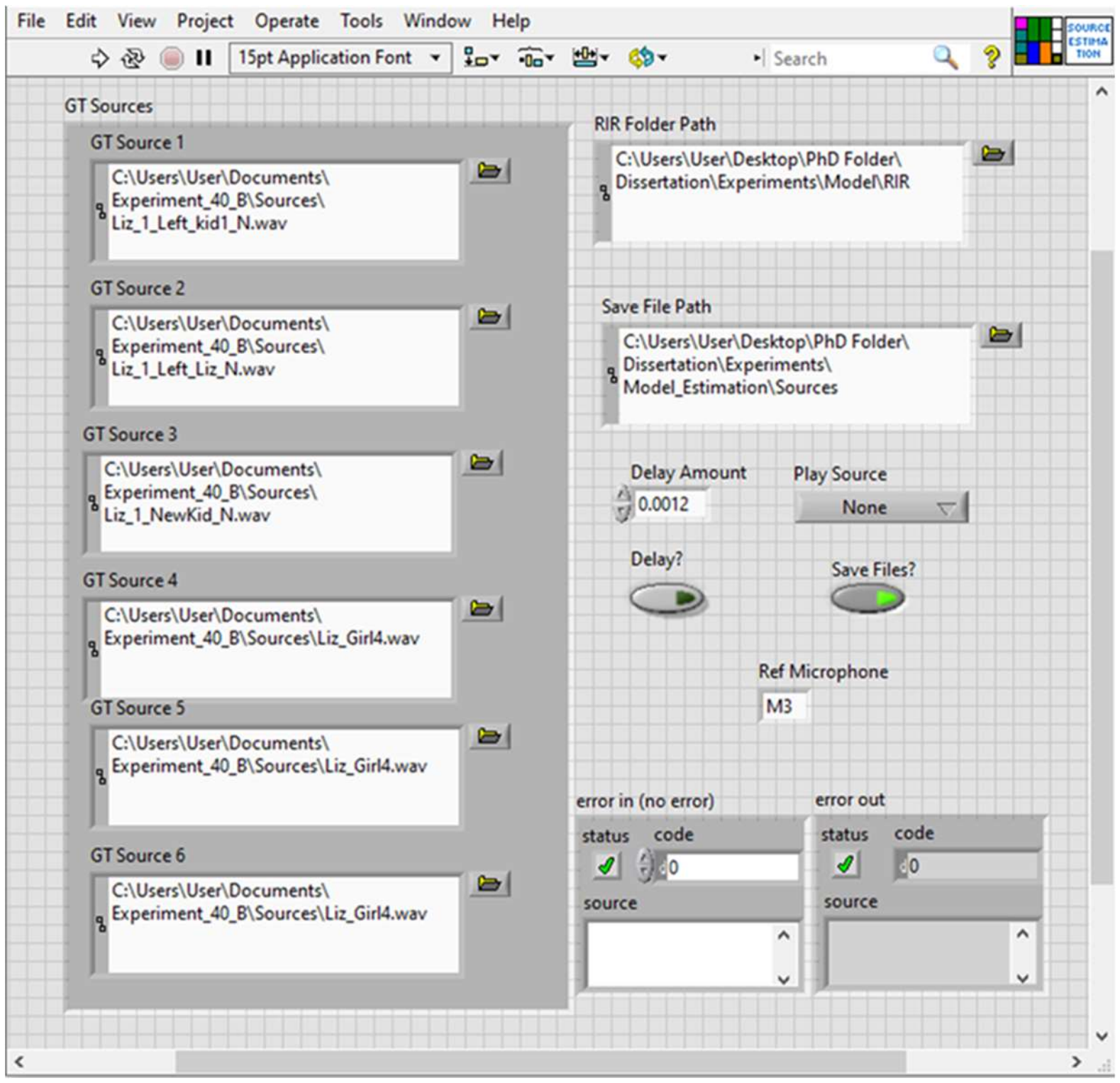

**Figure 56: Source Estimator Front Panel.**

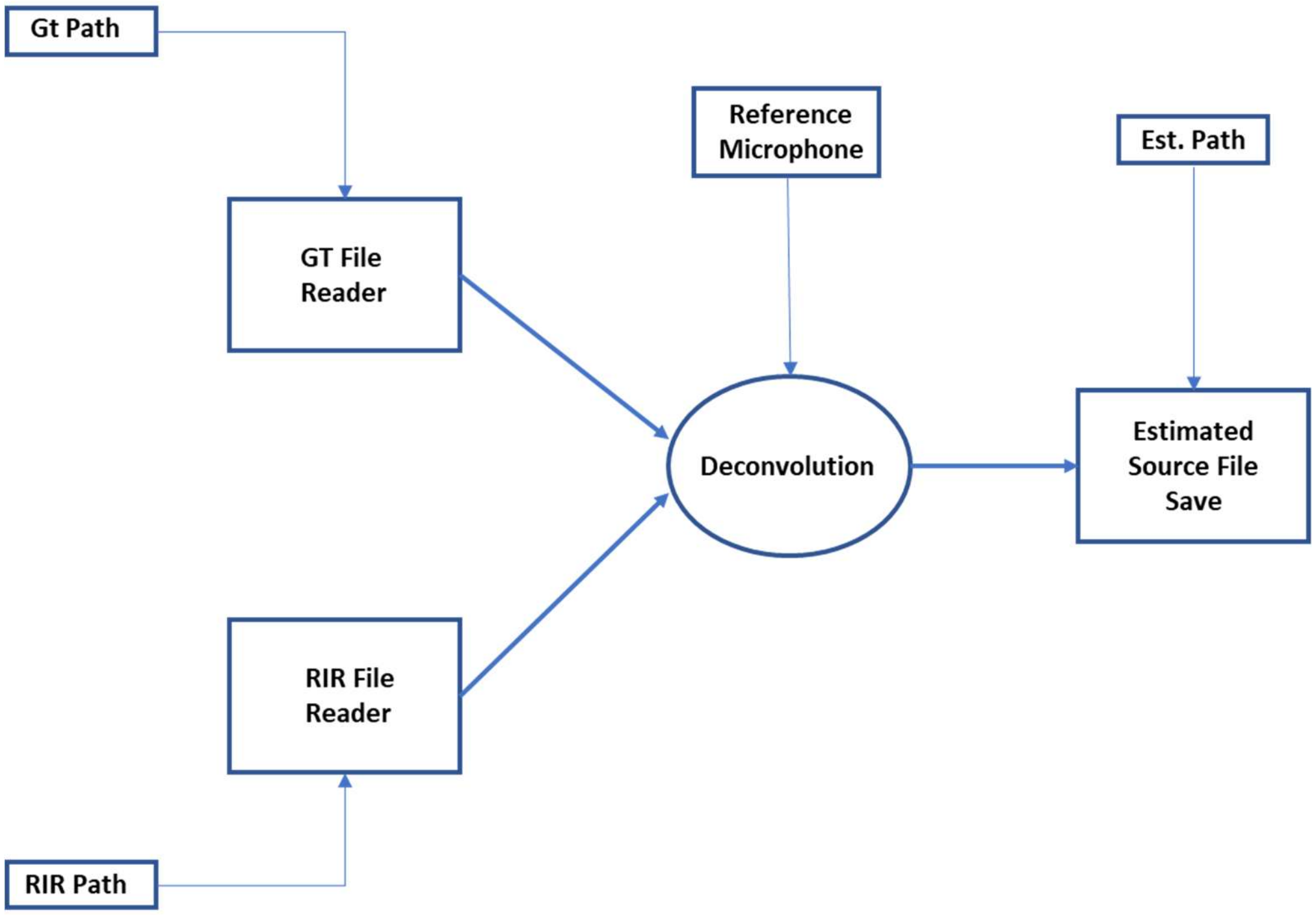

**Figure 57: Source Estimator Simplified Block Diagram.**

### *d) Cross-Correlation Model Calculator*

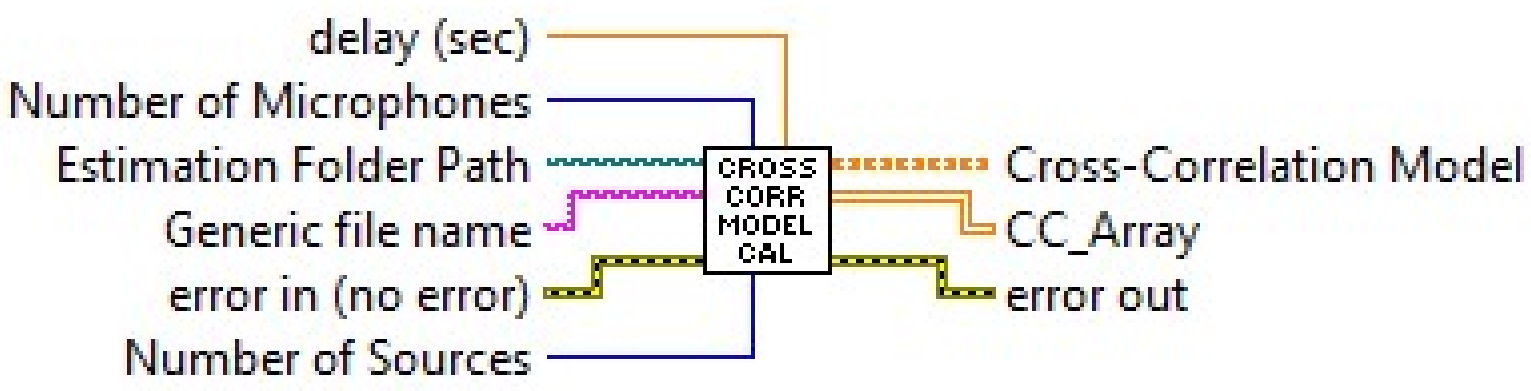

**Figure 58: Cross-Correlation Model Calculator Icon.**

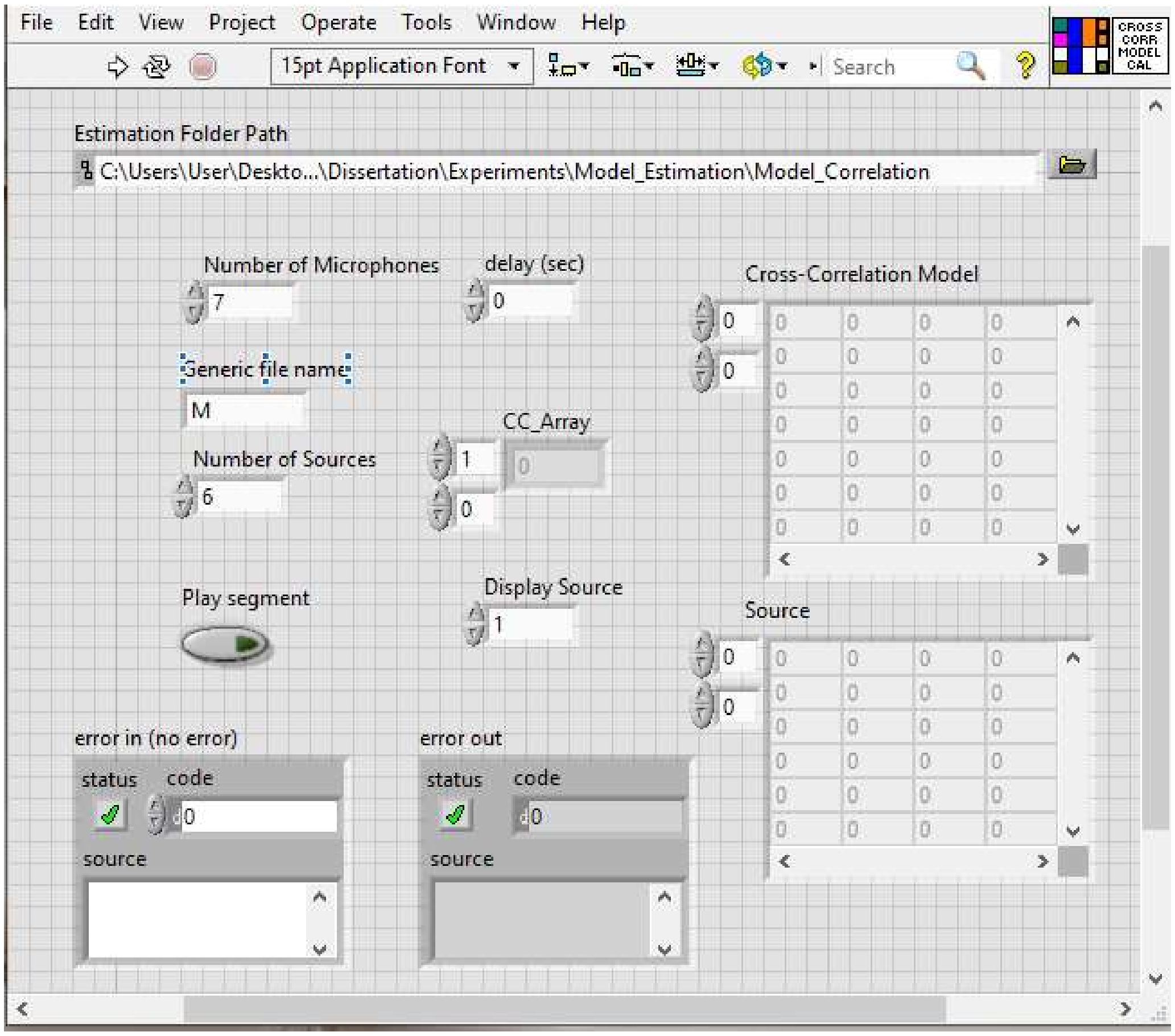

Figure 59: Cross-Correlation Model Calculator Front Panel.

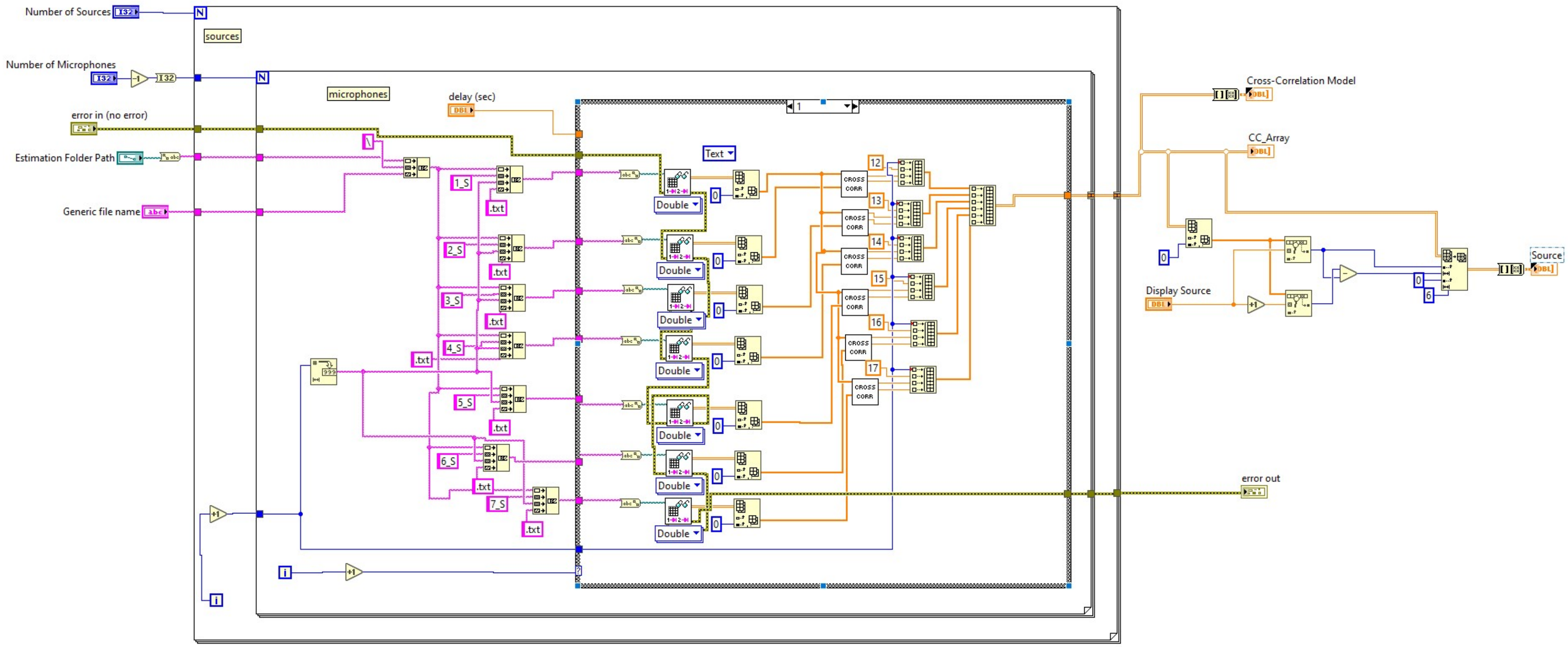

**Figure 60: Cross-Correlation Model Calculator Block Diagram.**

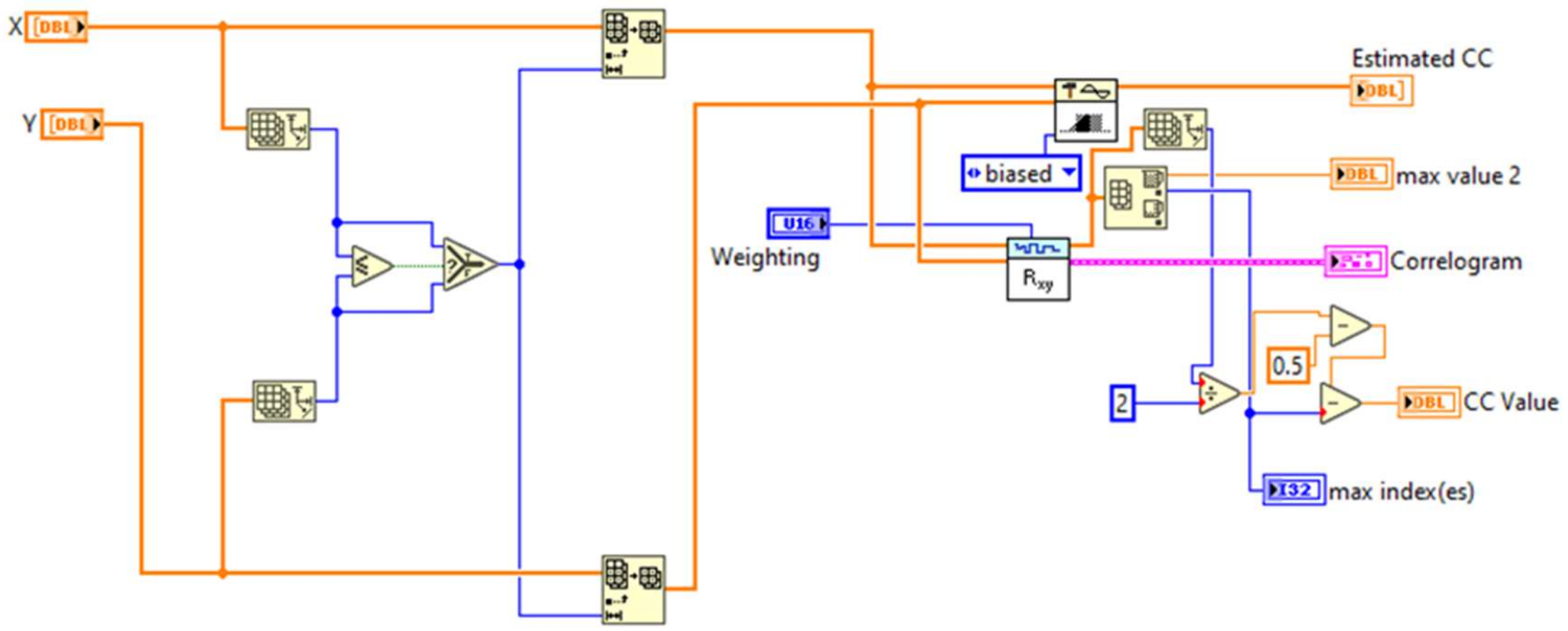

**Figure 61: Cross-Correlation Model Calculator. Cross-Correlator Sub-VI.**

Notes on this sub-VI: The cross-correlation results are indicated by the index where the max cross-correlation occur. This method makes the results independent of the sampling frequency. Also, this sub-VI truncates the largest input to make both input files the same size as the smallest one.

### *e) Model Classifier*

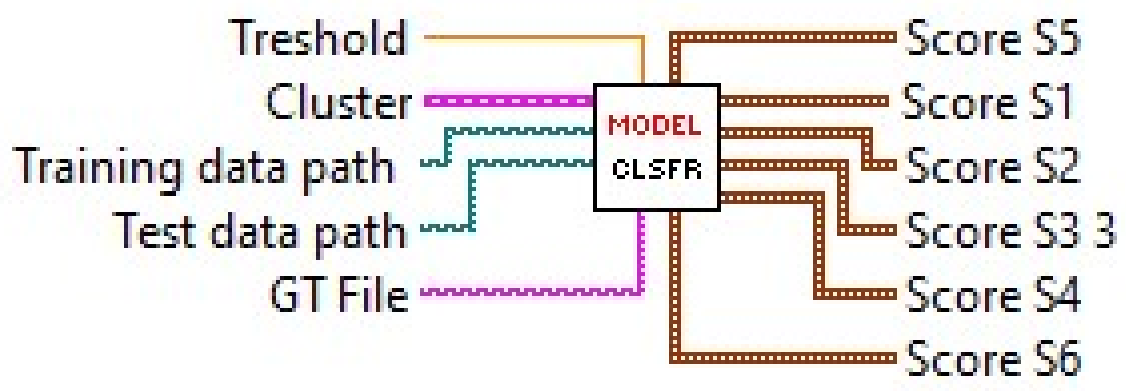

**Figure 62: Model Classifier Icon with Inputs and Outputs.**

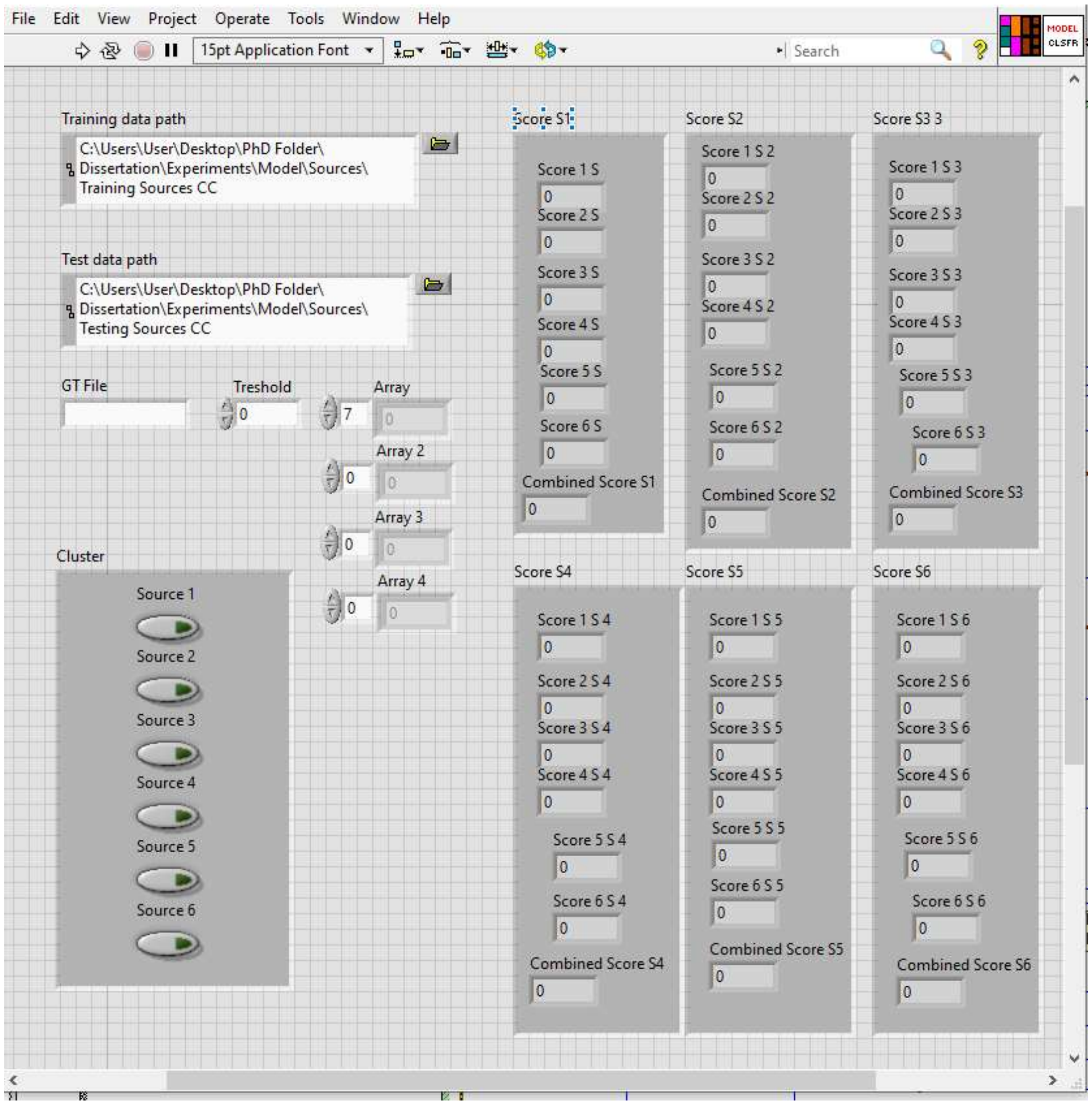

**Figure 63: Model Classifier Front Panel.**

### *f) Multi-Function Convolution and Correlator Visualizer*

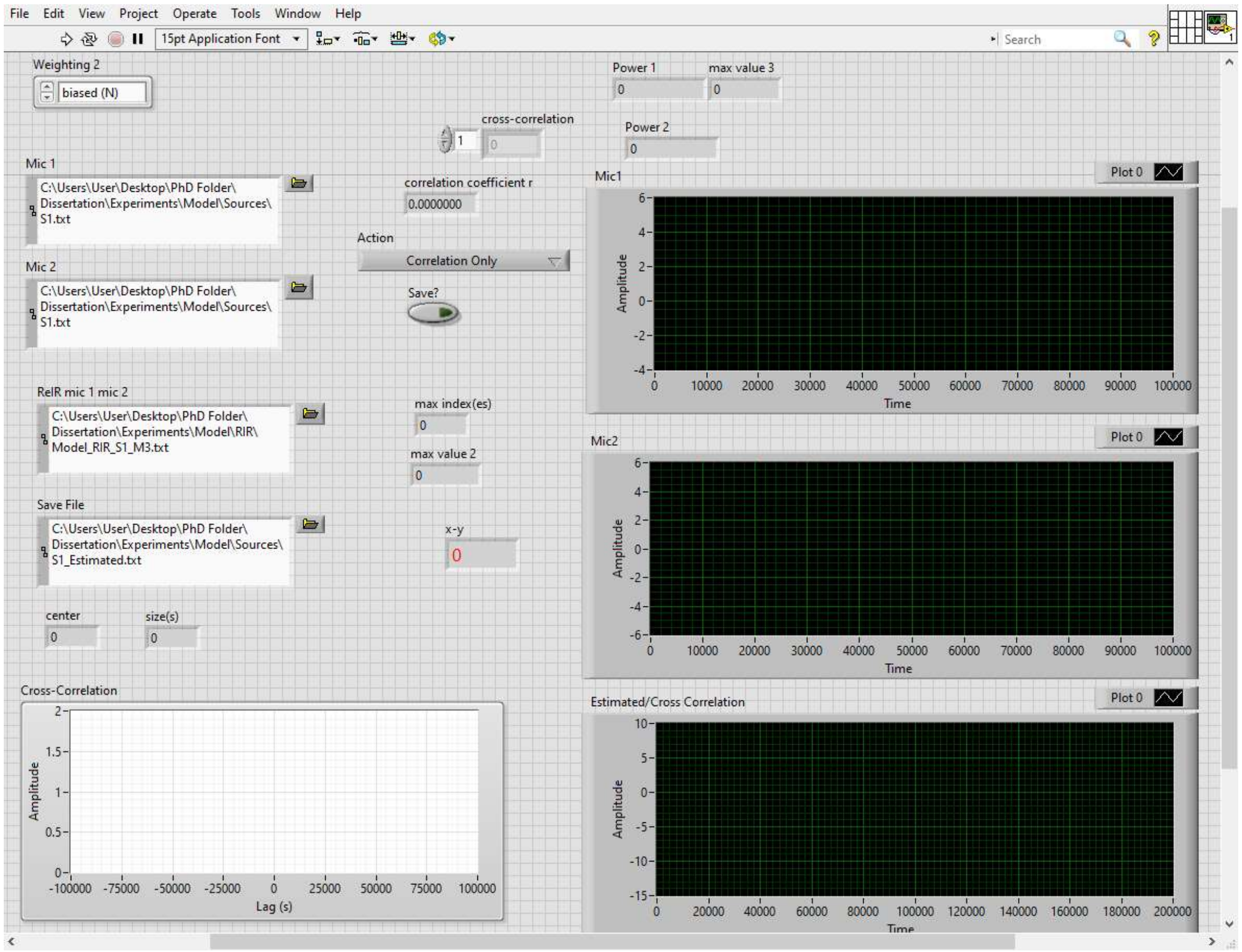

**Figure 64: Multi-Function Convolution and Correlator Visualizer Front Panel.**

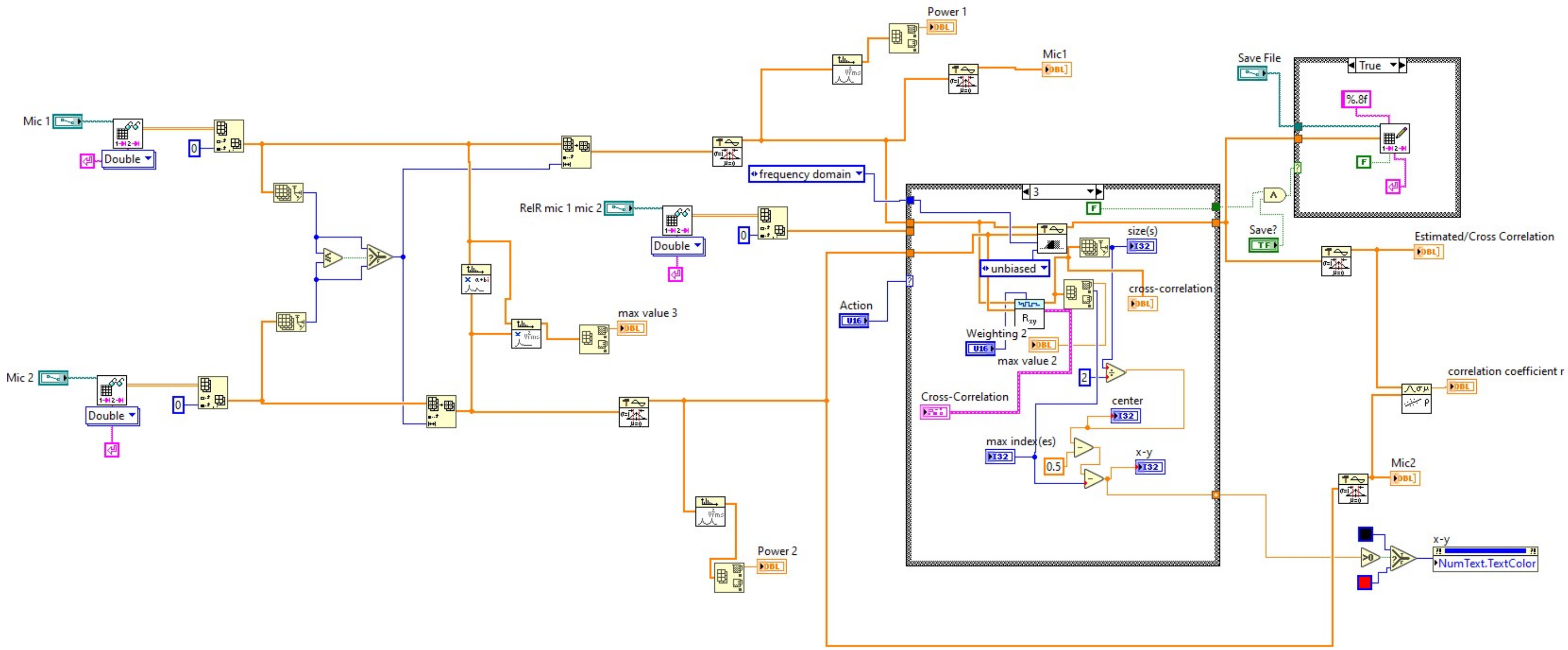

**Figure 65: Multi-Function Convolution and Correlator Visualizer Diagram.**

# Appendix C: Audio Lab Equipment Specifications

*a) Microphone Equipment:*

**Audio-Technica ATR3350xIs**

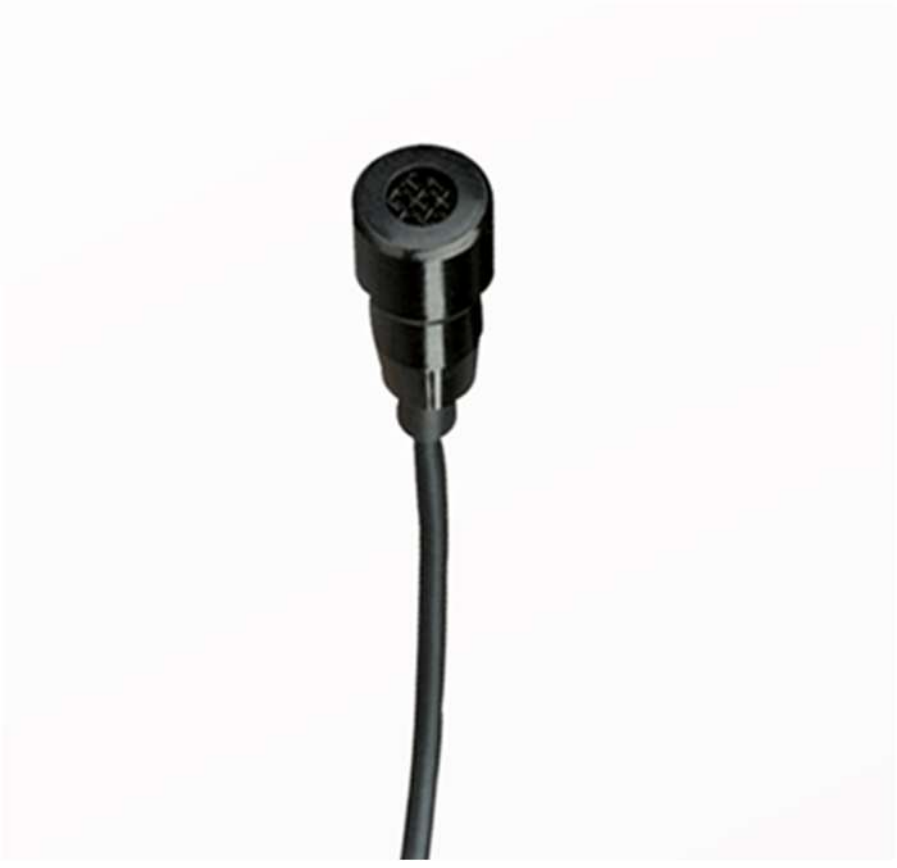

- Element: Condenser
- Polar Pattern: Omnidirectional
- Frequency Response: 50 – 18,000 Hz.
- Sensitivity: -54 db.
- Impedance: 1,000 ohms
- Power Source: Battery Type: LR44.

**Comica CVM-V020**

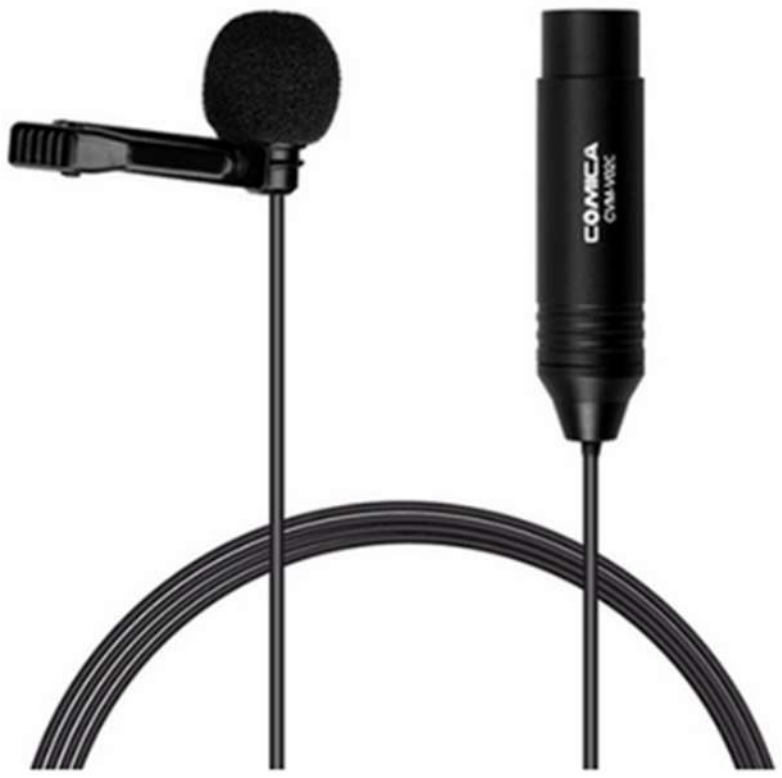

- Transducer: Back Electrets Condenser
- Directivity: Omnidirectional
- Frequency Range: 100Hz ~ 12KHz
- THD: ≤1%
- Sensitivity: 35dB ±3dB
- Signal/Noise Ratio: ≥60dB
- Power Source: 48V Phantom Powered

**Excelvan 700**

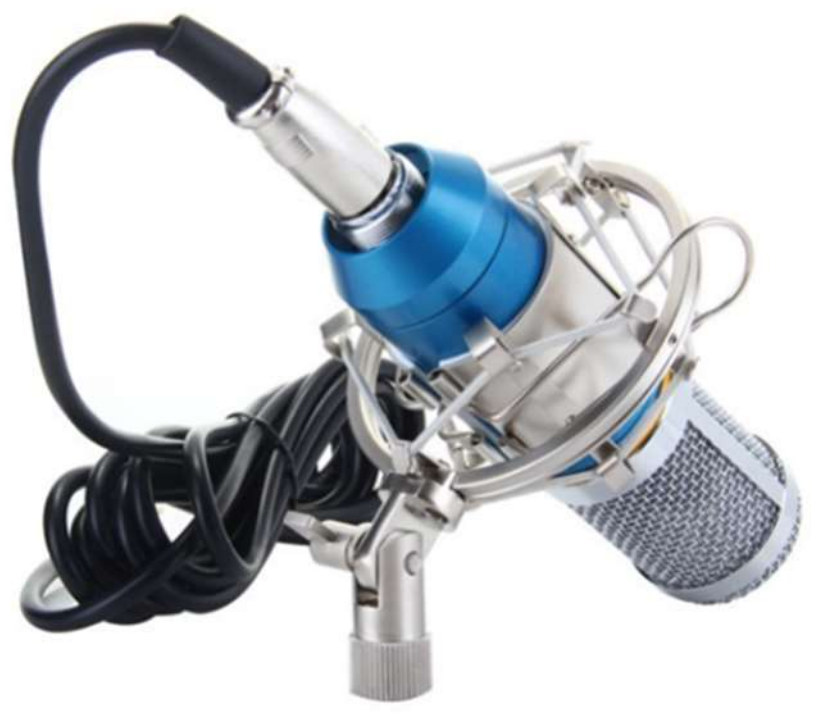

- Polar Pattern: Uni-directional
- Frequency Response: 20Hz-20kHz
- Sensitivity: 45dB±1dB
- Output Impedance:1500Ω±30%(at 1kHz)
- Load impedance: ≥1000 Ω
- Equivalent Noise level: 16dBA
- Power Source: 48V phantom power supply

*b) Audio Processing Equipment*

**TASCAM Model US-16x08**

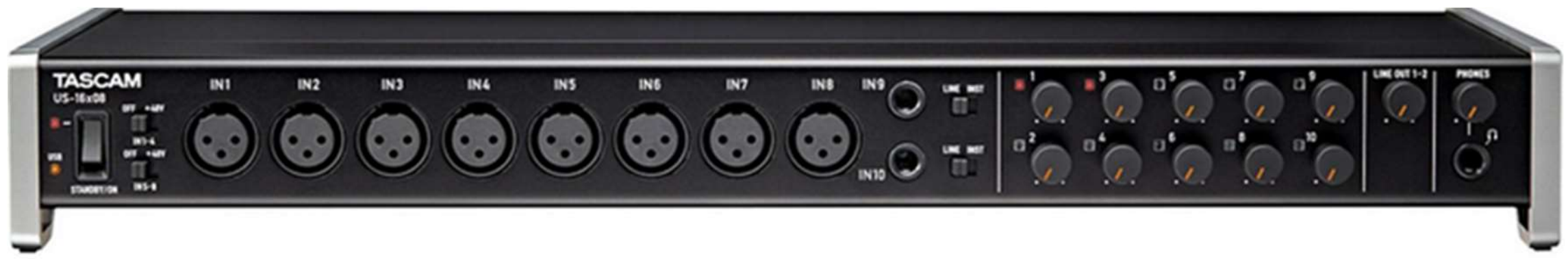

- Frequency response:
  - LINE OUT(BALANCED)
  - 44.1k/48k Hz  20Hz to 20kHz, ±0.3dB(JEITA)
  - 88.2k/96k Hz  20Hz to 40kHz, ±0.3dB(JEITA)
- THD   0.008% or less
- S/N ratio   100dB or more
- Crosstalk   100dB or more
- EIN   –125dBu or less
- Sampling frequency   44.1k/48k/88.2k/96k Hz
- Quantization bit rate   16/24-bit

- Analog audio inputs:
  - MIC IN(IN 1-8)
    - Connector XLR-3-31 (1: GND, 2: HOT, 3: COLD), BALANCED
    - Input impedance 2.4kΩ
    - Nominal input level
    - GAIN: MAX –68dBu (0.0003Vrms)
    - GAIN: MIN –12dBu (0.195Vrms)
    - Maximum input level +8dBu (1.947Vrms)
    - Gain 56dB
  - LINE IN (IN 9-10)
    - Connector 1/4" (6.3mm) TRS-jack (T: HOT, R: COLD, S: GND), BALANCED
    - Input impedance 10kΩ
    - Nominal input level
    - GAIN: MAX –41dBu (0.0069Vrms)
    - GAIN: MIN +4dBu (1.228Vrms)
    - Maximum input level +24dBu (12.182Vrms)
    - Gain 45dB

**AIWA Stereo Audio Amplifier**

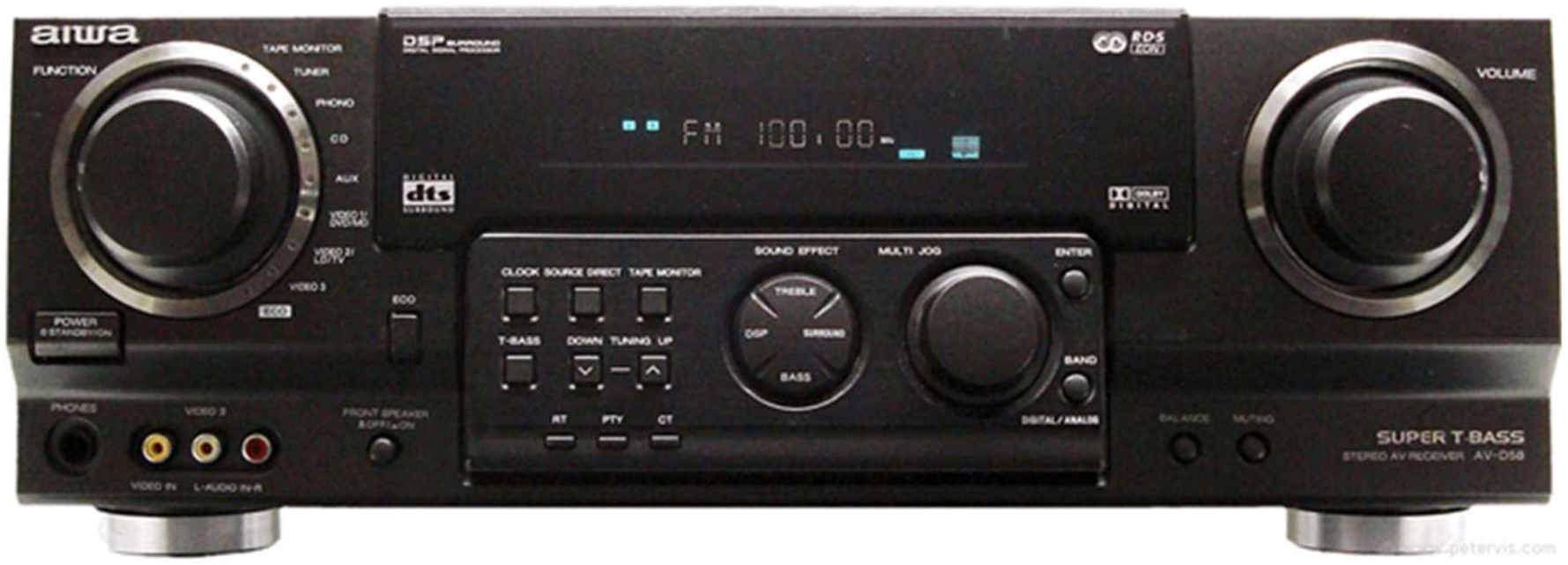

- Power output: 80 watts per channel into 8Ω (stereo)
- Surround output: 80W (front), 80W (center), 80W (rear)
- Frequency response: 20Hz to 20kHz
- Total harmonic distortion: 1%
- Input sensitivity: 2.5mV (MM), 300mV (line)
- Output: 300mV (line)

- Speaker load impedance: 8Ω (minimum)

c) ***Loudspeakers***

**Polk Audio RM6751**

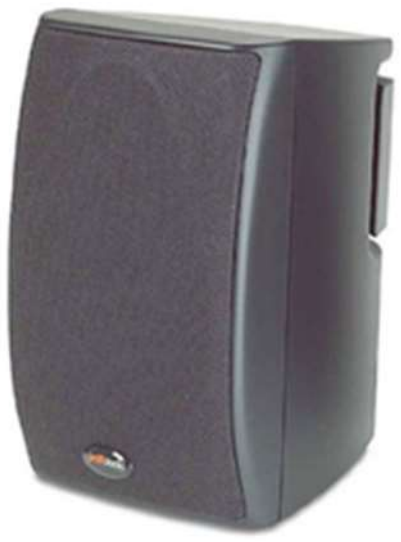

- Power Range: 20- 100 W
- Frequency Response: 40 Hz – 24 kHz
- Sensitivity: 89 @2.83Vrms dB
- Impedance (Ohms): 8